\documentclass[useAMS,usenatbib]{mn2e}

\usepackage{amsmath}
\usepackage{amssymb}
\usepackage{graphicx}
\usepackage{graphics}
\usepackage{float}
\usepackage{epstopdf}
\usepackage{multicol}
\usepackage{breakurl}
\usepackage{textcomp}
\usepackage{enumitem}
\usepackage{subfigure}
\usepackage[bf,labelsep=period,font=small]{caption}
\usepackage{subfloat}
\usepackage{subfigure}
\usepackage{gensymb}
\usepackage{supertabular}
\usepackage{longtable}
\usepackage[flushleft]{threeparttable}
\title[HD~35502's magnetic B5IVpe primary]{HD~35502: a hierarchical triple system 
with a magnetic B5IVpe primary\thanks{Based 
on spectropolarimetric observations obtained at the Canada-France-Hawaii Telescope 
(CFHT) which is operated by the National Research Council of Canada, the Institut 
National des Sciences de l'Univers (INSU) of the Centre National de la Recherche 
Scientifique of France, and the University of Hawaii, observations obtained using the 
Narval spectropolarimeter at the Observatoire du Pic du Midi (France), which is operated 
by the INSU, and observations obtained at the Dominion Astrophysical 
Observatory, NRC Herzberg, Programs in Astronomy and Astrophysics, National Research 
Council of Canada.}}
\author[J. Sikora et al.]
	{J.~Sikora,$^{1,2}$ G.~A.~Wade,$^{2}$ D.~A.~Bohlender,$^{3}$ M.~Shultz,$^{1,2}$ S.~J.~Adelman,$^{4}$
	\newauthor
	E.~Alecian,$^{6,7}$ D.~Hanes,$^{1}$ D.~Monin,$^{3}$ C.~Neiner,$^{5}$ and the MiMeS and
	\newauthor
	BinaMIcS Collaborations\\
$^{1}$Department of Physics, Engineering Physics \& Astronomy, Queen's University, Kingston, ON Canada, K7L 3N6\\
$^{2}$Department of Physics, Royal Military College of Canada, PO Box 17000 Kingston, Ontario K7K 7B4, Canada\\
$^{3}$Herzberg Astronomy and Astrophysics, National Research Council of Canada, 5071 West Saanich Road, Victoria, BC V9E 2E7, \\\hspace{2.5mm}Canada\\
$^{4}$Department of Physics, The Citadel, 171 Moultrie Street, Charleston, SC 29409, USA\\
$^{5}$LESIA, Observatoire de Paris, CNRS UMR 8109, UPMC, Universit\'{e} Paris Diderot, 5 place Jules Janssen, 92195 Meudon Cedex, \\\hspace{2.5mm}France\\
$^{6}$LESIA, Observatoire de Paris, PSL Research University, CNRS, Sorbonne Universit\'es, UPMC Univ. Paris 06, Univ. Paris Diderot, \\\hspace{2.5mm}Sorbonne Paris Cit\'e, 5 place Jules Janssen, 92195 Meudon, France\\
$^{7}$UJF-Grenoble 1/CNRS-INSU, Institut de Plan\'{e}tologie et d'Astrophysique de Grenoble (IPAG) UMR 5274, 38041 Grenoble, France}

\begin{document}

\date{Submitted 2015 Month XX. }

\pagerange{\pageref{firstpage}--\pageref{lastpage}} \pubyear{2002}

\maketitle

\label{firstpage}

\begin{abstract}
We present our analysis of HD~35502 based on high- and medium-resolution 
spectropolarimetric observations. Our results indicate that the magnetic B5IVsnp star 
is the primary component of a spectroscopic triple system and that it has an effective 
temperature of $18.4\pm0.6\,{\rm kK}$, a mass of $5.7\pm0.6\,M_\odot$, and a 
polar radius of $3.0^{+1.1}_{-0.5}\,R_\odot$. The two secondary components are found to 
be essentially identical A-type stars for which we derive effective temperatures 
($8.9\pm0.3\,{\rm kK}$), masses ($2.1\pm0.2\,M_\odot$), and radii 
($2.1\pm0.4\,R_\odot$). We infer a hierarchical orbital configuration for the system 
in which the secondary components form a tight binary with an orbital period of 
$5.66866(6)\,{\rm d}$ that orbits the primary component with a period of over 
$40\,{\rm yrs}$. Least-Squares Deconvolution (LSD) profiles reveal Zeeman 
signatures in Stokes $V$ indicative of a longitudinal magnetic field produced by the B 
star ranging from approximately $-4$ to $0\,{\rm kG}$ with a median uncertainty of 
$0.4\,{\rm kG}$. These measurements, along with the line variability produced by 
strong emission in H$\alpha$, are used to derive a rotational period of 
$0.853807(3)\,{\rm d}$. We find that the measured $v\sin{i}=75\pm5\,{\rm km\,s}^{-1}$ of the 
B star then implies an inclination angle of the star's rotation axis to the line of 
sight of $24^{+6}_{-10}\degree$. Assuming the Oblique Rotator 
Model, we derive the magnetic field strength of the B star's dipolar component 
($14^{+9}_{-3}\,{\rm kG}$) and its obliquity ($63\pm13\degree$). Furthermore, 
we demonstrate that the calculated Alfv\'{e}n radius ($41^{+17}_{-6}\,R_\ast$) and 
Kepler radius ($2.1^{+0.4}_{-0.7}\,R_\ast$) place HD~35502's central B star well 
within the regime of centrifugal magnetosphere-hosting stars.
\end{abstract}


\section{Introduction}

Magnetic B-type stars exhibiting strong emission \citep[e.g. $\sigma$ Ori E, HD 142184, 
HD 182180,][]{Landstreet1978,Grunhut2012b,Rivinius2013} serve as important testbeds 
for understanding how stellar winds interact with magnetic fields. Models such 
as the Rigidly Rotating Magnetosphere (RRM) model \citep{Townsend2005} provide a 
qualitative description of these systems \citep[e.g.][]{Townsend2005a,Krticka2009,
Oksala2010}; however, detailed comparisons with observations of $\sigma$ Ori E have 
uncovered important discrepancies which require explanations \citep{Oksala2012,
Oksala2015}. By relaxing the RRM model's condition that the magnetic field remains 
undistorted, \citet{Townsend2005} proposed the centrifugal breakout scenario in which 
the field loops episodically break and reconnect in response to an accumulating 
magnetospheric mass. Although magnetohydrodynamic simulations support this hypothesis 
\citep{ud-Doula2006_oth}, no observational evidence of the breakout events (e.g. 
optical flares) has yet been reported \citep{Townsend2013}.

While these tests of the current theoretical framework provide useful information, 
their conclusions are based on a relatively small number of case studies. Lately, this 
number has been increasing as demonstrated by the recent confirmation of HD 23478's 
centrifugal magnetosphere (CM) \citep{Sikora2015}, as well as the discovery of the 
candidate CM-host, HD 345439 \citep{Hubrig2015}. The latest addition to this 
particular subset of magnetic B-type stars, HD~35502, is the focus of this paper.

Over the past 60 years, the nature of HD~35502 has been redefined in various ways. 
Located within the Orion OB1 association \citep[likely within the OB1a 
subgroup,][]{Landstreet2007}, it was initially identified as a B5V star 
\citep{Sharpless1952, Crawford1958}. Higher resolution spectra later obtained by 
\citet{Abt1962} revealed both narrow and broad spectral lines, the latter of which 
being characterized with a $v\sin{i}$ of $290\,{\rm km\,s}^{-1}$. Moreover, He~{\sc i} lines 
were reported to be relatively weak; an analysis of early-type stars within Ori OB1 
carried out by \citet{Nissen1976} demonstrated that HD~35502's He abundance was 
approximately half that of the nearby chemically normal field stars. These results 
motivated its eventual reclassification as a B5IVsnp star \citep{Abt1977}.

HD~35502's magnetic field was first detected by \citet{Borra1981} and later 
confirmed by subsequent studies \citep{Bychkov2005,Glagolevskij2010}. Following the 
initial detection, it had been suggested that some of the unusual features apparent in 
its spectrum may be related to this strong field. In this paper, we use high-resolution 
spectra to provide a new interpretation of HD~35502 as a spectroscopic triple system 
whose primary component is a magnetic B-type star hosting a centrifugally supported 
magnetosphere.

In Section~\ref{sect_obs}, we discuss both the polarized and unpolarized 
spectroscopic observations used in this study. Section~\ref{sect_phys_param} 
focuses on our derivation of some of the physical parameters of the system including 
its orbital configuration, along with the effective temperatures, surface 
gravities, masses, radii, and projected rotational velocities of the three stellar 
components. The various analytical methods used to derive these parameters, such as the 
modelling of spectroscopic and photometric data, are also described. In Section 
\ref{P_rot} we discuss the evidence of rotational modulation from which we derive 
the rotational period of HD~35502's primary component. In Section~\ref{mag_field}, 
the magnetic field measurements of this component are derived along with the field 
geometry and strength. In Section~\ref{variability}, we discuss and characterize the 
magnetic B star's magnetosphere. Finally, our conclusions along with our 
recommendations for further analytical work to be carried out are summarized in Section 
\ref{conclusions}.

\begin{table*}
	\caption{ESPaDOnS and Narval spectropolarimetric observations. The SNRs per 
	$1.8\,\text{km/s}$ pixel are reported at $5400\,{\rm \AA}$. The fifth, sixth, 
	and seventh columns list the radial velocities of the three stellar components (see 
	Section~\ref{orb_sol}). The two right-most columns indicate the longitudinal 
	magnetic field derived from H$\beta$ (see Section~\ref{mag_field}) along with the 
	associated detection status: definite detection (DD), marginal detection (MD), and 
	no detection (ND) as outlined by \citet{Donati1997}.}
	\label{obs_tbl}
	\begin{center}
	\begin{tabular*}{1.91\columnwidth}{@{\extracolsep{\fill}}l c c c r r r r r}
		\hline
		\hline
		\noalign{\vskip0.5mm}
		HJD & Total Exp.    & SNR 		    & Instrument & \multicolumn{1}{c}{$v_{r,B}$}            & \multicolumn{1}{c}{$v_{r,A_1}$}          & \multicolumn{1}{c}{$v_{r,A_2}$}          & \multicolumn{1}{c}{$\langle B_z\rangle_{{\rm H}\beta}$} & Detection \\
			& Time (s)		& (pix$^{-1}$)	&			 & \multicolumn{1}{c}{$({\rm km\,s}^{-1})$} & \multicolumn{1}{c}{$({\rm km\,s}^{-1})$} & \multicolumn{1}{c}{$({\rm km\,s}^{-1})$} & \multicolumn{1}{c}{(kG)}			  		            & Status	\\
		\noalign{\vskip0.5mm}
		\hline	
		\noalign{\vskip0.5mm}
		2454702.138 & 1800 & 677 & ESPaDOnS & $19.4\pm1.3$ & $ 55.9\pm1.4$ & $  4.8\pm2.7$ & $-0.23\pm0.19$ & DD \\
        2455849.677 & 3600 & 662 & Narval   & $20.0\pm1.3$ & $-15.3\pm2.6$ & $ 71.3\pm1.7$ & $-0.03\pm0.16$ & DD \\
		2455893.623 & 3600 & 522 & Narval   & $21.2\pm1.4$ & $ -5.1\pm1.5$ & $ 61.3\pm2.9$ & $-2.05\pm0.20$ & DD \\
        2455910.518 & 3600 & 263 & Narval   & $22.7\pm2.0$ & $  0.7\pm2.0$ & $ 55.9\pm5.3$ & $-1.30\pm0.50$ & DD \\
		2455934.528 & 3600 & 587 & Narval   & $19.6\pm1.5$ & $-20.9\pm1.1$ & $ 76.4\pm2.0$ & $-2.12\pm0.19$ & DD \\
		2455936.534 & 3600 & 472 & Narval   & $18.6\pm1.4$ & $ 76.5\pm1.6$ & $-19.4\pm2.8$ & $-2.32\pm0.24$ & DD \\
		2455938.525 & 3600 & 564 & Narval   & $20.4\pm1.5$ & $ 18.1\pm5.5$ & $ 37.0\pm5.9$ & $ 0.08\pm0.19$ & ND \\
		2455944.500 & 3600 & 494 & Narval   & $20.9\pm1.4$ & $  1.9\pm2.5$ & $ 54.0\pm2.9$ & $ 0.16\pm0.23$ & ND \\
		2455949.429 & 3600 & 545 & Narval   & $20.5\pm1.3$ & $ 45.4\pm1.1$ & $ 11.2\pm2.5$ & $-0.95\pm0.20$ & DD \\
		2455950.472 & 3600 & 478 & Narval   & $21.4\pm1.8$ & $-12.5\pm2.4$ & $ 68.1\pm2.9$ & $-0.18\pm0.23$ & ND \\
		2455951.471 & 3600 & 431 & Narval   & $22.8\pm1.5$ & $-22.6\pm1.0$ & $ 77.7\pm2.0$ & $-1.00\pm0.27$ & DD \\
        2455966.376 & 3600 & 550 & Narval   & $19.8\pm1.5$ & $ 48.6\pm1.5$ & $  7.5\pm3.2$ & $-1.87\pm0.20$ & DD \\
        2455998.332 & 3600 & 397 & Narval   & $20.9\pm1.4$ & $ 53.6\pm2.6$ & $  3.4\pm3.4$ & $ 0.28\pm0.28$ & ND \\
		2455999.362 & 3600 & 402 & Narval   & $20.7\pm1.9$ & $ 81.8\pm2.9$ & $-25.2\pm1.0$ & $-1.64\pm0.28$ & DD \\
		2456001.309 & 3600 & 523 & Narval   & $20.0\pm1.5$ & $ -4.6\pm2.4$ & $ 60.0\pm1.6$ & $-2.39\pm0.23$ & DD \\
		2456003.329 & 3600 & 528 & Narval   & $21.0\pm1.6$ & $ 13.5\pm2.2$ & $ 42.8\pm1.9$ & $ 0.51\pm0.21$ & DD \\
		2456202.665 & 3600 & 604 & Narval   & $21.4\pm1.3$ & $ 65.4\pm3.4$ & $ -9.4\pm1.9$ & $-2.30\pm0.17$ & DD \\
		2456205.618 & 3600 & 494 & Narval   & $21.1\pm1.5$ & $-15.1\pm2.3$ & $ 69.5\pm2.8$ & $-0.34\pm0.22$ & DD \\
		2456224.646 & 3600 & 450 & Narval   & $23.0\pm1.3$ & $ 27.7\pm1.4$ & $ 27.7\pm1.4$ & $-0.52\pm0.24$ & ND \\
		2456246.505 & 3600 & 505 & Narval   & $20.5\pm1.3$ & $-15.6\pm2.6$ & $ 69.8\pm1.5$ & $-1.72\pm0.22$ & DD \\
		2456293.881 & 1600 & 710 & ESPaDOnS & $22.6\pm1.4$ & $ 79.4\pm2.5$ & $-24.9\pm0.9$ & $-1.39\pm0.20$ & DD \\
		2456295.787 & 1600 & 190 & ESPaDOnS & $21.0\pm2.6$ & $ 10.4\pm1.6$ & $ 44.1\pm1.4$ & $-3.06\pm0.71$ & ND \\
		2456295.808 & 1600 & 231 & ESPaDOnS & $22.8\pm2.3$ & $  9.3\pm1.3$ & $ 45.2\pm1.9$ & $-2.83\pm0.57$ & MD \\
		2456556.002 & 1600 & 582 & ESPaDOnS & $24.7\pm1.9$ & $ 43.0\pm1.9$ & $ 10.8\pm1.8$ & $-1.37\pm0.28$ & DD \\
		2456557.140 & 1600 & 670 & ESPaDOnS & $19.2\pm1.5$ & $-18.2\pm1.9$ & $ 70.9\pm1.6$ & $-3.27\pm0.20$ & DD \\
		2456560.077 & 1600 & 612 & ESPaDOnS & $20.8\pm1.4$ & $ 74.5\pm1.9$ & $-20.5\pm2.1$ & $-0.05\pm0.25$ & ND \\										
		\noalign{\vskip0.5mm}
		\hline \\
	\end{tabular*}
	\end{center}
\end{table*}

\section{observations}
\label{sect_obs}

\subsection{ESPaDOnS \& Narval spectropolarimetry}

Spectropolarimetric observations of HD~35502 were obtained over the course of 5 
years (Aug. 23, 2008 to Sept. 24, 2013) in the context of the MiMeS \citep{Wade2015} 
and BinaMIcS \citep{Alecian2015} surveys. Nineteen Stokes $V$ observations were 
obtained using the high-resolution ($R\simeq65\,000$) spectropolarimeter Narval 
installed at the T\'{e}lescope Bernard Lyot (TBL) over a wavelength range of 
approximately $3\,600-10\,000\,{\rm \AA}$. Ten Stokes $V$ spectra were also obtained 
using the twin instrument ESPaDOnS installed at the Canada-France-Hawaii Telescope 
(CFHT). Three of these observations exhibited signal-to-noise ratios (SNRs) 
$\lesssim100$ and were removed from the analysis. A median SNR of $522$ was obtained 
from the twenty six observations. Both the ESPaDOnS and Narval observations were 
reduced using the {\sc Libre-ESpRIT} pipeline \citep{Donati1997} yielding final Stokes 
$I$ and $V$ spectra \citep[for a detailed description of the reduction procedure, see 
e.g.][]{Silvester2012}. The Heliocentric Julian Dates (HJDs), total exposure times, and 
SNRs are listed in Table~\ref{obs_tbl}.

\subsection{dimaPol spectropolarimetry}

Twenty-four medium-resolution spectropolarimetric observations were obtained with 
dimaPol ($R\simeq10\,000$) installed at the Dominion Astrophysical Observatory (DAO) 
\citep{Monin2012} from Feb. 7, 2009 to Feb. 15, 2012. Two of these observations had 
SNRs $\lesssim100$ and were removed from the analysis. The remaining twenty-two Stokes 
$V$ observations of H$\beta$ were used to derive longitudinal field measurements; the 
HJDs, exposure times, SNRs, and longitudinal field measurements are listed in Table 
\ref{dao_tbl}.

\subsection{FEROS spectroscopy}

Thirty-two unpolarized spectra were acquired from Dec. 30, 2013 to Jan. 3, 2014 using 
the spectrograph FEROS mounted on the $2.2\,{\rm m}$ MPG/ESO telescope located at La 
Silla Observatory. The instrument has a resolving power of $R=48\,000$ across a 
wavelength range of $3\,600-9\,200\,\rm{\AA}$ \citep{Kaufer1999}. The spectra were 
reduced using the FEROS Data Reduction System. The pipeline automatically carries out 
bias subtraction, flat fielding, and extraction of the spectral orders; wavelength 
calibration is carried out using ThAr and ThArNe lamps. Uncertainties in the measured 
intensities were estimated from the root mean square (RMS) of the continuum intensity 
at multiple points throughout each spectrum \citep[e.g.][]{Wade2012b}. The HJDs, total 
exposure times, and SNRs are listed in Table~\ref{FEROS_tbl}.

\subsection{H$\alpha$ spectroscopy}

A total of 131 spectroscopic observations of H$\alpha$ covering various wavelength 
ranges from approximately $6\,300-6\,800\,{\rm \AA}$ are also used in this study. One 
hundred thirteen of these observations were obtained at the DAO from Nov. 26, 1991 to 
Feb. 4, 2012. Seven of the spectra were removed from the analysis on account of 
their SNRs being $\lesssim50$. Both the McKellar spectrograph installed at the $1.2\,{\rm m}$ Plaskett 
telescope and the spectrograph mounted at the Cassegrain focus of DAO's $1.8\,{\rm m}$ 
telescope were used to acquire the spectra.

The remaining eleven observations were obtained at CFHT from Nov. 21, 1991 to Oct. 
3, 1995 using the now decommissioned Coud\'{e} f/8.2 spectrograph.

\subsection{$uvby$ photometry}

149 $uvby$ photometric measurements were obtained from Jan. 27, 1992 to Mar. 13, 1994 
using the $0.75\,{\rm m}$ Four College Automated Photoelectric Telescope (FCAPT) on Mt. 
Hopkins, AZ. The dark count was first measured and then in each filter the 
sky-ch-c-v-c-v-c-v-c-ch-sky counts were obtained, where sky is a reading of the sky, 
ch that of the check star, c that of the comparison star, and v that of the variable 
star. No corrections have been made for neutral density filter differences among each 
group of variable, comparison, and check stars. HD~35575 was the comparison and 
HD~35008 the check (i.e. second comparison) star. The standard deviations of the 
ch-c values were 0.006 mag, except for u for which it was 0.008 mag. We adopted 
uncertainties of 0.005 mag for each measurement based on the highest precision 
typically achieved with FCAPT. Table~\ref{uvby_tbl} contains the complete list of 
photometry.

\begin{table}
	\caption{Spectropolarimetric observations of HD~35502 obtained with dimaPol. 
	Columns $1$ to $3$ list the HJDs, exposure times, and SNRs. Column $4$ lists the 
	longitudinal field measurements derived from the H$\beta$ Stokes $V$ profiles.}
	\label{dao_tbl}
	\begin{center}
	\begin{tabular*}{0.39\textwidth}{@{\extracolsep{\fill}}l c c r}
		\noalign{\vskip-0.2cm}
		\hline
		\hline
		\noalign{\vskip0.5mm}
		HJD & Total Exp. & SNR 		    & $\langle B_z\rangle_{\rm H\beta}$\\
		    & Time (s)	 & (pix$^{-1}$) & (kG) \\
		\noalign{\vskip0.5mm}
		\hline
		\noalign{\vskip0.5mm}
		2454869.764 & 4800 & 410 & $-1.95\pm0.27$ \\
        2454872.773 & 3600 & 240 & $-0.64\pm0.25$ \\
        2455109.047 & 5400 & 240 & $-2.88\pm0.51$ \\
        2455110.964 & 5400 & 260 & $-1.60\pm0.25$ \\
        2455167.855 & 6000 & 290 & $-2.27\pm0.21$ \\
        2455168.829 & 6000 & 320 & $-2.21\pm0.38$ \\
        2455169.881 & 7200 & 280 & $-1.72\pm0.43$ \\
        2455170.867 & 7200 & 190 & $ 0.31\pm0.31$ \\
        2455190.766 & 7200 & 230 & $-2.57\pm0.33$ \\
        2455191.781 & 7200 & 260 & $-2.29\pm0.30$ \\
        2455192.790 & 7200 & 130 & $-3.95\pm1.09$ \\
        2455193.742 & 7200 & 210 & $-1.61\pm0.33$ \\
        2455261.655 & 6000 & 290 & $-2.04\pm0.41$ \\
        2455262.685 & 6000 & 300 & $-2.31\pm0.26$ \\
        2455264.661 & 6000 & 270 & $-0.89\pm0.37$ \\
        2455580.803 & 6000 & 270 & $-0.07\pm0.38$ \\
        2455583.800 & 4800 & 140 & $-2.38\pm0.94$ \\
        2455594.711 & 7200 & 230 & $-2.39\pm0.60$ \\
        2455611.691 & 5700 & 210 & $-2.07\pm0.47$ \\
        2455904.864 & 6600 & 170 & $-3.60\pm0.56$ \\
        2455964.666 & 5400 & 190 & $-2.48\pm0.34$ \\
        2455972.644 & 7200 & 260 & $ 0.58\pm0.44$ \\
		\hline \\
		\noalign{\vskip-0.7cm}
	\end{tabular*}
	\end{center}
\end{table}

\section{Physical parameters}
\label{sect_phys_param}

Based on the high-resolution spectra obtained of HD~35502, three distinct sets of 
spectral lines are apparent: the strong and broad lines associated with a hot star 
and two nearly identifical components attributable to two cooler stars which are 
observed to change positions significantly. Based on HD~35502's reported spectral type, 
the bright, dominant component is presumed to be a hot B5 star \citep{Abt1977}; the 
weaker components are inferred to be two cooler A-type stars based on the presence of 
Fe~{\sc ii} lines and the absence of Fe~{\sc iii} lines. As will be shown in the next 
section, the lines of the A stars show velocity variations consistent with a binary 
system. Hence we conclude that HD~35502 is an SB3 system.

Some of the B star lines (the He~{\sc i} lines, in particular) appear to exhibit 
intrinsic variability. Such features are commonly found in magnetic He peculiar stars 
\citep[e.g.][]{Borra1983,Bolton1998,Shultz2015}.

\subsection{Orbital solution}
\label{orb_sol}

The radial velocity ($v_r$) of the central B star ($B$) in each observation was 
determined using spectral lines for which no significant contribution from the two A 
stars ($A_1$ and $A_2$) was apparent. C~{\sc ii}$\,\lambda4267$ was found to be both 
relatively strong (with a depth of $10$ per cent of the continuum) and only weakly 
variable. The H$\alpha$ spectra encompassed a limited range of wavelengths with few 
lines from which $v_r$ could be accurately determined. We used 
C~{\sc ii}$\,\lambda6578$ and He~{\sc i}$\,\lambda6678$ in order to estimate $v_r$ from 
all of the spectra (spanning a $22$ year period). However, measurements made from 
He~{\sc i}$\,\lambda6678$ were subject to systematic errors associated with strong 
variability (see Section~\ref{variability}). Moreover, the shallower depth of 
C~{\sc ii}$\,\lambda6578$ ($<4$ per cent of the continuum) and its blending with 
H$\alpha$ resulted in both a decrease in precision and a larger scatter in $v_r$ 
compared with those values derived from C~{\sc ii}$\,\lambda4267$.

The radial velocities were calculated by fitting a rotationally-broadened Voigt 
function to the C~{\sc ii}$\,\lambda4267$, C~{\sc ii}$\,\lambda6578$, and 
He~{\sc i}$\,\lambda6678$ lines. The uncertainties were estimated through a 
bootstrapping analysis involving the set of normalized flux measurements ($I/I_c$) 
spanning each line. A random sample of 61 per cent of the data points was selected to 
be removed at each iteration. These points were then replaced by another set that was 
randomly sampled from $I/I_c$. The fitting routine was then repeated on this new data 
set. 1000 iterations of the bootstrapping routine were carried out and a probability 
distribution was obtained for each fitting parameter. The uncertainties in each of the 
fitting parameters were then taken as the $3\sigma$ standard deviations associated 
with each probability distribution.

The value of $v_r$ inferred from C~{\sc ii}$\,\lambda4267$ was found to exhibit 
a median uncertainty of $1.5\,{\rm km\,s}^{-1}$ and a standard deviation of 
$1.4\,{\rm km\,s}^{-1}$. Larger uncertainties were derived using 
He~{\sc i}$\,\lambda6678$ and C~{\sc ii}$\,\lambda6578$ ranging from 
$1-53\,{\rm km\,s}^{-1}$. Similarly, $v_r$ inferred from C~{\sc ii}$\,\lambda6578$ and 
He~{\sc i}$\,\lambda6678$ yielded larger standard deviations of $9$ and 
$18\,{\rm km\,s}^{-1}$, respectively. In the case of He~{\sc i}$\,\lambda6678$, the 
decrease in precision and increase in scatter relative to the more stable 
C~{\sc ii}$\,\lambda4267$ measurements is likely the result of the intrinsic line 
variability. No significant $v_r$ variability was detected using 
C~{\sc ii}$\,\lambda4267$ ($\langle v_r\rangle=21\pm2\,{\rm km\,s}^{-1}$), 
C~{\sc ii}$\,\lambda6578$ ($\langle v_r\rangle=30\pm15\,{\rm km\,s}^{-1}$), or 
He~{\sc i}$\,\lambda6678$ ($\langle v_r\rangle=30\pm8\,{\rm km\,s}^{-1}$).

\begin{figure}
	\centering
	\includegraphics[width=0.999\columnwidth]{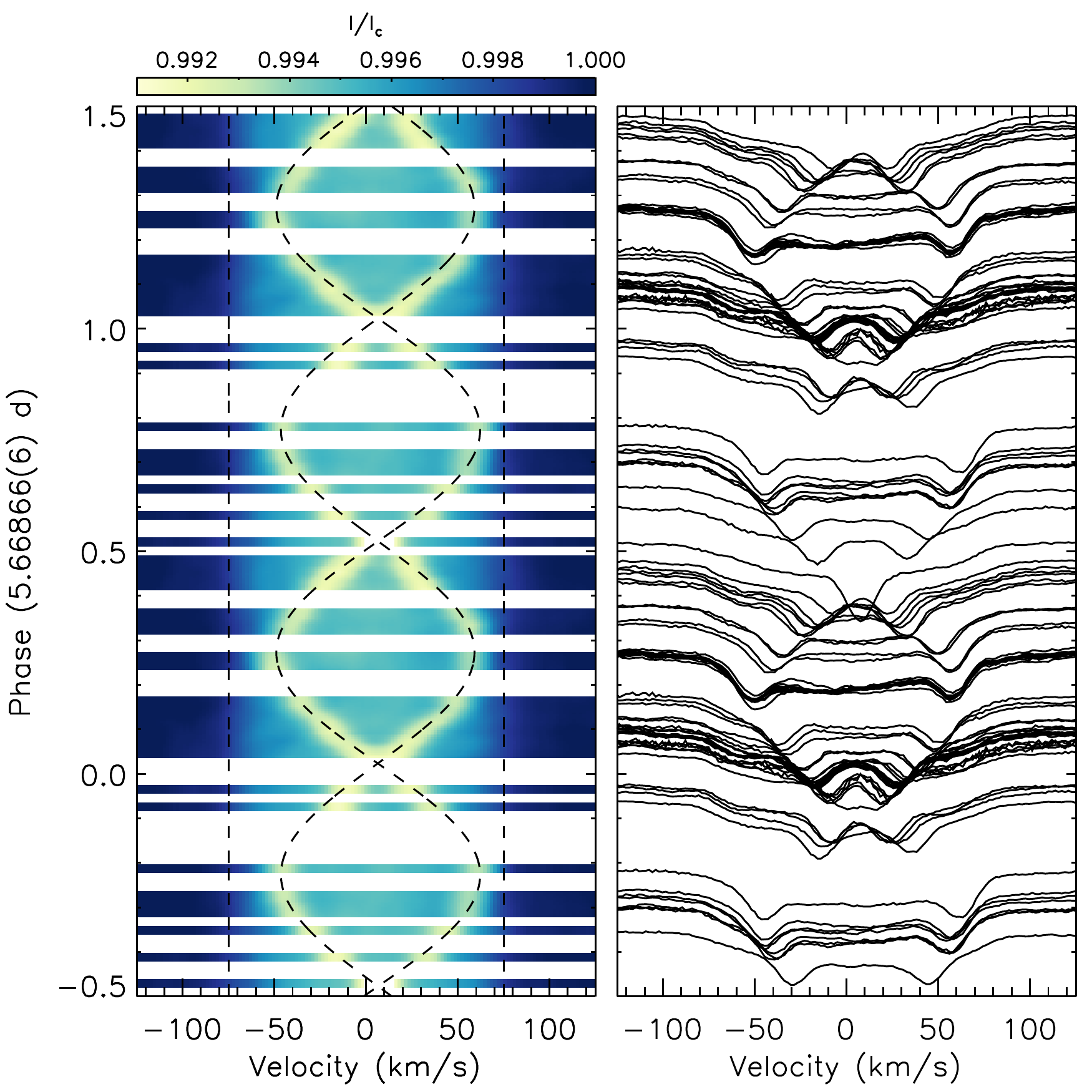}
	\caption{ESPaDOnS, Narval, and FEROS Stokes $I$ LSD profiles (right) and dynamic 
	Stokes $I$ LSD profiles (left) generated such that the three spectral components 
	are emphasized. The observations are phased by the A star orbital period of 
	$5.66866(6)\,{\rm d}$. The vertical dashed black lines indicate the surface of the B 
	star located at $v=\pm v\sin{i}=\pm75\,{\rm km\,s}^{-1}$. The dashed black 
	sinusoids correspond to the fits obtained for $v_r$ of the two A stars.}
	\label{lsd_full}
\end{figure}

The radial velocities of the two A stars were calculated from Stokes I profiles 
produced using the Least Squares Deconvolution (LSD) method 
\citep{Donati1997,Kochukhov2010}. The LSD line mask used to carry out the procedure was 
compiled using data taken from the Vienna Atomic Line Database (VALD) 
\citep{Kupka2000}. In order to isolate the A stars from the dominant B star component 
in the Stokes I LSD profiles, we used a line list associated with an $8000\,{\rm K}$ 
star having a surface gravity of $\log{g}=4.0$ (cgs) and a microturbulence of 
$v_{\rm mic}=0$. Fig.~\ref{lsd_full} shows the LSD profiles generated using a 
different line mask in which both the A and B star components are apparent. The 
majority of the radial velocities were then determined by simultaneously fitting two 
Gaussians to the sharp components of the Stokes I profiles. In the case of the Narval 
observation obtained at ${\rm HJD}=2456224.646$, the sharp line profiles were 
completely blended and the radial velocities were estimated by fitting a single 
Gaussian and adopting the resultant velocity for both components. The $v_r$ errors 
were estimated using a $1000$ iteration bootstrapping analysis. We note that the 
contribution of the B star to the Stokes $I$ LSD profiles was generally weak; however, 
in certain observations, small contributions were present which resulted in small 
deformations in the continuum between the two A star profiles. In these cases, the 
A star line appearing closest to these deformations was more affected than the other 
A star thereby yielding slightly higher uncertainties in the fitting procedure. The 
values of the two A stars' radial velocities were found to range from $-30.4$ to 
$78.6\,{\rm km\,s}^{-1}$ with an average uncertainty of $2.4\,{\rm km\,s}^{-1}$.

\begin{table*}
	\caption{Unpolarized spectra obtained using FEROS. The SNRs per 
	$2.8\,{\rm km\,s}^{-1}$ pixel listed in column 3 are 
	estimated from the RMS of the continuum near $\lambda=5400\,{\rm \AA}$. The fourth, 
	fifth, and sixth columns list the radial velocities of the three stellar 
	components (see Section~\ref{orb_sol}).}
	\label{FEROS_tbl}
	\begin{center}
	\begin{tabular*}{1.5\columnwidth}{@{\extracolsep{\fill}}l c c r r r}
		\hline
		\hline
		\noalign{\vskip0.5mm}
		HJD & Total Exp.    & RMS & \multicolumn{1}{c}{$v_{r,B}$}            & \multicolumn{1}{c}{$v_{r,A_1}$}           & $v_{r,A_2}$           \\
			& Time (s)		& SNR & \multicolumn{1}{c}{$({\rm km\,s}^{-1})$} & \multicolumn{1}{c}{$({\rm km\,s}^{-1})$} & $({\rm km\,s}^{-1})$ \\
		\noalign{\vskip0.5mm}
		\hline	
		\noalign{\vskip0.5mm}
		2456656.568 & 300 & 271 & $19.5\pm1.6$ & $ 75.4\pm2.4$ & $-25.4\pm1.5$ \\
		2456656.609 & 600 & 315 & $20.9\pm1.4$ & $ 76.0\pm2.1$ & $-26.2\pm1.1$ \\
		2456658.535 & 600 & 262 & $19.2\pm1.9$ & $ 10.3\pm3.1$ & $ 39.2\pm2.9$ \\
		2456658.542 & 124 & 145 & $19.2\pm2.7$ & $ 10.1\pm1.8$ & $ 39.2\pm1.5$ \\
		2456658.679 & 300 & 215 & $20.5\pm1.8$ & $  2.9\pm1.8$ & $ 46.5\pm2.5$ \\
		2456658.683 & 300 & 200 & $21.8\pm1.8$ & $  2.8\pm1.7$ & $ 46.5\pm3.0$ \\
		2456658.686 & 300 & 195 & $22.3\pm1.5$ & $  2.8\pm1.7$ & $ 46.8\pm2.0$ \\
		2456658.690 & 300 & 178 & $20.2\pm2.0$ & $  2.2\pm2.1$ & $ 47.0\pm2.4$ \\
		2456658.694 & 300 & 215 & $20.9\pm1.8$ & $  2.3\pm2.3$ & $ 47.2\pm2.7$ \\
		2456658.698 & 300 & 248 & $21.4\pm1.5$ & $  2.1\pm2.0$ & $ 47.4\pm2.9$ \\
		2456658.701 & 300 & 219 & $20.4\pm1.7$ & $  1.9\pm1.8$ & $ 47.6\pm2.1$ \\
		2456658.702 & 300 & 181 & $21.3\pm1.6$ & $  1.6\pm2.3$ & $ 47.6\pm1.6$ \\
		2456658.709 & 300 & 286 & $21.8\pm1.9$ & $  1.3\pm1.5$ & $ 48.1\pm2.2$ \\
		2456658.713 & 300 & 251 & $21.5\pm3.3$ & $  1.3\pm2.3$ & $ 48.1\pm2.4$ \\
		2456658.763 & 600 & 252 & $19.7\pm1.6$ & $ -1.5\pm4.7$ & $ 51.1\pm2.4$ \\
		2456659.628 & 600 & 267 & $18.4\pm1.6$ & $-29.3\pm1.3$ & $ 77.8\pm2.3$ \\
		2456659.671 & 300 & 222 & $19.1\pm1.5$ & $-29.9\pm1.2$ & $ 78.1\pm1.7$ \\
		2456659.677 & 400 & 284 & $19.6\pm1.6$ & $-30.1\pm1.5$ & $ 78.3\pm1.8$ \\
		2456659.682 & 300 & 233 & $19.9\pm1.9$ & $-30.0\pm1.6$ & $ 78.5\pm2.3$ \\
		2456659.685 & 300 & 209 & $19.7\pm1.7$ & $-29.8\pm1.3$ & $ 78.0\pm1.6$ \\
		2456659.689 & 300 & 299 & $18.8\pm1.8$ & $-29.9\pm1.7$ & $ 78.0\pm1.6$ \\
		2456659.693 & 300 & 245 & $17.5\pm2.9$ & $-30.3\pm1.9$ & $ 78.1\pm2.0$ \\
		2456659.697 & 300 & 310 & $19.9\pm1.7$ & $-30.0\pm1.1$ & $ 78.1\pm1.3$ \\
		2456659.700 & 300 & 240 & $20.5\pm1.7$ & $-30.1\pm2.0$ & $ 78.2\pm1.8$ \\
		2456659.704 & 300 & 222 & $20.4\pm1.6$ & $-30.1\pm1.6$ & $ 78.1\pm1.9$ \\
		2456659.708 & 300 & 231 & $20.7\pm2.0$ & $-30.2\pm1.3$ & $ 78.2\pm1.3$ \\
		2456659.746 & 600 & 300 & $20.5\pm1.5$ & $-29.8\pm1.3$ & $ 78.0\pm1.4$ \\
		2456660.614 & 600 & 249 & $20.6\pm1.6$ & $ -4.4\pm3.6$ & $ 53.2\pm4.3$ \\
		2456660.653 & 600 & 276 & $21.6\pm1.6$ & $ -2.4\pm3.1$ & $ 51.7\pm4.1$ \\
		2456660.724 & 600 & 265 & $21.5\pm1.6$ & $  1.1\pm2.7$ & $ 48.1\pm2.9$ \\
		2456660.763 & 600 & 299 & $19.7\pm1.7$ & $  2.8\pm2.2$ & $ 46.5\pm3.1$ \\
		2456660.801 & 600 & 250 & $21.5\pm2.0$ & $  5.1\pm1.9$ & $ 43.9\pm2.6$ \\								
		\noalign{\vskip0.5mm}
		\hline
	\end{tabular*}
	\end{center}
\end{table*}

The spectral characteristics of the two A-type components are nearly identical; 
therefore, it is not possible to unambiguously attribute a particular line profile in 
each spectrum to a particular star. Nevertheless, the importance of this ambiguity can 
be reduced by making simplifying assumptions.

First, we assumed that the two A stars are gravitationally bound and therefore orbit a 
common center of mass (having a radial velocity $v_{\rm cm}$) with a period 
$P_{\rm orb}$. Furthermore, we assumed that the orbits are circular implying that the 
A star $v_r$ variations are purely sinusoidal and described by
\begin{equation}\label{eqn:vr_sin}
v_{r,i}(t)=v_{\rm cm}+K_i\sin{(2\pi t/P_{\rm orb}+\phi_i)}
\end{equation}
where $K_i$ and $\phi_i$ are the semi-amplitude and phase shift of the 
$i^{\rm th}$ A-type component, respectively. The fact that the radial velocities are 
observed to oscillate symmetrically about a constant average radial velocity of 
$\langle v_{r,A}\rangle=(v_{r,1}+v_{r,2})/2=25\pm3\,{\rm km\,s}^{-1}$ suggests that 
(1) $K_1=K_2$ and (2) $|\phi_1-\phi_2|=\pi$. With these assumptions, we applied the 
following procedure:
\begin{enumerate}[leftmargin=0.8cm,labelwidth=16pt]
	\item define a grid of possible orbital periods;
	\item define an amplitude and phase shift of the radial velocity variations based 
	on the maximum observed $v_r$ separation;
	\item for every period, determine which sinusoidal model the blue and red 
	shifted spectral lines must be associated with in order to minimize the 
	residuals.
\end{enumerate}
The two components in each observation were then identified using whichever period 
returned the minimal residual fit. A traditional period fitting routine (e.g. 
Lomb-Scargle) could then be applied to the $v_r$ time series of each star separately 
thereby yielding more precise periods, amplitudes, and phase shifts for each model.

\begin{figure}
	\centering
	\includegraphics[width=1\columnwidth]{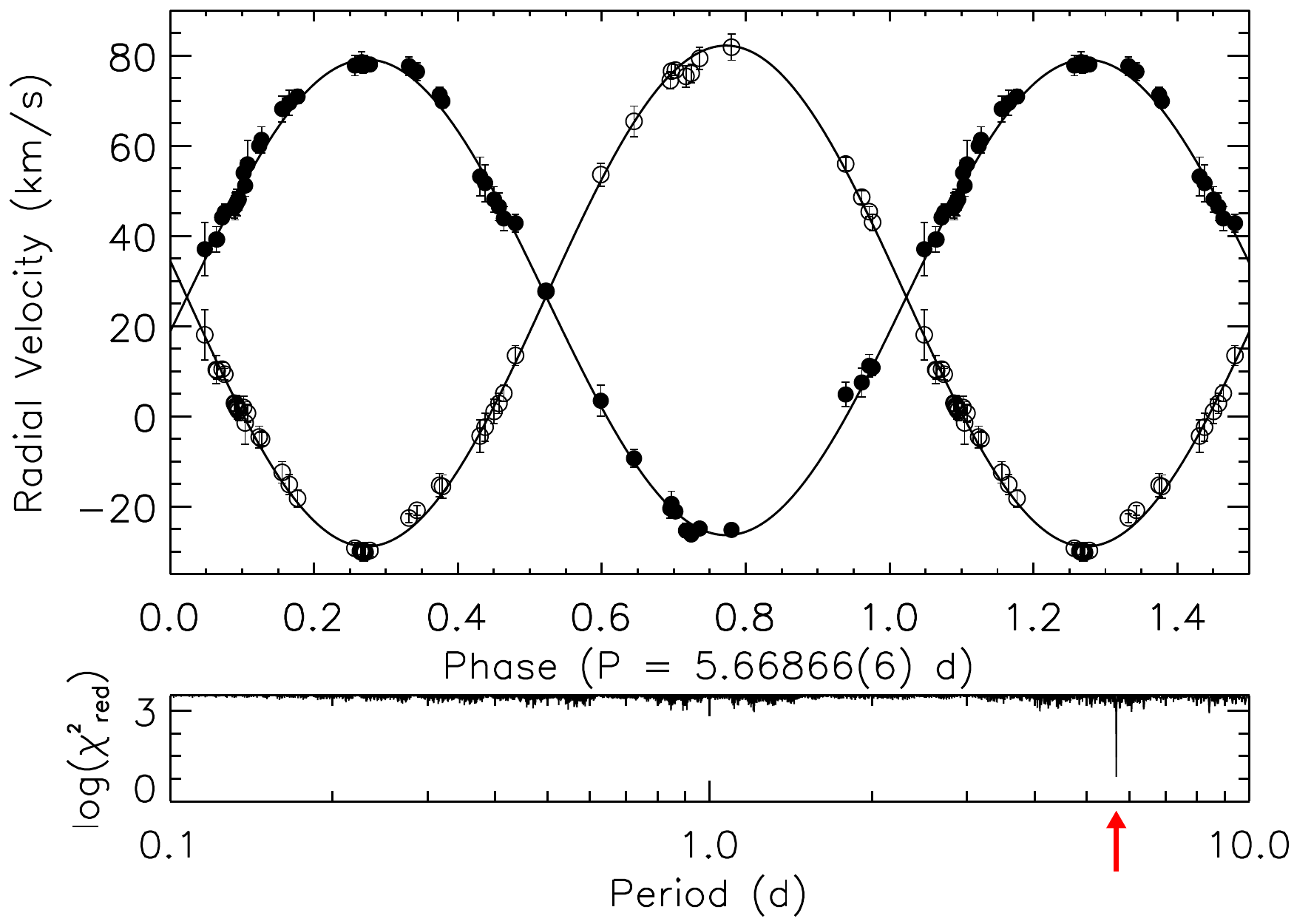}
	\caption{\emph{Top:} The nearly sinusoidal fits to the radial velocities of the two 
	A stars phased by a period of $5.66866\,\text{d}$. \emph{Bottom:} The reduced 
	$\chi^2$ distribution yielded by the period fitting routine applied to one of the A 
	stars. The $5.66866\,\text{d}$ period is indicated by the red arrow.}
	\label{A_orbit}
\end{figure}

We chose a grid of periods ranging from $0.1$ to $10\,{\rm d}$ in increments of 
$10^{-5}\,{\rm d}$ ($\sim1\,{\rm s}$). The amplitudes ($K_1=K_2$) and phase shifts 
($\phi_1=|\phi_2-\pi|$) were defined by the maximum $v_r$ separation of 
$109\,{\rm km\,s}^{-1}$ (i.e. phase $0.994$ where this phase corresponds to the phase 
of the B star's maximum longitudinal magnetic field derived in Section 
\ref{mag_field}). The analysis then involves assigning radial velocities to each of the 
A star components based on a best-fitting period of $5.6687{\rm d}$. An alternative 
means of identifying the orbital period uses the fact that the quantity 
$|v_{r,1}-v_{r,2}|$ varies with a period of $2P_{\rm orb}$, as outlined by 
\citet{Hareter2008}. Applying this method yields a similar value of 
$P_{\rm orb}=5.6680(6)\,{\rm d}$.

With the radial velocities of the two A stars correctly assigned to each individual 
component, a more precise analysis of the binary orbital parameters was carried out 
using {\sc orbitx}, a {\sc fortran} code later adapted to {\sc idl} which determines 
the best-fitting $P_{\rm orb}$, time of periastron passage ($T$), eccentricity ($e$), 
longitude of the periastron ($\omega$), semi-amplitudes of each component's 
radial velocities ($K_1$ and $K_2$), and the radial velocity of the center of mass 
($\gamma$) \citep{Tokovinin1992}. This calculation yielded 
$P_{\rm orb}=5.66866(6)\,{\rm d}$, $T=2456658.172\pm1.652$, 
$e=0.003_{-0.003}^{+0.006}$, $\omega=82\pm105\degree$, 
$K_1=55.5\pm0.4\,{\rm km\,s^{-1}}$, $K_2=52.7\pm0.4\,{\rm km\,s^{-1}}$, and 
$\gamma=26.5\pm0.2\,{\rm km\,s^{-1}}$. These results imply a mass ratio of 
$M_1/M_2=1.05\pm0.02$, a projected total mass of $(M_1+M_2)\sin^3{i}=0.186\pm0.008$, 
and a projected semi-major axis of $a\sin{i}=0.0564\pm0.0008$. These values are 
listed in Table~\ref{orbital_tbl}. Fig.~\ref{A_orbit} shows the radial velocities of 
the A stars phased by the $5.66866\,{\rm d}$ orbital period and compared with the radial velocities computed 
from the orbital solution.

Comparing $\langle v_{r,A}\rangle=25\pm3\,{\rm km\,s}^{-1}$ with the average B star 
$v_r$ of $20.5\pm1.6\,{\rm km\,s}^{-1}$ and noting that no significant variability in 
$\langle v_{r,A}\rangle$ was detected over the $22$ year observing period implies a 
very long orbital period of the A binary about the B star. A lower limit for this 
period is derived in Section~\ref{SED_fit}.

\begin{table}
	\caption{Orbital parameters of the A+A binary.}
	\label{orbital_tbl}
	\begin{center}
	\begin{tabular*}{0.72\columnwidth}{@{\extracolsep{\fill}}l r}
		\hline
		\hline
		\noalign{\vskip1mm}
		$P_{\rm orb}\,({\rm d})$        & $5.66866(6)$ \vspace{0.8mm}\\
		$T$                             & $2456658.172\pm1.652$\vspace{0.8mm}\\
		$e$                             & $0.003^{+0.006}_{-0.003}$ \vspace{0.8mm}\\
		$\omega\,(\degree)$             & $82\pm105$ \vspace{0.8mm}\\
		$K_1\,{\rm (km/s)}$             & $55.5\pm0.4$ \vspace{0.8mm}\\
		$K_2\,{\rm (km/s)}$             & $52.7\pm0.4$ \vspace{0.8mm}\\
		$\gamma\,{\rm (km/s)}$          & $26.5\pm0.2$ \vspace{0.8mm}\\
		$M_1/M_2$                       & $1.05\pm0.02$ \vspace{0.8mm}\\
		$(M_1+M_2)\sin^3{i}\,(M_\odot)$ & $0.186\pm0.008$ \vspace{0.8mm}\\
		$a\sin{i}\,{\rm (AU)}$        & $0.0564\pm0.0008$ \vspace{0.8mm}\\
		\hline
	\end{tabular*}
	\end{center}
\end{table}

\subsection{SED fitting}
\label{SED_fit}

Photometric fluxes of HD~35502 have been measured throughout the UV, visible, and near 
infrared spectral regions thereby allowing the temperatures and radii of the three 
stellar components to be constrained. Ultraviolet measurements were previously obtained 
at four wavelengths -- $1565\,\rm{\AA}$, $1965\,\rm{\AA}$, $2365\,\rm{\AA}$, and 
$2749\,\rm{\AA}$ -- by the $S2/68$ instrument on board the $TD1$ satellite 
\citep{Thompsons1978}. Photometry spanning the visible spectrum were taken from the 
Geneva Observatory's catalogue of $U$, $B$, $V$, $B_1$, $B_2$, $V_1$, and $G$ 
filters \citep{Rufener1981}. Additionally, infrared observations obtained by 2MASS 
($J$, $H$, and $Ks$ filters) \citep{Skrutskie2006} and WISE ($W1$ and $W2$ filters) 
\citep{Wright2010} were used. The reported Geneva, 2MASS, and WISE magnitudes were 
converted to the flux units of ${\rm ergs\,s^{-1}\,cm^{-2}\,{\AA}^{-1}}$ using the zero 
points reported by \citet{Rufener1988}, \citet{Cohen2003}, and \citet{Wright2010}. 

The reported photometric measurements of HD~35502 include the contributions 
from each of the three stellar components. This renders an SED fitting analysis 
particularly susceptible to degenerate solutions; however, speckle interferometry 
measurements obtained by \citet{Balega2012} provide additional photometric constraints 
on the system. They detected magnitude differences of $1.45\pm0.02\,{\rm mag}$ and 
$1.21\pm0.02\,{\rm mag}$ using filters centered on $5500\,\rm{\AA}$ and 
$8000\,\rm{\AA}$, respectively. The sources were reported to have angular separations 
of $69\pm1\,{\rm mas}$ and $68\pm1\,{\rm mas}$ in the two filters. The speckle 
companion is also identifed in observations obtained by \citet{Horch2001} with a 
consistent angular separation of $\rho<59\,{\rm mas}$.

In conjunction with the distance to HD~35502, the angular separations may be used to 
determine the associated linear separation. A distance of $d=430\pm120\,\rm{pc}$ was 
inferred from the $2.35\pm0.68\,\rm{mas}$ Hipparcos parallax \citep{VanLeeuwen2007}. 
However, assuming HD~35502 to be a member of the Orion OB1a subassociation, we 
inferred a moderately more precise value of $400\pm80\,{\rm pc}$ based on the 
subassociation's reported average distance modulus of 
$\langle {\rm dm}\rangle=8.00\pm0.46\,{\rm mag}$ \citep{Brown1994}. The projected 
linear separation between the two speckle sources was then found to be 
$27\pm5\,\rm{AU}$. The minimum orbital period of the A star binary system around the B 
star can then be approximated by assuming upper limit masses of $8\,M_\odot$ and 
$3\,M_\odot$ for the B and A stars, respectively (the actual masses are derived in 
Section~\ref{hrd}). This implies an orbital period of $P_{\rm orb}\gtrsim40\,\rm{yrs}$, 
which is consistent with the fact that no significant variations were detected in 
either the B star radial velocities or the A star binary's systemic radial velocity.

\begin{figure}
	\centering
	\includegraphics[width=1.0\columnwidth]{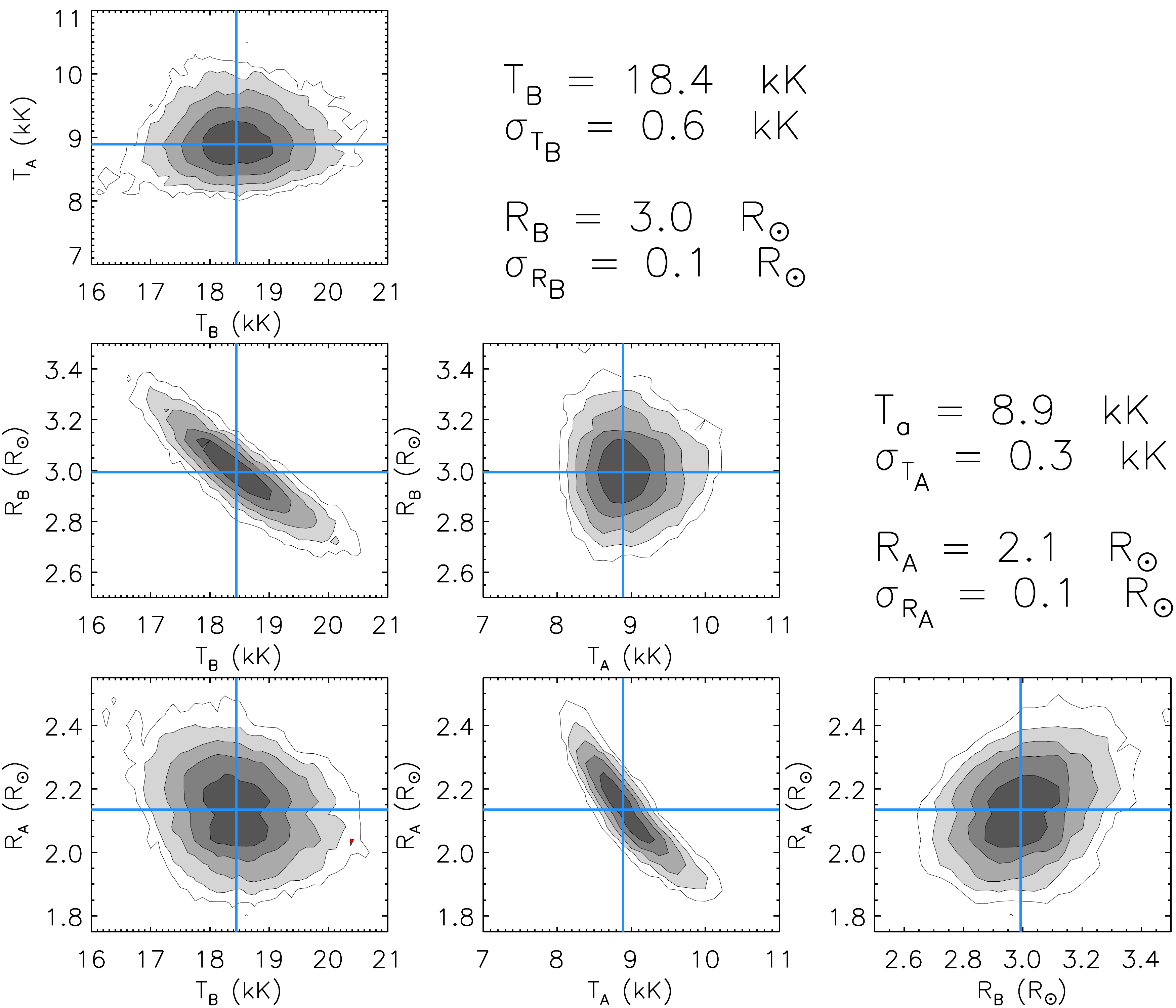}
	\caption{Marginalized posterior probability distributions returned by the MCMC 
	algorithm that was applied to the SED fitting. Each frame demonstrates the 
	correlations that are apparent between various parameters, while the vertical 
	and horizontal blue lines indicate the value of each parameter associated with 
	the maximum likelihood solution (i.e. $T_B$, $T_A$, $R_B$, and $R_A$). The 
	contours approximately correspond to $1-3\sigma$ confidence regions in 
	increments of $0.5\sigma$; $\sigma_{T_B}$, $\sigma_{T_A}$, $\sigma_{R_B}$, and 
	$\sigma_{R_A}$ indicate the value of each parameter's $1\sigma$ 
	region.}
	\label{SEDa}
\end{figure}

\begin{figure}
	\centering
	\includegraphics[width=1.0\columnwidth]{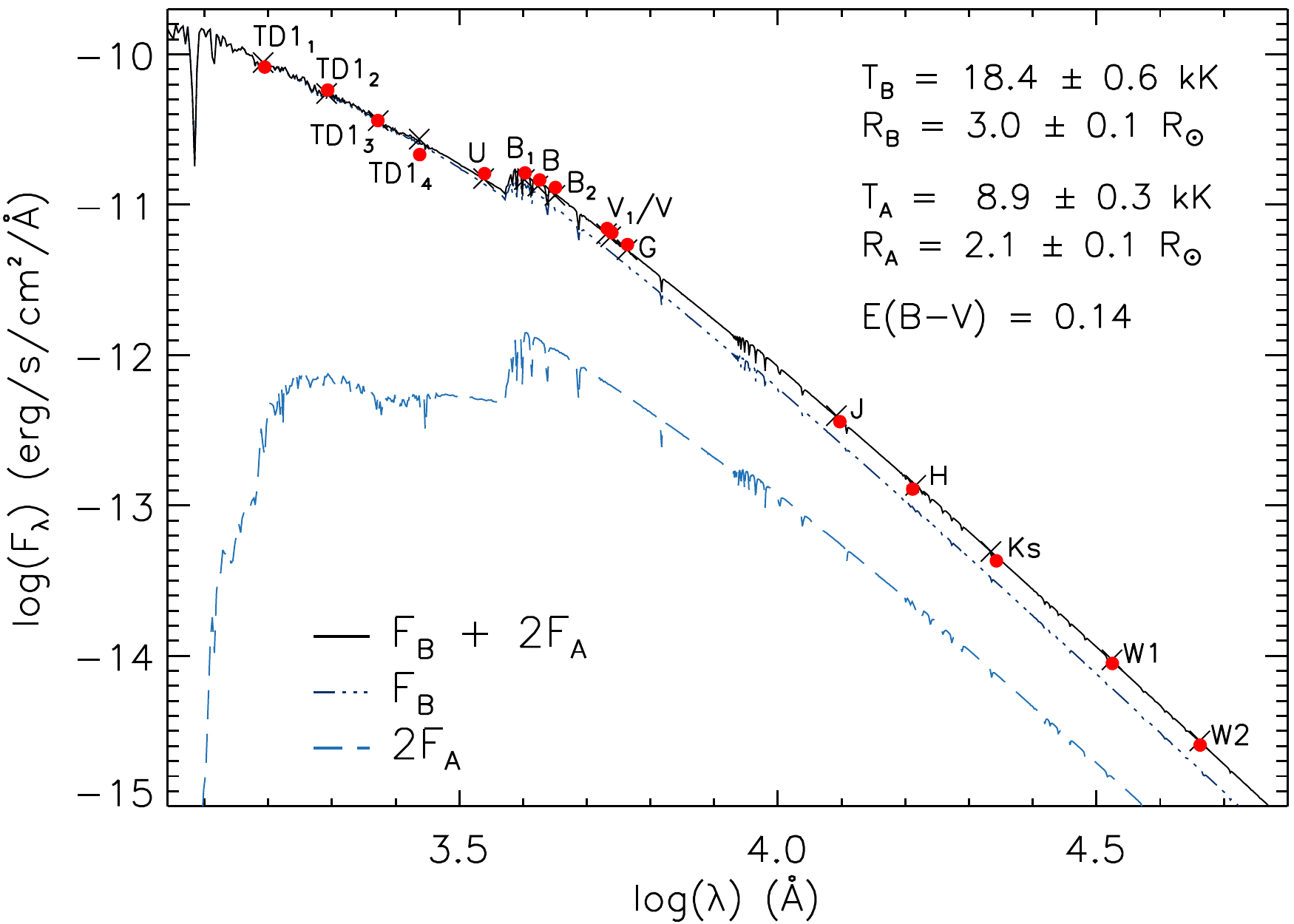}
	\caption{Comparisons between the observed photometry (red points) and the model 
	SED (solid black curve). The dashed blue and dot-dashed black curves correspond to 
	the model SEDs of the composite A star components (i.e. $2F_A(\lambda)$) and 
	the B star, respectively. The black crosses indicate the flux obtained by 
	multiplying the model SED by the transmission function of the associated 
	filter.}
	\label{SEDb}
\end{figure}

The observed photometry was fit using {\sc atlas9} synthetic spectral energy 
distributions (SEDs) generated from the atmospheric models of \citet{Castelli2004}. 
The grid consists of models with effective temperatures ranging from 
$3.5-50.0\,\rm{K}$ and surface gravities spanning $\log{g}=0.5-5.0\,\rm{(cgs)}$, as 
described in detail by \citet{Howarth2011}. This grid was linearly interpolated in 
order to produce models with a uniform temperature and surface gravity resolution of 
$125\,\rm{K}$ and $0.01\,\rm{dex}$ for $T_{\rm eff}=5-25\,\rm{kK}$ and 
$\log{g}=3.0-4.75\,\rm{(cgs)}$. All of the SEDs were then multiplied by the 
transmission functions associated with each of the narrow band filters: TD1 $UV$ 
\citep{Carnochan1982}, Geneva \citep{Rufener1988}, 2MASS \citep{Cohen2003}, and WISE 
\citep{Wright2010}.

Modelling the photometry of un-resolved multi-star systems using synthetic SEDs 
requires a large number of fitting parameters and therefore the solution is expected 
to be highly degenerate. The contribution to the total flux from each of the three 
stellar components depends on, among other factors, their effective temperatures, 
surface gravities, and radii. In order to reduce the number of solutions, we adopted a 
solar metallicity and a microturbulence velocity of $v_{\rm mic}=0\,{\rm km\,s}^{-1}$. 
As with many Bp stars, HD~35502's primary exhibits chemical spots on its surface (see 
Section~\ref{variability}); however, on average, a solar metallicity may be adopted.

The high-resolution spectra of HD~35502 obtained by Narval, ESPaDOnS, and FEROS 
suggest that the two cooler A star components are approximately identical in terms of 
their $T_{\rm eff}$, $\log{g}$, and line-broadening parameters (see 
Section~\ref{line_fit}). If we assume that the two A stars contribute identically to 
the SED, the number of independent models required in the fitting routine is reduced 
from three to two thereby resulting in a total of six free parameters: $T_{\rm eff}$, 
$\log{g}$, and the stellar radius, $R$, for both the B star and the (identical) A 
stars.

The effective temperature of a star inferred from fitting model SEDs to photometry is 
highly dependent on the assumed colour excess, $E(B-V)$. Given HD~35502's probable 
location within the Orion OB1a association \citep{Landstreet2007}, the extinction 
caused by gas and dust is expected to be significant. Indeed, \citet{Sharpless1952} 
and \citet{Lee1968} report values (without uncertainties) of $E(B-V)$ of $0.13$ and 
$0.14$, respectively. We used the method of \citet{Cardelli1989} with an adopted to 
selective total extinction ratio of $R(V)=3.1$ in order to deredden the observed 
photometry. Small differences in the resulting best-fitting parameters of $<3$ per cent 
were found by using an $E(B-V)$ of $0.13$ or $0.14$. Although we investigated how our 
analysis was affected by varying the colour excess from $0.0-0.2$, the final effective 
temperatures are reported after assuming $E(B-V)=0.14$.

We found that a Markov Chain Monte Carlo (MCMC) fitting routine provided a suitable 
means of determing the most probable solution while simultaneously revealing any 
significant degeneracies. This was carried out by evaluating the likelihood function 
yielded by a set of randomly selected fitting parameters drawn from a prior 
probability \citep[see e.g.][]{WallJenkins2003}. For each iteration, the derived 
likelihood is compared with that produced by the previous iteration. If a new solution 
is found to yield a higher quality of fit (higher likelihood), these parameters are 
adopted, otherwise, the previous solution is maintained. In order to broadly sample 
the solution space, the MCMC algorithm is designed to adopt poorer fitting solutions 
at random intervals thereby preventing a local (but not global) maximum likelihood 
from being returned.

Uniform prior probability distributions (flat priors) were defined for $T_{\rm eff}$, 
$\log{g}$, and $R$, where the latter was constrained within $1.0-10.0\,R_\odot$. The 
two speckle observations \citep{Balega2012} were then included in the total prior 
probability as monochromatic flux ratios (i.e. magnitude differences) at 
$5500\,{\rm \AA}$ and $8000\,{\rm \AA}$. We assumed that the reported 
$0.02\,{\rm mag}$ uncertainties correspond to $1\sigma$ significance. The marginalized 
posterior probability distributions produced after $10^6$ iterations in the Markov 
Chain are shown in Fig.~\ref{SEDa}.

\begin{figure*}
	\centering
	\includegraphics[width=2.1\columnwidth]{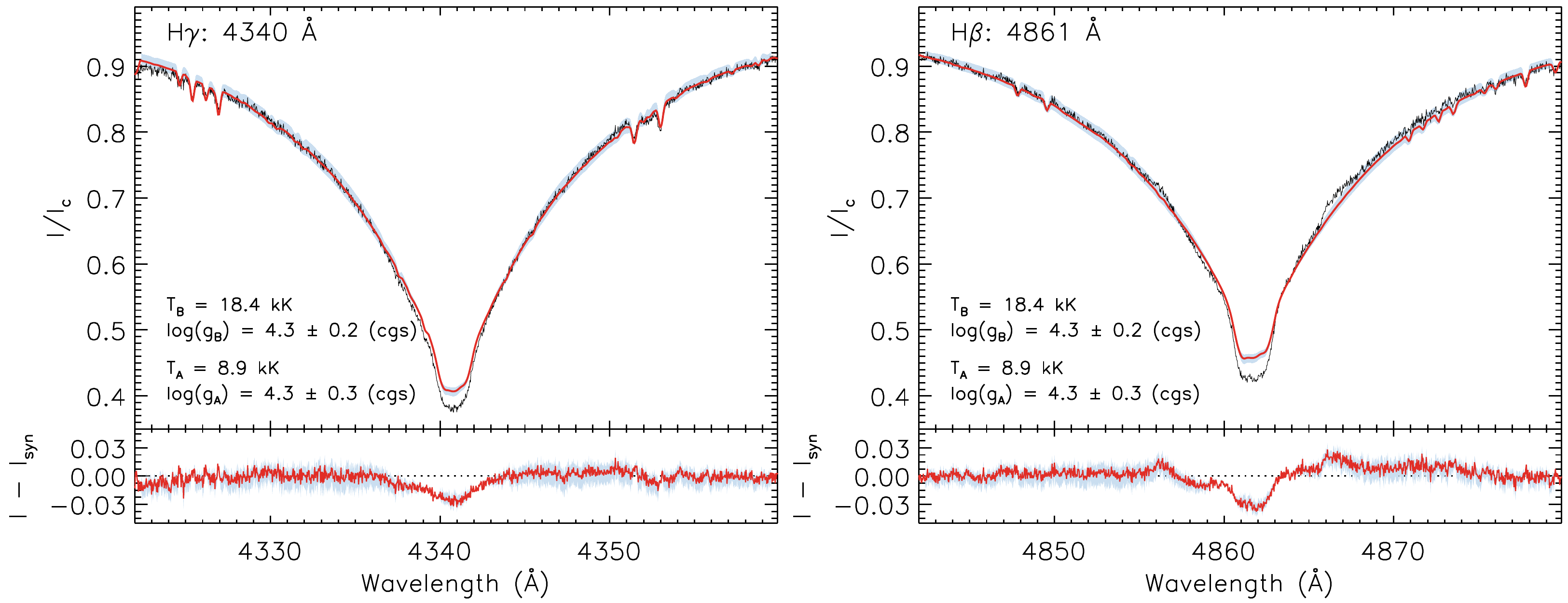}
	\caption{\emph{Top:} Comparisons between the best-fitting synthetic (red) and 
	observed (black) H lines, H$\gamma$ (left) and H$\beta$ (right). The filled blue 
	region indicates the total uncertainty associated with $\log{g_B}$ and $\log{g_A}$. 
	The observations occur at phase $0.49$ in the B star's rotational period (see 
	Section~\ref{P_rot}) when the weakest emission is visible in the wings of H$\beta$. 
	\emph{Bottom:} The residuals associated with the model spectrum (red).}
	\label{B_lines}
\end{figure*}

\begin{figure}
	\centering
	\includegraphics[width=0.99\columnwidth]{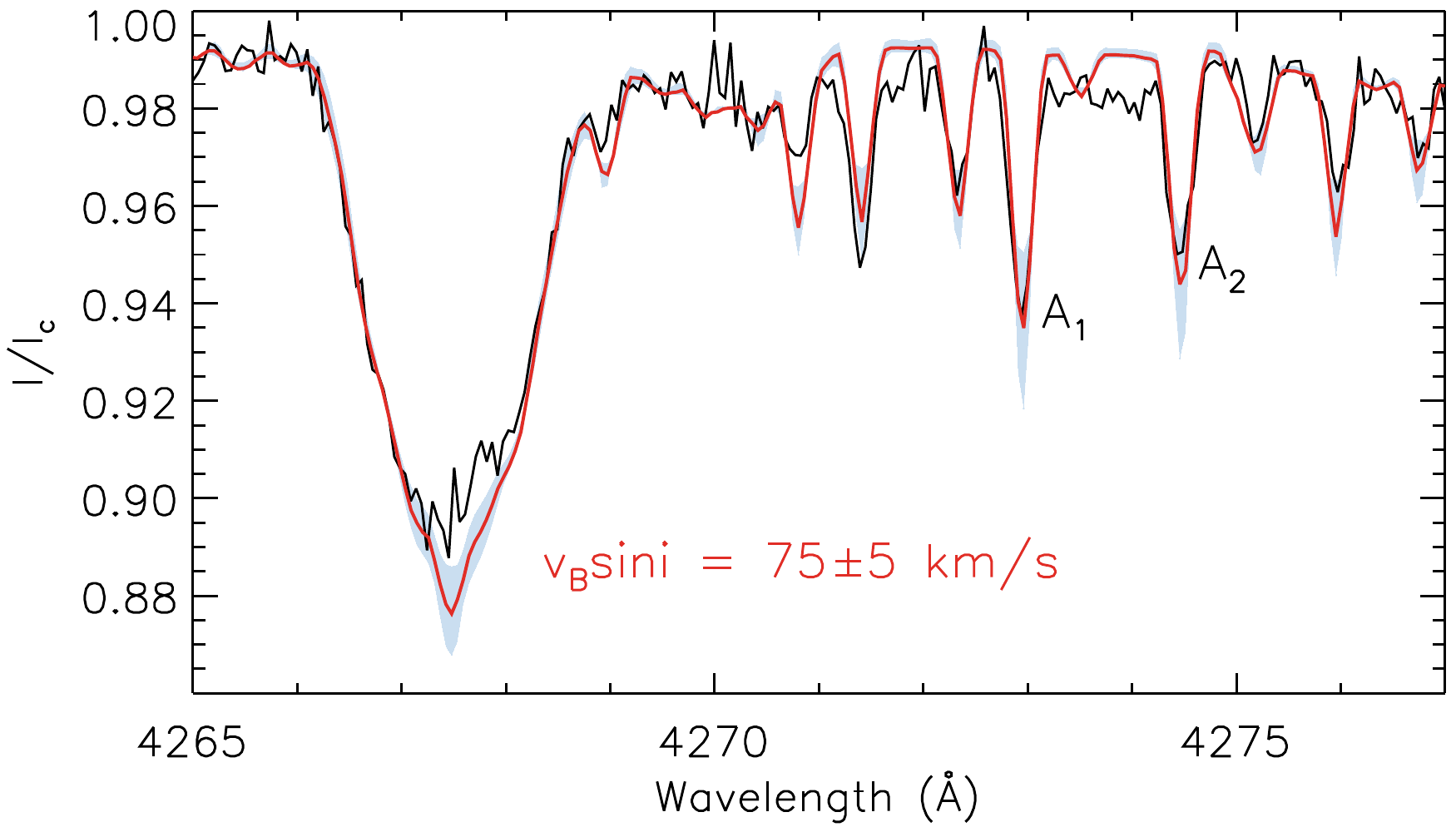}
	\caption{A subsample of the wavelength region used to determine the luminosity 
	ratio, $L_B/L_A$. The red curve corresponds to the best-fitting model spectrum; the 
	filled blue region indicates the total uncertainty associated with a $v_B\sin{i}$ 
	of $\pm5\,{\rm km\,s}^{-1}$ as determined by fitting various metal lines. The 
	black curve shows the observed spectrum. The Fe~{\sc i}$\,\lambda4273$ lines of 
	the two A star components are labeled as `A$_1$' and `A$_2$'.}
	\label{B_vsini}
\end{figure}

The most probable effective temperatures for the B and A star models were found to be 
$18.4\pm1.2\,{\rm kK}$ and $8.9\pm0.6\,{\rm kK}$, respectively, where the 
uncertainties correspond to the $93^{\rm rd}$ percentile (approximately $2\sigma$). 
The fitting parameters used to derive the stellar radii, $R_B$ and $R_A$, depend on the 
distance to HD~35502 (i.e. as a scaling factor given by $R_\ast^2/d^2$). Although 
the posterior probability distributions for $R_B$ and $R_A$ both yield $2\sigma$ 
uncertainties of $0.2\,R_\odot$, the consideration of the relatively large 
distance uncertainty ($d=400\pm80\,{\rm pc}$) implies larger uncertainties of 
$\delta R_B=0.6\,R_\odot$ and $\delta R_A=0.4\,R_\odot$. The most probable radii and 
their uncertainties found from the MCMC analysis are then given by 
$R_B=3.0\pm0.6\,R_\odot$ and $R_A=2.1\pm0.4\,R_\odot$. The derived temperatures 
and stellar radii are listed in Table~\ref{param_tbl}. The analysis was insensitive 
to changes in $\log{g}$ as indicated by an essentially flat posterior probability 
distribution; therefore, no definitive surface gravity can be reported. Comparisons 
between the observed photometry and the best-fitting model are shown in Fig.~\ref{SEDb}, 
where we have adopted $\log{g}=4.3$ for both the A and B models as derived in 
Section~\ref{line_fit}.

The model B star flux ($F_B$) and the model binary A star flux ($2F_A$) can be 
used to verify our initial assumption that the two components detected in the speckle 
observations do indeed correspond to the central B star and the A star binary system. 
The models can be compared with the speckle observations by calculating the flux 
ratios, $2F_A(\lambda)/F_B(\lambda)$, at the speckle observation wavelengths. Both 
$F_A(\lambda)$ and $F_B(\lambda)$ are integrated over wavelength intervals of 
$200\,{\rm \AA}$ and $1000\,{\rm \AA}$ \citep[i.e. the FWHM of the filters used by][]
{Balega2012} centered at $5500\,{\rm \AA}$ and $8000\,{\rm \AA}$, respectively. We 
then obtain magnitude differences of $\Delta m_{\rm syn}(5500\,{\rm \AA})=1.46$ and 
$\Delta m_{\rm syn}(8000\,{\rm \AA})=1.23$. These values yield a negligible 
discrepancy with the speckle observations of $1$ per cent at $\lambda=5500\,{\rm \AA}$ 
and $2$ per cent at $\lambda=8000\,{\rm \AA}$.

\begin{table}
	\caption{Stellar parameters of HD~35502.}
	\label{param_tbl}
	\begin{center}
	\begin{tabular*}{0.35\textwidth}{@{\extracolsep{\fill}}l r}
		\hline
		\hline
		\noalign{\vskip1mm}
		Sp. Type$^{1}$ & B5IVsnp+A+A\\
		${\rm \pi}$ (mas)$^{2}$ & $2.35\pm0.68$\\
		$\langle {\rm dm}\rangle$ (mag)$^{3}$ & $8.00\pm0.46$\\
		$d$ (pc) & $400\pm80$\vspace{2mm}\\
		Photometry &\\
		\hline
		\noalign{\vskip1mm}
		$V$ (mag)$^{4}$ & $7.331\pm0.004$\vspace{0.8mm}\\
		$E(B-V)$ (mag)$^{5}$ & $0.14$\vspace{2mm}\\
		B Star Parameters &\\
		\hline
		\noalign{\vskip1mm}
		$T_{\rm eff}\,(\text{kK})$ & $18.4\pm0.6$\vspace{0.8mm}\\
		$\log(g)\,(cgs)$ & $4.3\pm0.2$\vspace{0.8mm}\\
		$v\sin{i}\,({\rm km\,s}^{-1})$ & $75\pm5$\vspace{0.8mm}\\
		$\log{L/L_\odot}$ & $3.0\pm0.3$\vspace{0.8mm}\\
		$M/M_\odot$ & $5.7\pm0.6$\vspace{0.8mm}\\
		$R_p/R_\odot$ & $3.0^{+1.1}_{-0.5}$\vspace{0.8mm}\\
		$R_{\rm eq}/R_\odot$ & $3.1^{+1.8}_{-0.4}$\vspace{0.8mm}\\
		$\tau_{\rm age}\,(\text{Myr})$ & $20\pm20$\vspace{0.8mm}\\
		$P_{\rm rot}\,(\text{d})$ & $0.853807(3)$\vspace{2mm}\\
		A Star Parameters &\\
		\hline
		\noalign{\vskip1mm}
		$T_{\rm eff}\,(\text{kK})$ & $8.9\pm0.3$\vspace{0.8mm}\\
		$\log(g)\,(cgs)$ & $4.3\pm0.3$\vspace{0.8mm}\\
		$v\sin{i}\,({\rm km\,s}^{-1})$ & $12\pm2$\vspace{0.8mm}\\
		$\log{L/L_\odot}$ & $1.4\pm0.3$\vspace{0.8mm}\\
		$M/M_\odot$ & $2.1\pm0.2$\vspace{0.8mm}\\
		$R/R_\odot$ & $2.1\pm0.4$\vspace{0.8mm}\\
		$\tau_{\rm age}\,(\text{Myr})$ & $<630$\vspace{0.8mm}\\
		\hline		
	\end{tabular*}\par
	\begin{tablenotes}
	\small
	\item Table references: $^1$\citet{Abt1977}, $^2$\citet{VanLeeuwen2007}, 
	$^3$\citet{Brown1994}, $^4$\citet{Rufener1981}, $^5$\citet{Lee1968}.
	\end{tablenotes}
	\end{center}
\end{table}

\subsection{Spectral line fitting}
\label{line_fit}

Several properties of HD~35502's three stellar components may be estimated through 
comparisons with synthetic spectra (e.g. the surface gravity, line broadening 
characteristics, etc.). We carried this out using local thermodynamic equilibrium (LTE) 
models generated with {\sc synth3} \citep{Kochukhov2007a}. The code computes 
disc-integrated spectra using spectral line data provided by VALD \citep{Kupka2000} 
obtained using an {\rm extract stellar} request for a specified effective 
temperature, surface gravity, and microturbulence velocity in conjunction with 
{\sc atlas9} atmospheric models \citep{Kurucz1993_oth}. The synthetic spectra can then 
be convolved with the appropriate functions in order to account for instrumental and 
rotational broadening effects.

The ESPaDOnS, Narval, and FEROS observations were normalized using a series of 
polynomial fits to the continuum. The relatively shallow ($\approx5$ per cent of the 
continuum) and narrow lines produced by the two A stars made the spectral line 
modelling inherently uncertain. For instance, the typical root mean square of 
the continuum near the A stars' Mg~{\sc i}$\,\lambda4703$ lines was found to be 
approximately $14$ per cent of the line depth. Thus, the SNRs of the majority of the A 
star lines were relatively low. This was mitigated to some extent by binning the 
observed spectra with a bin width of $\approx0.03\,{\rm \AA}$ (i.e. $2$ pixels). In 
order to account for the instrumental profile of the ESPaDOnS and Narval observations, 
the synthetic spectra were convolved with a Gaussian function assuming a resolving 
power of $R=65\,000$; similarly, the FEROS spectra were fit after convolving the 
synthetic spectra assuming $R=48\,000$.

The quality of fit yielded by the total normalized synthetic spectrum ($F_{\rm tot}$) 
depends not only on $T_{\rm eff}$ of the three models but also on their (relative) 
luminosities: ${F_{\rm tot}=(\sum_iL_iF_i)/\sum_iL_i}$, where $L_i$ and $F_i$ are the 
luminosities and synthetic spectra of the $i^{\rm th}$ component. We adopted the 
$18.4\,{\rm kK}$ and $8.9\,{\rm kK}$ values associated with the B and two A stars 
obtained from the SED fitting (Section~\ref{SED_fit}). Moreover, we assumed that the 
luminosities of the two A stars are equal, thereby reducing the number of degrees of 
freedom in the spectral line fitting analysis.

With $T_{\rm eff}$ specified, the stellar luminosities can be estimated through various 
methods. We found that the best results were obtained by letting the luminosity ratio 
of the B and A star models, $L_B/L_A$, be a free parameter and subsequently finding the 
minimum $\chi^2$ fit for a given surface gravity ($\log{g}$) and rotational broadening 
($v\sin{i}$). This was carried out using the observed spectra for which the two A stars 
were most widely separated in wavelength (phase $0.994$ in Fig.~\ref{A_orbit}) in the 
wavelength range of $4200-4300\,{\rm \AA}$. This region was chosen because of the 
presence of the strong and essentially non-variable C~{\sc ii} line produced by the B 
star along with many A star lines of various elements (e.g. Fe, Ti, Cr, Mn). Most 
importantly, this wavelength range is free of B star lines exhibiting obvious chemical 
abundance anomalies and variability such as those observed from He. A subsample of 
this region containing C~{\sc ii}$\,\lambda4267$ is shown in Fig.~\ref{B_vsini}. This 
technique yielded $L_B/L_A=5.2$, which is consistent with the median value implied by 
the SED fitting analysis -- within $4000\,{\rm \AA}\le\lambda\le6000\,{\rm \AA}$ -- of 
$8.4^{+4.1}_{-2.4}$. Ultimately, using an $L_B/L_A$ of $5.2$ instead of $8.4$ produced 
a moderate increase of $1.1\sigma$ in the best-fitting solution's overall quality of 
fit.

$\log{g}$ and $v\sin{i}$ of HD~35502's three components were then fit using various B 
and A star lines while recalculating the best-fitting $L_B/L_A$ for every change in the 
parameters. As a result of the presumed chemical peculiarities and line variability of 
the B star (see Section~\ref{variability}) and the limited number of lines, 
$\log{g_B}$ could not be reliably constrained using He or metal lines (e.g. Mg~{\sc i} 
and Mg~{\sc ii}). Instead, we relied upon the wings of the strong and broad Balmer 
lines. In particular, the observations of H$\beta$, H$\gamma$, and H$\delta$ obtained 
at a phase of $0.49$ in the B star's rotational period (the phase of minimum emission) 
were used in order to minimize the effects of emission. Several metal lines were used 
to constrain $v_B\sin{i}$ such as C~{\sc ii}$\,\lambda4267$, 
S~{\sc ii}$\,\lambda5640$, and Fe~{\sc ii}$\,\lambda5780$. $\log{g_B}$ and $v_B\sin{i}$ 
were found to be $4.3\pm0.2\,{\rm (cgs)}$ and $75\pm5\,{\rm km\,s}^{-1}$, respectively. 
Examples of the best-fitting model spectra are shown in Fig.~\ref{B_lines}; the 
adopted range in $v_B\sin{i}$ is shown in Fig.~\ref{B_vsini}.

\begin{figure}
	\centering
	\includegraphics[width=0.99\columnwidth]{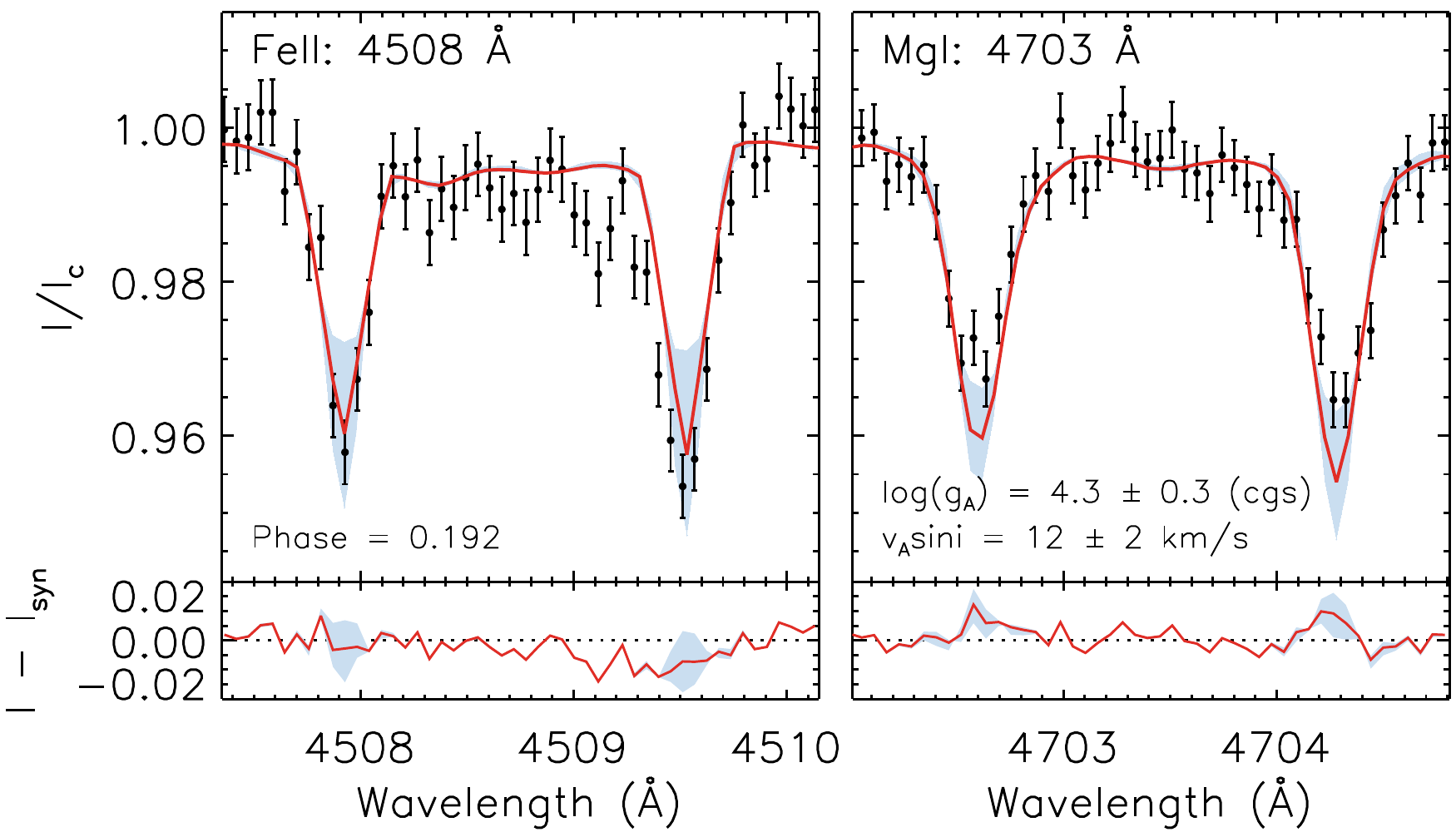}
	\caption{Comparisons between best-fitting model (red) and observed (black points) 
	Fe~{\sc ii} (left) and Mg~{\sc i} (right) lines used to constrain the surface 
	gravity and rotational broadening of the two A stars; the filled blue region 
	indicates the total uncertainties associated with $v_A\sin{i}$ and $\log{g_A}$. 
	The phase corresponds to the maximum observed separation between the two binary 
	components.}
	\label{A_lines}
\end{figure}

The surface gravity and rotational broadening of the two A stars were fit 
simultaneously using several Fe~{\sc ii} and Mg~{\sc i} lines. Their best-fitting 
$\log{g}$ and $v\sin{i}$ values were found to be $4.3\pm0.3\,{\rm (cgs)}$ and 
$12\pm2\,{\rm km\,s}^{-1}$, respectively. Two examples of the modelled A star lines are 
shown in Fig.~\ref{A_lines}.

\subsection{Hertzsprung-Russell Diagram}
\label{hrd}

The masses, ages, and polar radii of HD~35502's three stellar components may be 
estimated by comparing their positions on the Hertzsprung-Russell diagram (HRD) with 
theoretical isochrones.

In order to determine the B star's luminosity, we used the $18.4\pm0.6\,{\rm kK}$ 
effective temperature and $3.0\pm0.6\,R_\odot$ radius derived from the three star SED 
fit discussed in Section~\ref{SED_fit}. The Stefan-Boltzmann law then yields a 
luminosity of $\log{L/L_\odot}=3.0^{+0.4}_{-0.5}$. Similarly, the position of the A 
stars on the HRD can be identified using $T_{\rm eff}=8.9\pm0.3\,{\rm kK}$ and 
$R_A=2.1\pm0.4\,R_\odot$. We calculate an A star luminosity of 
$\log{L/L_\odot}=1.4\pm0.3$. The HRD positions of the B and A stars are shown 
in Fig.~\ref{hrd}.

The masses ($M$) and polar radii ($R_p$) associated with a given $T_{\rm eff}$ and 
$\log{L/L_\odot}$ were determined using a grid of Geneva model isochrones generated by 
\citet{Ekstrom2012}. The grid is calculated for the evolutionary timescale beginning 
with the zero-age main sequence up until the core carbon-burning phase for masses of 
$0.8-120\,M_\odot$. The microturbulence velocity was fixed at 
$v_{\rm mic}=0.0\,{\rm km\,s}^{-1}$ and a solar metallicity of $Z=0.014$ was assumed. 
In the case of HD~35502's central B star, the ratio of the angular velocity to the 
critical angular velocity, $\Omega/\Omega_c$, is known to be significant based on the 
$0.853807(3)\,{\rm d}$ rotational period (see Section~\ref{P_rot}). Its position on the 
HRD was therefore compared against several additional grids calculated using 
$\Omega/\Omega_c=0.4-0.9$ in increments of $0.1$ \citep{Georgy2013a}. While no 
significant difference in the inferred $M$ was apparent (i.e. $<4$ per cent), 
$R_{\rm p}$ was found to decrease by as much as $15$ per cent.

In order to select the most accurate grid of isochrones and thus, the most accurate 
$R_{\rm p}$, $\Omega/\Omega_c$ must first be estimated. Since $\Omega_c$ depends on 
both the mass and polar radius, it was calculated using the parameters derived from 
each grid of isochrones. Using $P_{\rm rot}$ inferred in Section~\ref{P_rot} to 
determine $\Omega$, a range of $\Omega/\Omega_c$ values were found. The appropriate 
grid was then chosen based on whichever $\Omega/\Omega_c$ most closely agreed with the 
$\Omega/\Omega_c$ associated with the isochrone grid. A calculated $\Omega/\Omega_c$ 
of $0.53$ yielded the best agreement; we found 
$R_{{\rm p},B}=3.0^{+1.1}_{-0.5}\,R_\odot$ and $M_B=5.7\pm0.6\,M_\odot$ using 
the $\Omega/\Omega_c=0.5$ isochrones. These results imply an equatorial radius 
of $R_{{\rm eq},B}=3.2^{+1.6}_{-0.6}\,R_\odot$. Using von Zeipel's law 
\citep{VonZeipel1924}, we estimate that the ratio of $T_{\rm eff}$ between the pole 
and the equator is approximately $1.02$.

The two A stars' $M$ and $R_{\rm p}$ were inferred using the $\Omega/\Omega_c=0.0$ 
isochrone grid. The $T_{\rm eff}$ and $\log{L/L_\odot}$ derived from the three star 
SED fit then yielded $R_{{\rm p},A}=2.0^{+0.8}_{-0.5}\,R_\odot$ and 
$M_A=2.1\pm0.2\,M_\odot$. We note that both $R_{{\rm p},B}$ and $R_{{\rm p},A}$ are 
consistent with $R_B$ and $R_A$ derived in Section~\ref{SED_fit}.

\begin{figure}
	\centering
	\includegraphics[width=0.999\columnwidth]{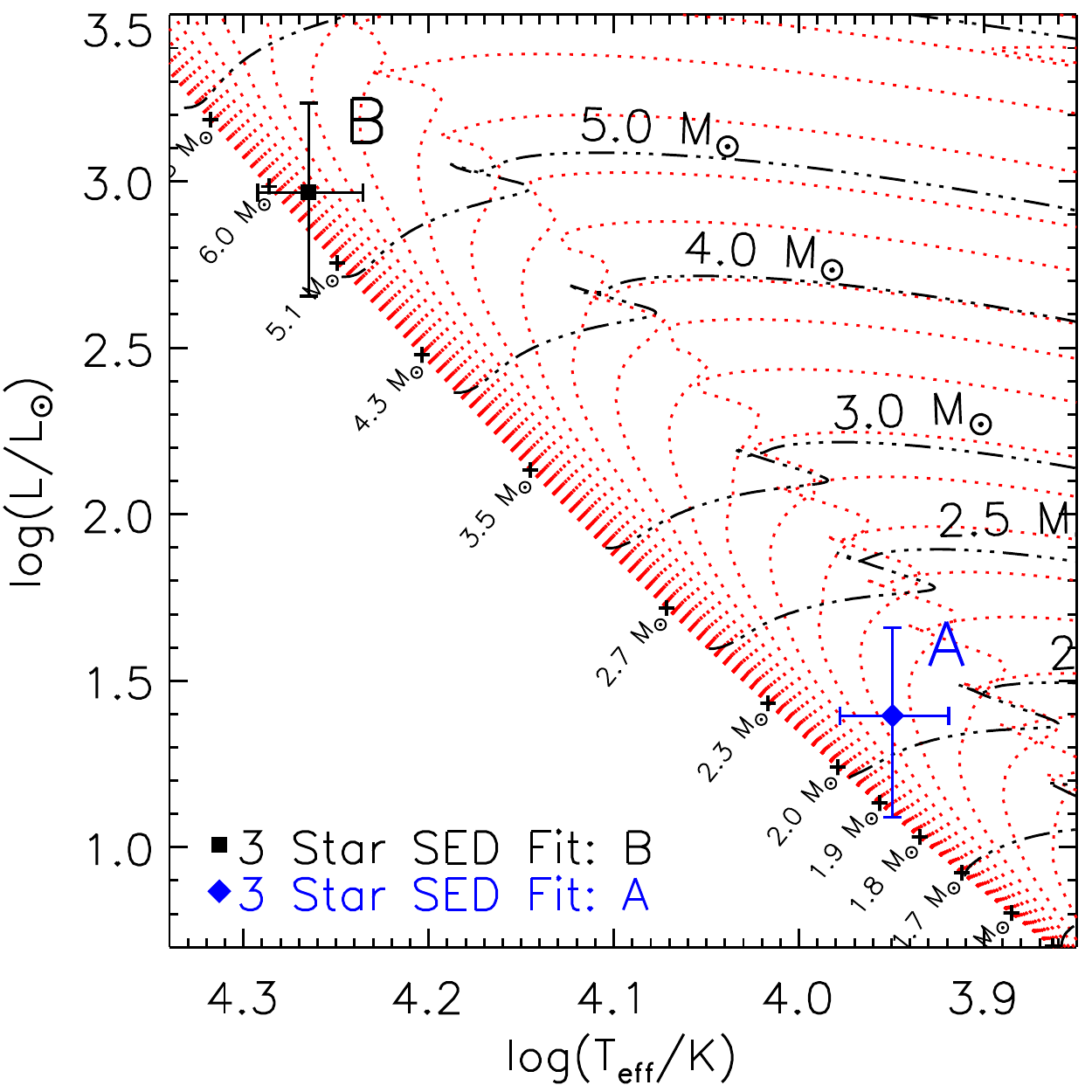}
	\caption{The positions of HD~35502's A and B star components are indicated by 
	the blue diamond and black square. The evolutionary tracks (black dot-dashed 
	lines) and isochrones (red dotted lines) assume a non-rotating star of solar 
	metallicity \citep{Ekstrom2012}.}
	\label{hrd}
\end{figure}

\begin{figure*}
	\centering
	\includegraphics[width=1.9\columnwidth]{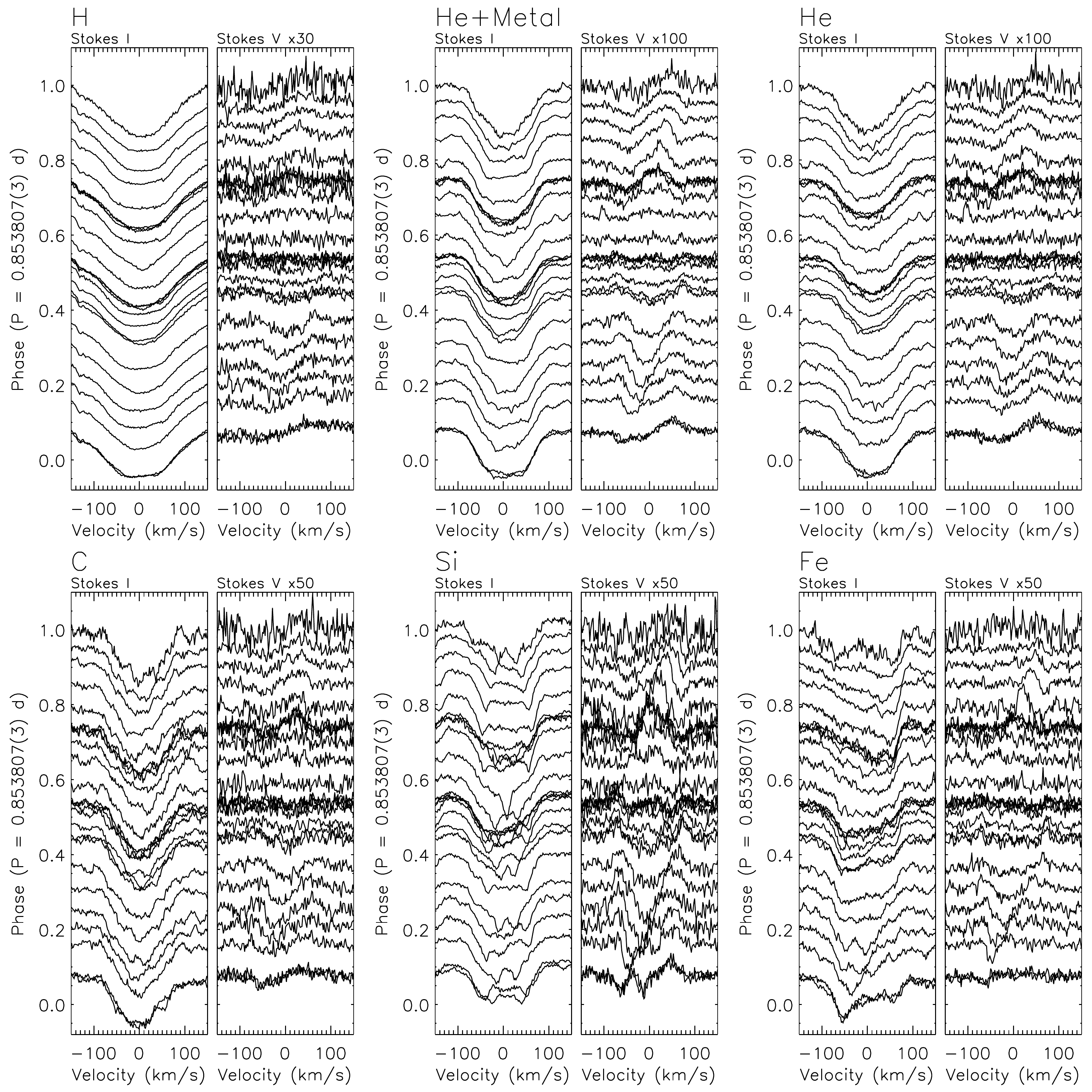}
	\caption{H, He+metal, He, C, Si, and Fe Stokes $I$ and $V$ LSD profiles generated 
	from a $T_{\rm eff}=26\,{\rm kK}$, $\log{g}=4.0\,{\rm (cgs)}$ line mask. The 
	profiles are all phased by the B star's rotational period of $0.853807\,{\rm d}$.}
	\label{multi_phased_lsd}
\end{figure*}

\section{Rotational period}
\label{P_rot}

Several observed properties of HD~35502 exhibit periodic variability with varying 
significance. In order to correctly interpret the origin of these variations, it is 
crucial to identify the periods, the phases at which the maxima and minima occur, 
and amplitudes with which they occur. This was carried out using the same procedure 
discussed in Section~\ref{orb_sol} in which the orbital solution of the A star binary 
was derived.

We first assumed a sinusoidal fit to the data, $f(t)$, given by 
$f(t)=C_0+C_1\sin{(2\pi[t-t_0]/P+C_2)}$, where $P$ is the period of variability, $t$ is 
the observation's HJD, $t_0$ is the HJD corresponding to phase $0$, and $C_0$, $C_1$, 
and $C_2$ are fitting parameters. A $\chi^2$ distribution was then generated using 
periods ranging from $0.1$ to $10.0\,{\rm d}$ in increments $\sim1\,{\rm s}$. The 
best-fitting period was inferred from the minimal $\chi^2$ solution and the $3\sigma$ 
$\chi^2$ interval was taken as the associated uncertainty. The uncertainties in the 
three fitting parameters were estimated using a $1\,000$ iteration bootstrapping 
analysis. The statistical significance of each derived period was evaluated by 
comparing the quality of the sinusoidal fit to that yielded by a constant fitting 
function given by $f(t)=C_0$, where $C_0$ is a time-independent fitting parameter. The 
difference between the minimal $\chi^2$ values associated with the constant fit 
($\chi^2_{\rm const}$) and sinusoidal fit ($\chi^2_{\rm sin}$) were then calculated. 
Any sinusoidal fit having $\chi^2_{\rm const}-\chi^2_{\rm sin}\geq3\sigma$ was 
considered to be statistically significant.

Various periods were found when this procedure was applied to the longitudinal 
field measurements ($\langle B_z\rangle$, see Section~\ref{mag_field}) along with the 
multiple photometry and equivalent width (EW) measurements (see 
Section~\ref{variability}). The analyses of nearly all datasets yielded statistically 
significant variability, with over half corresponding to a unique period near 
$0.85\,{\rm d}$. They were found to be equal to one another within 
$\approx10\,{\rm sec}$ with typical uncertainties $\lesssim10\,{\rm s}$ and were 
therefore averaged to obtain a period of $0.85382(7)\,{\rm d}$. However, when the 
H$\alpha$ EWs were phased with this period, the oldest measurements showed an 
$\approx0.1$ phase offset relative to the more recent measurements. This discrepancy 
was resolved by adopting the best-fitting ephemeris derived from the H$\alpha$ EWs of 
\begin{equation}\label{Prot_eqn}
JD=2456295.812850\pm0.853807(3)\cdot E
\end{equation}
where the reference JD ($2456295.812850(3)$) corresponds to the epoch of 
$\langle B_z\rangle$ maximum magnitude. Therefore, while the general accuracy of the 
rotational period is established by the diverse photometric, spectroscopic, and 
magnetic data sets, the adopted value and its precision correspond to those implied by 
the H$\alpha$ EWs.

The periodic variability of $\langle B_z\rangle$ can be explained, in part, as a 
consequence of a stable oblique magnetic field configuration that is modulated by the 
star's rotation. Similarly, rotationally-modulated variations exhibited by the 
equivalent widths of various spectral lines can be produced by at least two mechanisms: 
(1) non-uniform distributions of chemicals on the stellar surface and (2) hot plasma 
accumulating in the star's magnetosphere resulting in emission and absorption. All of 
these phenomena are commonly exhibited by magnetic B-type stars 
\citep[e.g.][]{Landstreet1978,Leone2010,Bohlender2011}. Therefore, we conclude that the 
ephemeris given by Eqn. \ref{Prot_eqn} is the B star's rotational period.

\section{Magnetic field}
\label{mag_field}

Zeeman signatures produced by a magnetic star in the HD~35502 system were 
detected in circularly polarized (Stokes $V$) ESPaDOnS and Narval observations. They 
were found to be coincident with the B star's spectral lines regardless of the 
inferred velocities of the two A stars. We therefore assumed the detected field to be 
entirely produced by the B star.

The Stokes $V$ Zeeman signature associated with H$\beta$ yielded 18 definite 
detections (DDs), 1 marginal detection (MD), and 7 non-detections (NDs) based on the 
detection criterion outlined by \citet{Donati1997}. The SNRs of the observed 
signatures were optimized using the LSD procedure \citep{Donati1997,Kochukhov2010} 
introduced in Section~\ref{orb_sol}. A master line mask containing He and metal lines 
was generated using data obtained from VALD \citep{Kupka2000} with a specified 
$T_{\rm eff}$, $\log{g}$, and microturbulence velocity ($v_{\rm mic}$). All Balmer 
lines were also removed along with any regions affected by atmospheric absorption 
(i.e. telluric lines). Several single element line masks were subsequently generated 
from the He+metal mask by retaining only specific chemical elements including He, C, 
Si, Fe, and Mg. Clearly, the magnitudes of $\langle B_z\rangle$ derived using different 
elements will be affected by any non-uniform distribution of chemicals across the 
star's surface. Therefore, our analysis also includes measurements obtained using H 
lines (both from LSD profiles and H$\beta$) which do not typically exhibit non-solar 
abundances or non-homogeneous surface distributions (i.e. chemical spots). A H line 
mask was generated in which the H lines exhibiting moderate emission (e.g. H$\beta$ 
and H$\alpha$) were removed. The resultant mask contained three H~{\sc i} lines: 
H~{\sc i}$\,\lambda3970$, H~{\sc i}$\,\lambda4102$, and H~{\sc i}$\,\lambda4340$.

Two approaches were used to isolate the B star lines from the A star lines using 
LSD. The first method used a mask generated with $T_{\rm eff}=15\,{\rm kK}$, 
$\log{g}=4.0\,{\rm (cgs)}$, and $v_{\rm mic}=0\,{\rm km\,s^{-1}}$ which yielded LSD 
profiles with clear Stokes $I$ contributions from all three stellar components. The 
narrow line components associated with the two A stars were then fit by Gaussian 
functions which were subsequently subtracted from the Stokes $I$ profiles. We found 
that this method could not be consistently applied to all observations. Moreover, the 
quality of the Gaussian fits was dramatically reduced when applied to the C, Si, and 
Fe line masks.

The second method used the same $\log{g}$ and $v_{\rm mic}$ with a significantly 
higher temperature of $T_{\rm eff}=26\,{\rm kK}$. This yielded LSD profiles with 
minimal contributions from the two A stars. The Stokes $I$ and $V$ LSD profiles 
generated using the $T_{\rm eff}=26\,{\rm kK}$ line mask for H, He+metal, He, C, Si, 
and Fe are shown in Fig.~\ref{multi_phased_lsd}, phased according to 
Eqn.~\ref{Prot_eqn}. Aside from the Si and Fe LSD profiles, no strong contributions 
from the A star Stokes $I$ profiles can be discerned.

Each ESPaDOnS and Narval spectropolarimetric observation includes a diagnostic null 
which may be used to evaluate the significance of any polarized signal. No spurious 
signals were detected in any of the diagnostic null profiles.

\begin{figure}
	\centering
	\includegraphics[width=0.99\columnwidth]{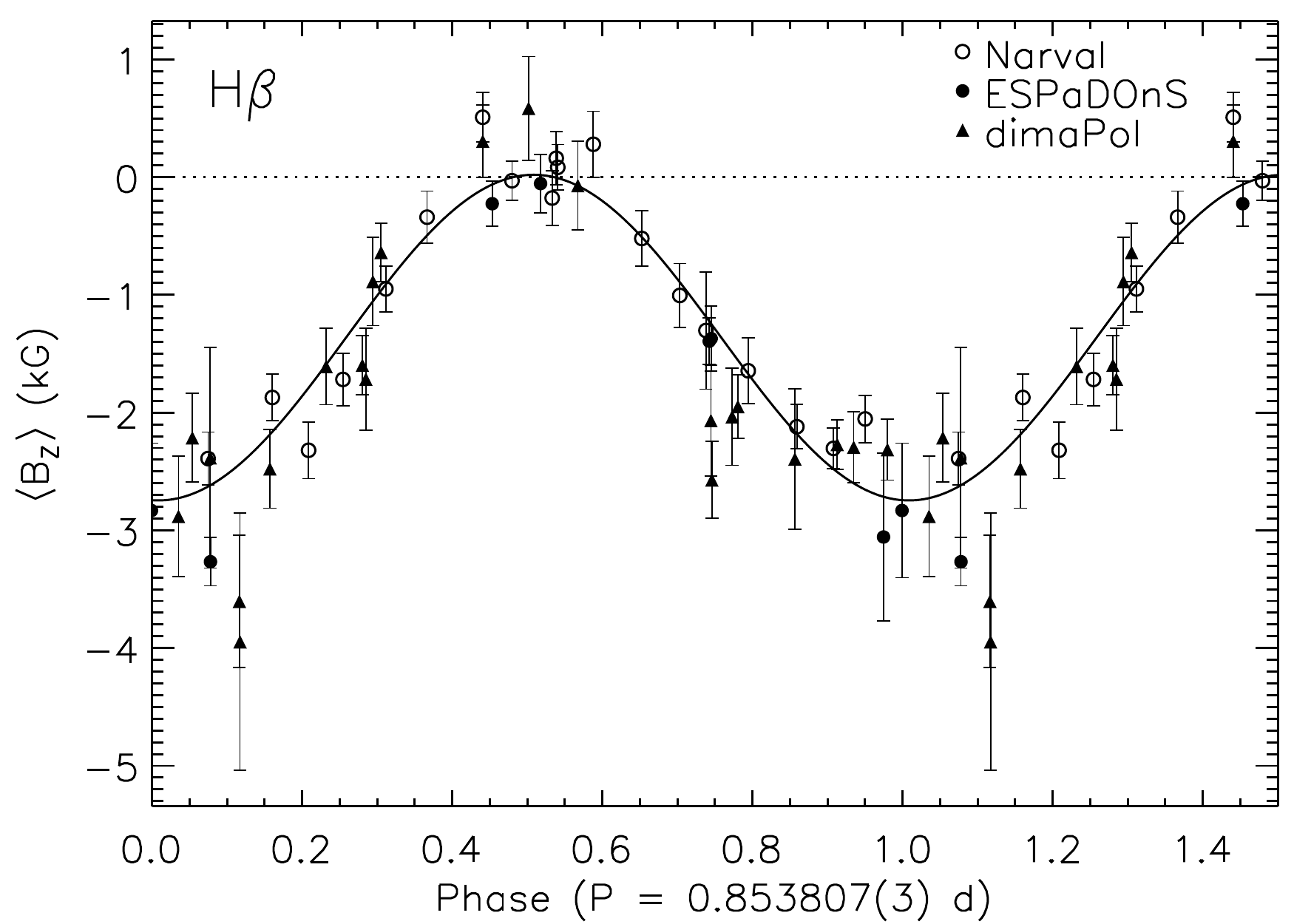}
	\caption{H$\beta$ longitudinal magnetic field measurements phased by the rotational 
	period (Eqn.~\ref{Prot_eqn}). The measurements obtained from Narval (open circles), 
	ESPaDOnS (filled circles), and dimaPol (filled triangles) are shown along with the 
	best fitting sinusoid (solid black).}
	\label{bz_Hbeta_phase}
\end{figure}

$\langle B_z\rangle$ was inferred from each of the Stokes $I$ and $V$ LSD profiles, as 
well as from H$\beta$, using equation (1) of \citet{Wade2000}. We used a wavelength of 
$500\,{\rm nm}$ with a Land\'{e} factor of $1.2$ for the He and metal mask measurements 
and a Land\'{e} factor of unity for the H mask and H$\beta$ measurements. The Doppler 
shift produced by the B star's radial velocity of $\approx20\,{\rm km\,s}^{-1}$ was 
subtracted from each LSD profile. The Stokes $I$ and $V$ profiles were then normalized 
to the continuum intensity at a velocity of $v=-125\,{\rm km\,s}^{-1}$, where the 
average Stokes $V$ intensity is approximately zero. An integration range of 
$v\in[-110,110]\,{\rm km\,s}^{-1}$ was then used in the calculation of 
$\langle B_z\rangle$ for each of the LSD profiles (i.e. H, He+metal, He, C, Si, and 
Fe); for the H LSD profiles and H$\beta$ line, this integration range corresponds to 
the width of the Doppler core. The values of $\langle B_z\rangle_{{\rm H}\beta}$ 
inferred from the Narval, ESPaDOnS, and dimaPol observations, along with the status of 
their detections, are listed in Tables~\ref{obs_tbl} and~\ref{dao_tbl}. 
$\langle B_z\rangle$ derived from the H, He, and metal LSD profiles are listed in 
Table~\ref{bz_full_tbl}.

High resolution spectropolarimetry is essentially insensitive to the polarization in 
the wings of the Balmer lines. As an example of how the Doppler cores of Balmer lines 
may be used to infer $\langle B_z\rangle$, see Fig. 2 of \citep{Landstreet2015}, who 
explain the method in some detail. Our measurements obtained in the context of the 
current paper, as well as those obtained by \citep{Sikora2015}, demonstrate that this 
method results in longitudinal field intensity and variability in good agreement with 
other approaches.

\begin{figure*}
	\centering
	\includegraphics[width=2.1\columnwidth]{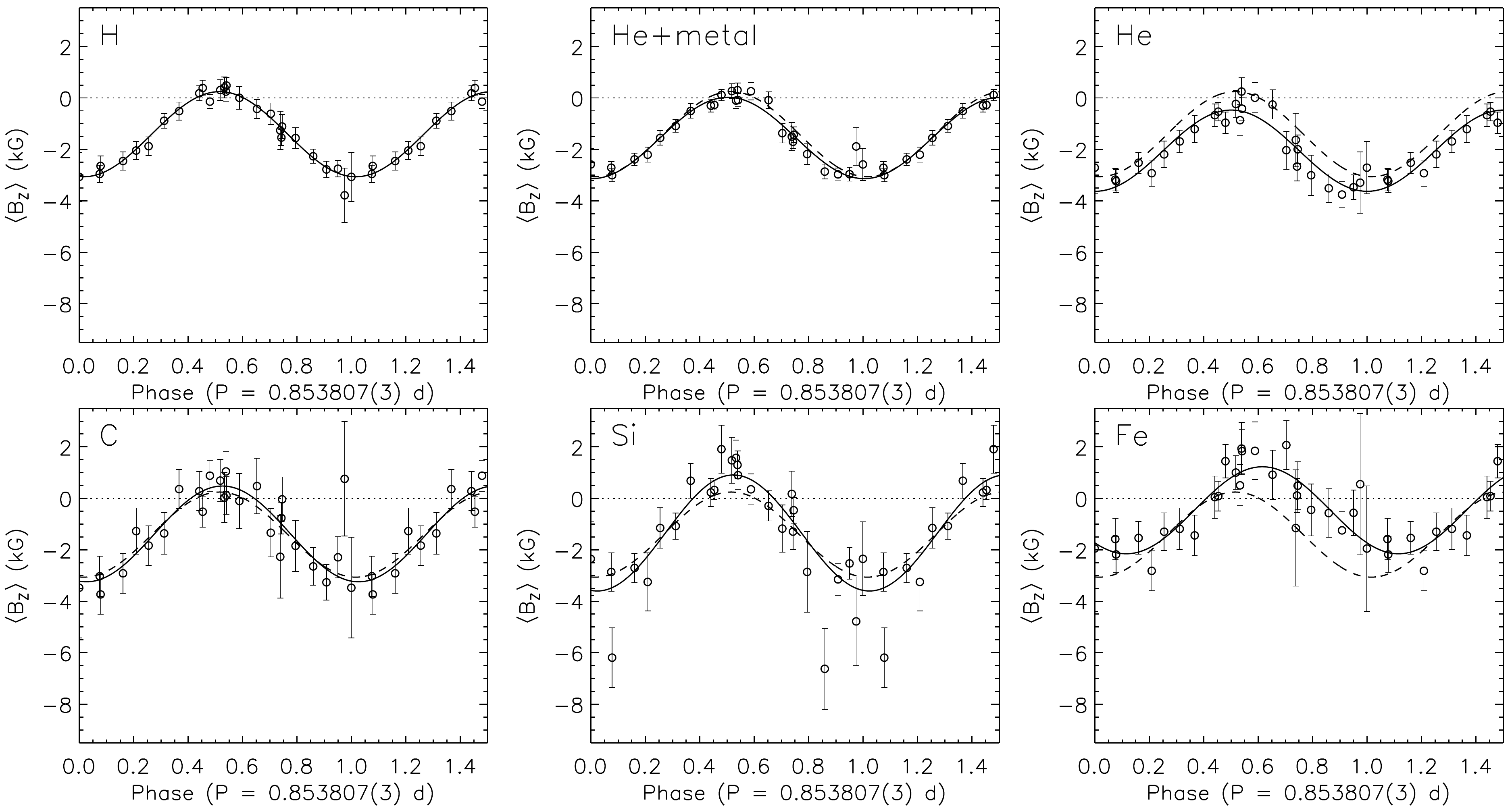}
	\caption{Longitudinal magnetic field measurements phased by the rotational period 
	(Eqn.~\ref{Prot_eqn}). The measurements obtained from the H, He+metal, He, C, Si, 
	and Fe line masks are shown along with their best fitting sinusoids (solid black). 
	The fit to $\langle B_z\rangle_{\rm H}$ is shown as dashed black curves in the 
	other panels.}
	\label{bz_phase}
\end{figure*}

All of the $\langle B_z\rangle$ measurements were found to exhibit statistically 
significant varitions with best-fitting periods ranging from 
$0.85380-0.85389\,{\rm d}$. Only the $\langle B_z\rangle$ values obtained using the Fe 
LSD profiles ($\langle B_z\rangle_{\rm Fe}$) yielded more than one period. 
Figures~\ref{bz_Hbeta_phase} and~\ref{bz_phase} show the $\langle B_z\rangle$ 
measurements inferred from H$\beta$ and the LSD profiles, respectively, phased by the 
B star's rotational period (Eqn.~\ref{Prot_eqn}). It is clear that the scatter of the 
$\langle B_z\rangle_{\rm He}$, $\langle B_z\rangle_{\rm Si}$, and 
$\langle B_z\rangle_{\rm Fe}$ measurements is significantly larger than that yielded 
by $\langle B_z\rangle_{\rm H}$, $\langle B_z\rangle_{\rm He+metal}$, and 
$\langle B_z\rangle_{\rm C}$. This is likely caused by the presence of He, Si, and Fe 
chemical spots which are commonly observed on the surfaces of Bp stars. As discussed in 
Section~\ref{variability}, we find strong evidence for He and Si spots.

The mean and amplitude of the phased $\langle B_z\rangle$ measurements are defined by 
the fitting parameters $B_0$ and $B_1$ associated with the sinusoidal fitting function 
$\langle B_z\rangle=B_0+B_1\sin{(2\pi\theta+\phi)}$, where $\theta$ is the phase 
calculated using Eqn.~\ref{Prot_eqn} and $\phi$ is the phase shift. The most precise 
$B_0$ and $B_1$ values -- as indicated by the uncertainties estimated using a $1\,000$ 
iteration bootstrapping analysis -- were derived using H$\beta$, and the H and 
He+metal LSD profiles. They were found to be consistent within $2\sigma$. The lowest 
uncertaintes were obtained from the $\langle B_z\rangle_{\rm H}$ measurements, which 
yielded a mean and amplitude of $B_0=-1.41\pm0.11\,{\rm kG}$ and 
$B_1=1.64\pm0.16\,{\rm kG}$ where the uncertainties correspond to $3\sigma$.

If we assume that the field is characterized by an important dipole component, the 
sinusoidal variations in $\langle B_z\rangle$ imply that the dipole's axis of symmetry 
is inclined (i.e. has an obliquity angle $\beta$) with respect to the star's rotational 
axis. This interpretation, first described by \citet{Stibbs1950}, is known as the 
Oblique Rotator Model (ORM). Under the assumptions of the ORM, $\beta$ can be 
calculated from equation (3) of \citet{Preston1967} which depends on 
$r\equiv|\langle B_z\rangle|_{\rm min}/|\langle B_z\rangle|_{\rm max}$ and the 
inclination angle, $i$, of the star's axis of rotation. The value of $i$ can be 
determined using $P_{\rm rot}$ given by Eqn.~\ref{Prot_eqn}, 
$v\sin{i}=75\pm5\,{\rm km\,s}^{-1}$ derived in Section~\ref{line_fit}, and 
$R_{\rm eq}=3.2^{+1.6}_{-0.6}\,R_\odot$ listed in Table~\ref{param_tbl}. We obtained a 
value of $i=24^{+8}_{-9}\,\degree$. The value of $r$ was determined from $B_0$ and 
$B_1$. Using the values inferred from the $\langle B_z\rangle_{\rm H}$ measurements, 
we obtained $r=0.08^{+0.09}_{-0.07}$. Finally, the obliquity angle was found to be 
$\beta=63\pm13\,\degree$ using Eqn. (3) of \citet{Preston1967}.

In addition to the obliquity, the strength of the magnetic field's dipole component, 
$B_p$, can be calculated by inverting equation (1) of \citet{Preston1967} and 
letting $t=0$ correspond to $\langle B_z\rangle_{\rm max}$. We used a linear limb 
darkening constant that was averaged over the values derived by \citet{vanHamme1993} 
for the $U$, $B$, $V$, $R$, and $I$ bandpasses. These specific filters were selected 
because of the approximate correspondance with the ESPaDOnS and Narval wavelength 
range. A value of $u=0.265$ was obtained after interpolating the published table for an 
effective temperature and surface gravity of $18.4\,{\rm kK}$ and $4.3\,{\rm (cgs)}$. 
$i$, $\beta$, and $u$ then yield $B_p=14^{+9}_{-3}\,{\rm kG}$. Similar obliquity 
angles and dipolar field strengths are derived using H$\beta$ along with the He+metal, 
He, C, and Si LSD profiles. $\langle B_z\rangle_{\rm Fe}$ exhibits significantly 
weaker and more uncertain values of $\beta=33^{+36}_{-27}\,\degree$ and 
$B_p=8^{+6}_{-3}$.

\section{Emission and variability}
\label{variability}

Hot magnetic B-type stars are commonly found to exhibit spectral line variability 
either as a result of chemical spots \citep[e.g.][]{Kochukhov2015,Yakunin2015} or from 
the presence of a hot plasma beyond the stellar surface \citep[e.g.][]{Landstreet1978}. 
Furthermore, photometric variability correlated with both of these phenomena, as well 
as with strong, coherent magnetic fields has been previously reported 
\citep[e.g.][]{Shore1990a,Oksala2010}.

Along with the $uvby$ photometric measurements listed in Table~\ref{uvby_tbl}, we also 
analyzed \emph{Hipparcos} Epoch Photometry for variability. The catalogue 
\citep{Perryman1997} contains 98 observations of HD~35502 which were obtained over a 
period of $3.1\,{\rm yrs}$. Three of these measurements have multiple quality flags 
reported and were therefore removed from our analysis. The remaining measurements have 
an average of $7.331\,{\rm mag}$, a standard deviation of $0.011\,{\rm mag}$, and an 
average uncertainty of $0.009\,{\rm mag}$. 

\begin{figure}
	\centering
	\includegraphics[width=0.99\columnwidth]{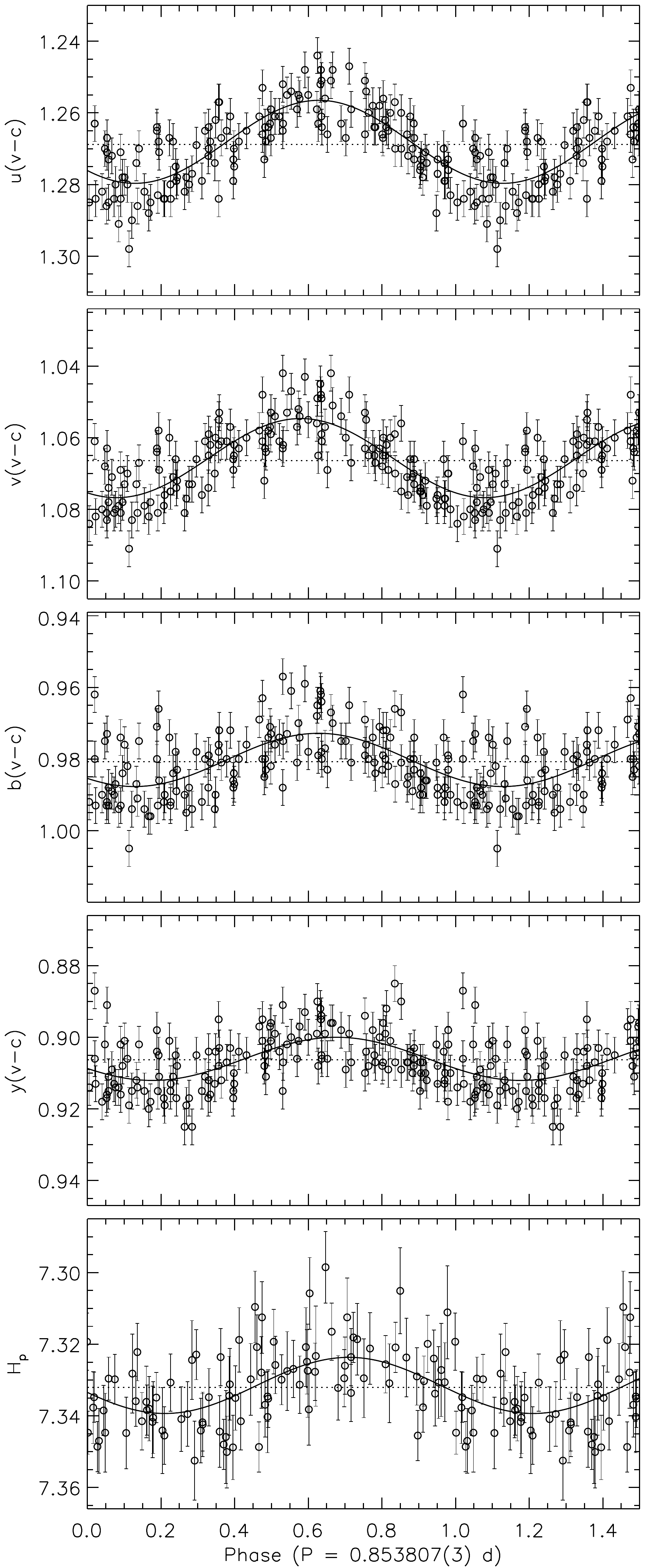}
	\caption{Photometric observations obtained using various filters and phased by the 
	B star's rotational period of $0.853807\,{\rm d}$. The top four panels show the 
	$u$(v-c), $v$(v-c), $b$(v-c), and $y$(v-c) magnitude differences between HD~35502 
	(`v') and the non-variable comparison star, HD~35575 (`c'). The bottom panel shows 
	the \emph{Hipparcos} Epoch Photometry measurements. The horizontal dotted lines 
	indicate the best constant fit to the data.}\label{photometry_plt}
\end{figure}

The period searching routine described in Section \ref{P_rot} was applied to both the 
$uvby$ and \emph{Hipparcos} data sets. All of the $uvby$ measurements were found to 
exhibit statistically significant variability; however, only $u$(v-c) and $v$(v-c) 
yielded unique periods of $0.8537(3)\,{\rm d}$. The analysis of the \emph{Hipparcos} 
magnitudes ($H_p$) resulted in a best-fitting period of $0.8630(2)\,{\rm d}$ along 
with five other statistically significant periods ranging from $0.46$ to 
$1.7\,{\rm d}$. Fig.~\ref{photometry_plt} shows the sinusoidal fits to $u$(v-c), 
$v$(v-c), $b$(v-c), $y$(v-c), and $H_p$ obtained when phased by the B star's 
$0.853807\,{\rm d}$ rotational period given by Eqn.~\ref{Prot_eqn}.

The variability of the spectral lines associated with HD~35502's central B star is 
most easily detected by calculating EWs. We carried this out for a number of lines for 
which no significant absorption produced by the two A stars was evident. This included 
He~{\sc i}, C~{\sc ii}, and Si~{\sc iii} lines. The EWs of the He and metal 
lines were calculated using integration ranges of $[-100,100]\,{\rm km\,s}^{-1}$ and 
were normalized to the continuum just outside these limits. The Balmer line EWs 
(H$\alpha$, H$\beta$, and H$\gamma$) were measured by normalizing to the flux at 
$|v|\gtrsim700\,{\rm km\,s}^{-1}$ and integrating over a velocity range of 
$[-600,600]\,{\rm km\,s}^{-1}$. The uncertainties in the EW measurements were then 
estimated using a bootstrapping analysis with $1\,000$ iterations. All of the 
calculated EWs and uncertainties are listed in and Tables~\ref{halpha_ew_tbl} and 
\ref{ew_tbl}.

The contributions of the two A stars to the total measured Balmer line EWs were 
approximated by comparing the synthetic EWs associated with the {\sc synth3} models 
discussed in Section~\ref{line_fit}. We found that the total synthetic spectrum 
(including the B star and the two A stars) yielded EWs of $5.6\,{\rm \AA}$ averaged 
over H$\alpha$, H$\beta$, H$\gamma$, and H$\delta$. A similar calculation applied to 
the single B star model yielded average Balmer line EWs of $4.7\,{\rm \AA}$ suggesting 
that the presence of the two A stars increase the EW measurements by a factor of 
$\approx1.2$.

Statistically significant variations were detected from EW measurements of H$\alpha$, 
H$\beta$, H$\gamma$, He~{\sc i}$\,\lambda4713$, and Si~{\sc iii}$\,\lambda4553$. 
We note that telluric absorption lines were not removed or minimized in the 
calculation of these EWs. A range of best-fitting periods were derived; however, only 
the H$\alpha$ and He~{\sc i}$\,\lambda4713$ EW measurements yielded unique periods of 
$0.853807(3)\,{\rm d}$ and $0.85377(3)\,{\rm d}$, respectively. The strongest 
variability was measured from H$\alpha$ for which an amplitude of 
$0.51\pm0.03\,{\rm \AA}$ was derived. Similar variability -- both in terms of the phase 
of maximum emission and the best-fitting period -- was also detected in H$\beta$ 
although, at a much lower amplitude of $0.08\pm0.01\,{\rm \AA}$. The phased H$\alpha$ 
and H$\beta$ EWs exhibit a maximum emission at a phase of $0.99\pm0.04$ and 
$0.0\pm0.3$, respectively, and are therefore in phase with the 
$\langle B_z\rangle$ measurements. The He~{\sc i}$\,\lambda4713$ EWs are approximately 
in anti-phase with respect to the $\langle B_z\rangle$ variation with minimum 
absorption occuring at a phase of $0.5\pm0.1$.

\begin{figure}
	\centering
	\includegraphics[width=0.99\columnwidth]{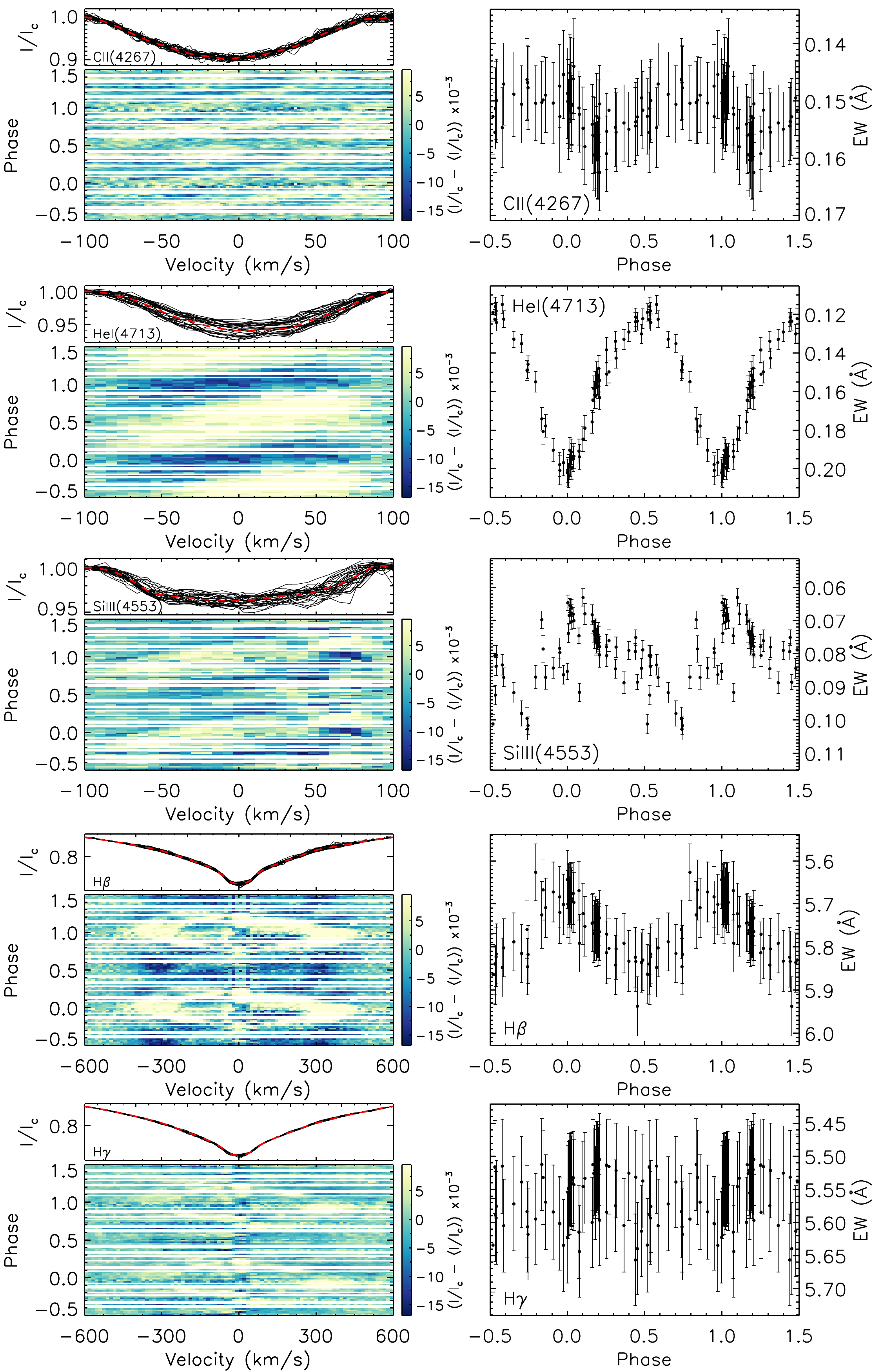}
	\caption{\emph{Left:} Dynamic spectra of various H~{\sc i}, He~{\sc i}, C~{\sc ii},
	and Si~{\sc iii} lines. A low-pass filter has been applied to the 
	spectrum in order to minimize the A star contributions and tellurics. Each set of 
	observations are compared with the average spectrum (dashed red). \emph{Right:} 
	Measured equivalent widths associated with the lines shown in the dynamic spectra 
	plots. All of the measurements are phased by the B star's rotational period of 
	$0.853807\,{\rm d}$.}
	\label{ew_He_metal}
\end{figure}

Along with EWs, dynamic spectra were also computed by comparing various spectral 
lines with their respective average normalized intensity ($\langle I/I_c\rangle$). 
Both the dynamic spectra and the EWs of C~{\sc ii}$\,\lambda4267$, 
He~{\sc i}$\,\lambda4713$, Si~{\sc iii}$\,\lambda4553$, H$\beta$, and H$\gamma$ are 
shown in Fig. \ref{ew_He_metal}. It is evident that He~{\sc i}$\,\lambda4713$, 
Si~{\sc iii}$\,\lambda4553$, and to a lesser extent, C~{\sc ii}$\,\lambda4267$, show 
absorption features crossing from negative to positive velocities. These features 
suggest the presence of chemical spots on the B star's surface. The most obvious spot 
is associated with He~{\sc i}, which exhibits a maximum absorption at a phase of 
$0.0\pm0.1$ and is therefore coincident with the epoch of maximum $\langle B_z\rangle$ 
magnitude. Assuming that the star's magnetic field consists of a strong 
dipole component as discussed in Section~\ref{mag_field}, this result suggests that He 
is more concentrated near the field's negative pole. Enhanced He abundances on the 
surfaces of magnetic Bp stars have been commonly reported to coincide with either the 
magnetic equator or magnetic poles \citep[e.g.][]{Neiner2003,Bohlender2011,
Grunhut2012b,Rivinius2013}.

\begin{figure}
	\centering
	\includegraphics[width=0.99\columnwidth]{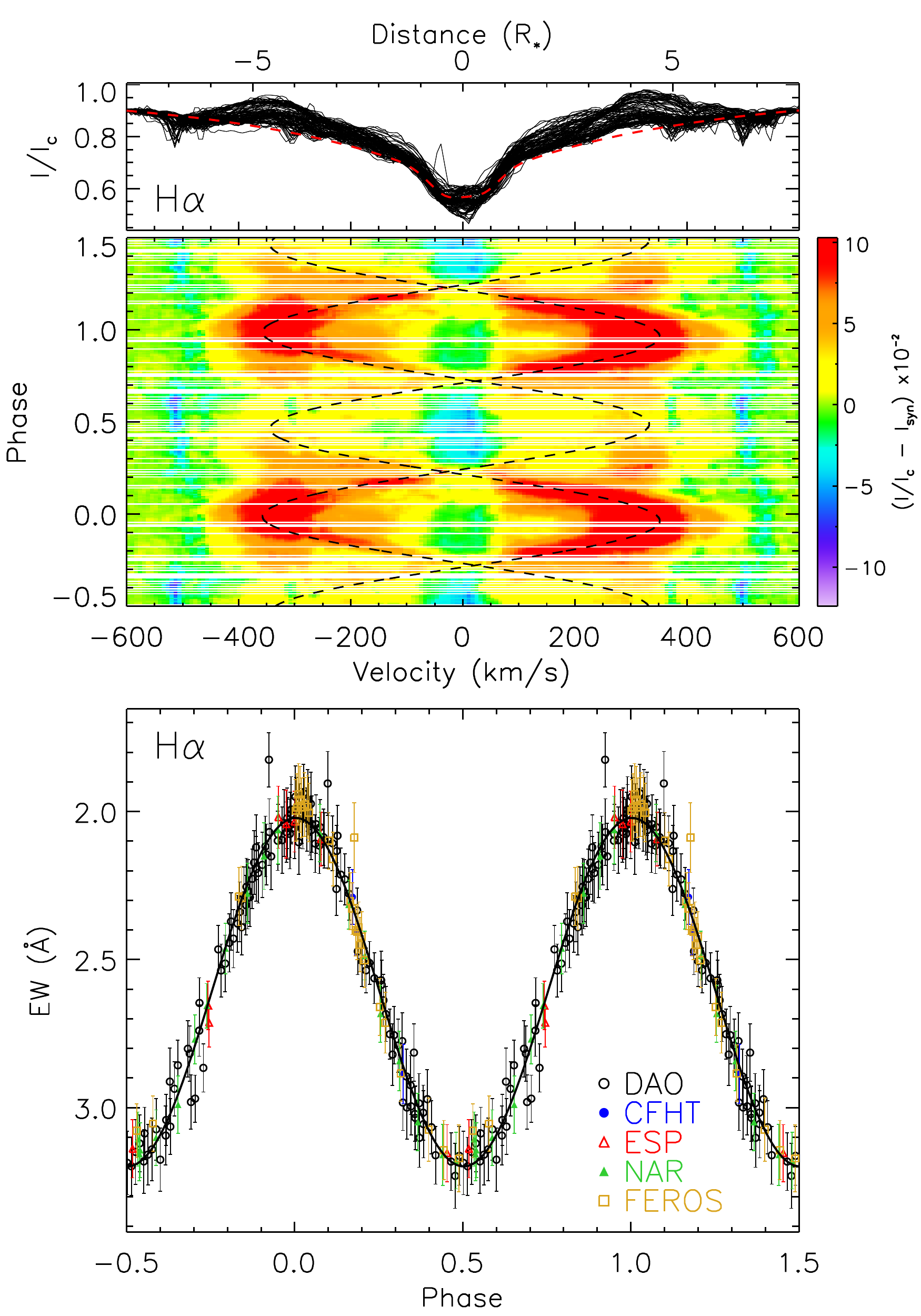}
	\caption{\emph{Top:} Observed H$\alpha$ spectrum ($I/I_c$, solid black curve) 
	compared with the average synthetic spectrum (dotted red). \emph{Middle:} Dynamic 
	spectra of $I/I_c-I_{\rm syn}$ where $I_{\rm syn}$ is the synthetic spectrum. A 
	low-pass filter has been applied to the spectrum in order to minimize the A star 
	contributions and tellurics. \emph{Bottom:} Measured equivalent widths of H$\alpha$ 
	using ESPaDOnS (open red triangles), Narval (filled green triangles), FEROS (yellow 
	squares), CFHT f/8.2 spectrograph (filled blue circles), and DAO (open black 
	circles) observations. The black curve indicates the best-fitting sinusoid. Both 
	the EWs and dynamic spectra are phased by the B star's rotational period of 
	$0.853807\,{\rm d}$.}
	\label{ew_Halpha}
\end{figure}

A similar plot of the dynamic spectrum and EWs of H$\alpha$ is shown in 
Fig.~\ref{ew_Halpha}, where the additional DAO and CFHT spectra are also included. In 
order to reduce normalization errors, all of the H$\alpha$ spectra were consistently 
normalized using a linear fit to the measured flux at velocities of 
$\pm600\,{\rm km\,s}^{-1}$. The observed spectra are compared with the synthetic 
spectrum ($I_{\rm syn}$) discussed in Section~\ref{line_fit} rather than the average 
observed spectrum. $I_{\rm syn}$ includes the contributions from the A stars which move (in 
velocity space) relative to the B star throughout the B star's rotational period. 
Therefore, this method results in a greater contrast between the emission and 
absorption features associated only with the B star. Strong, nearly symmetrical 
emission peaks are observed at a distance of $\approx4\,R_\ast$ at a phase of $0.0$. 
The intensity of this emission is observed to decrease by a factor of $\approx2$ at a 
phase of $0.5$. Similarly, the ratio between the maximum core emission (at phase 
$0.25$) and minimum core emission (at phase $0.5$) is also found to be $\approx2$.

The standard interpretation of the broad H$\alpha$ emission peaks that are associated 
with a small number of magnetic B-type stars is that they are produced by two dense 
clouds of hot plasma, trapped in the magnetic field above the stellar surface, which 
co-rotate with the star \citep[e.g.][]{Walborn1976,Landstreet1978}. Under the 
assumption that the cloud is optically thin, one would expect the same blue shifted 
emission feature to be observed half a rotational cycle later shifted towards redder 
wavelengths (e.g. in Fig.~\ref{ew_Halpha}, the blue emission peak occuring at phase 
0.0 should reappear red shifted at phase 0.5). The fact that the strength of both the 
blue and red shifted emission peaks decrease between phase 0.0 and phase 0.5 suggests 
that the plasma clouds are, to an extent, optically thick. The relatively large 
decrease in emission is currently unprecedented amongst the known CM hosting stars; 
however, a more moderate decrease in HR~5907's H$\alpha$ emission is shown in 
Fig.~15 of \citet{Grunhut2012b}.

Adopting the standard interpretation, the trajectories of the H$\alpha$-emitting clouds 
may be approximately inferred by fitting the velocities at which the peak emission is 
found on either side of the H$\alpha$ core as a function of rotational phase. The 
resulting fits suggest that the plasma clouds follow nearly circular trajectories as 
indicated by the two dashed curves shown in Fig.~\ref{ew_Halpha}. The mechanism by 
which this plasma is confined is discussed in the following section.

\section{Magnetosphere}
\label{sect_magnetosphere}

As described by \citet{ud-Doula2008_oth}, various characteristics of a star's 
magnetosphere may be inferred by comparing two parameters: the Kepler radius, $R_K$, 
and the Alfv\'{e}n radius, $R_{\rm Alf}$. $R_K$ is the radius at which the 
gravitational force is balanced by the centrifugal force in a reference frame that is 
co-rotating with the star. $R_{\rm Alf}$ characterizes the point within which the 
magnetic field dominates over the wind and approximately corresponds to the extent of 
the closed field loops \citep{ud-Doula2002_oth,ud-Doula2008_oth}. Their ratio, 
$R_{\rm Alf}/R_K$, can therefore be used to define a magnetosphere as either dynamical 
($R_{\rm Alf}/R_K<1$) or centrifugal ($R_{\rm Alf}/R_K>1$) \citep{Petit2013}. It also 
serves as an indicator of the volume of the magnetosphere: those stars having 
comparatively larger $R_{\rm Alf}/R_K$ will be capable of confining the emitted wind at 
larger radii. Furthermore, since a stronger field would be capable of confining more 
mass, a correlation between the Alfv\'{e}n radius and the magnetosphere's density may 
be expected.

Using the mass and rotational period of HD~35502's B star, we find a Kepler radius 
of $R_K=2.1^{+0.4}_{-0.7}\,R_\ast$, where $R_\ast$ is the stellar radius at the 
magnetic equator. We approximate $R_\ast$ using $R_{\rm eq}$ since this corresponds 
to the stellar radius at the latitude where the plasma is expected to accumulate. The 
Alv\'{e}n radius is estimated using equation (9) of \citet{ud-Doula2008_oth} for a 
dipole magnetic field. This expression requires the calculation of the wind confinement 
parameter, $\eta_\ast$, which in turn depends on the dipole magnetic field strength, 
the equatorial radius, the terminal wind speed ($V_\infty$), and the wind mass loss 
rate in the absence of a magnetic field ($\dot{M}_{B=0}$). Following the recipe 
outlined by \citet{Vink2000}, $\dot{M}_{B=0}$ and $V_\infty$ are derived for a B star 
having $12.5<T_{\rm eff}\leq22.5\,{\rm kK}$ using $V_\infty/V_{\rm esc}=1.3$, where 
$V_{\rm esc}$ is the escape velocity. We obtain 
$\dot{M}_{B=0}=(1.3^{+6.0}_{-1.0})\times10^{-10}\,M_\odot/{\rm yr}$ and 
$V_\infty=1100^{+40}_{-110}\,{\rm km\,s}^{-1}$. Finally, $\eta_\ast$ is found to be 
$(2.6^{+7.9}_{-1.3})\times10^6$ using the value of $B_p$ derived from 
$\langle B\rangle_{\rm H}$, which then yields $R_{\rm Alf}=41^{+17}_{-6}\,R_\ast$. The 
magnetospheric parameters associated with both the H and He+metal longitudinal field 
measurements derived in Section~\ref{mag_field} are listed in Table~\ref{magneto_tbl}.

Given that the hot plasma surrounding HD~35502's B star is co-rotating with the star at 
a distance of $\approx4\,R_\ast$, i.e. between $R_K$ and $R_{\rm Alf}$, it is likely 
that the plasma is being confined by the strong magnetic field. Similar examples of 
magnetic B-type stars producing H emission well beyond the stellar radius (at distances 
of $\approx2-4\,R_\ast$) have been previously reported \citep[e.g.][]{Bohlender2011,
Oksala2012,Grunhut2012b}. In each of these cases, the star's Alfv\'{e}n radius exceeds 
its Kepler radius by approximately an order of magnitude \citep{Petit2013,Shultz2014}. 
Using the $R_{\rm Alf}$ value obtained from the $\langle B\rangle_{\rm H}$ measurements, we 
derived an $R_{\rm Alf}$ to $R_K$ ratio of $19^{+20}_{-5}$. Therefore, the fact that we observe 
strong H$\alpha$ emission is in agreement with this $R_{\rm Alf}/R_K\gtrsim10$ empirical 
limit.

The magnetic confinement-rotation diagram compiled by \citet{Petit2013} allows $R_K$ 
and $R_{\rm Alf}$ to be understood within the broader context of all known O and B stars that 
host magnetospheres. Our characterization of the magnetosphere hosted by HD~35502's B 
star suggests that it is well within the centrifugal magnetosphere regime. Only 
two other stars have been discovered exhibiting similar $R_{\rm Alf}$ and $R_K$ values within 
the derived $R_{\rm Alf}>40\,R_\ast$ and $1.5\leq R_K\leq2.5\,R_\ast$. Although approximately 
four other stars have lower limits of $R_{\rm Alf}$ and $R_K$ that are consistent 
with HD~35502's, only HD$\,182180$ \citep{Rivinius2013} and HD$\,142184$ 
\citep{Grunhut2012b} have reported upper and lower uncertainties. These two examples 
have similar effective temperatures, surface gravities, radii, and masses to HD~35502's 
magnetic B star. However, HD$\,182180$ and HD$\,142184$ are slightly faster rotators 
($P_{\rm rot}\approx0.5\,{\rm d}$) and host magnetic fields with weaker dipolar 
components ($B_p\approx10\,{\rm kG}$).

\begin{table}
	\caption{Magnetospheric parameters derived from $\langle B_z\rangle_{\rm H}$ and 
	$\langle B_z\rangle_{\rm He+metal}$ measurements.}
	\label{magneto_tbl}
	\begin{center}
	\begin{tabular*}{0.9\columnwidth}{@{\extracolsep{\fill}}l c c}
		\noalign{\vskip1mm}
		\hline
		\noalign{\vskip1mm}
		                                          & H$_{\rm LSD}$ & He+metal$_{\rm LSD}$ \vspace{0.8mm}\\
		\hline
		\noalign{\vspace{0.8mm}}
		$i\,(\degree)$                            & $24^{+8}_{-9}$                      & $24^{+8}_{-9}$                      \vspace{0.8mm}\\
		$\beta\,(\degree)$                        & $63\pm13$                           & $66^{+9}_{-13}$                     \vspace{0.8mm}\\
		$B_p\,{\rm (kG)}$                         & $14^{+9}_{-3}$                      & $15^{+10}_{-4}$                     \vspace{0.8mm}\\
		$\dot{M}_{B=0}\,(M_\odot\,{\rm yr}^{-1})$ & $(1.3^{+6.0}_{-1.0})\times10^{-10}$ & $(1.3^{+6.0}_{-1.0})\times10^{-10}$ \vspace{0.8mm}\\
		$V_\infty\,({\rm km\,s}^{-1})$            & $1100^{+40}_{-110}$                 & $1100^{+40}_{-110}$                 \vspace{0.8mm}\\
		$\eta_\ast$                               & $(2.6^{+7.9}_{-1.3})\times10^6$     & $(3.0^{+6.5}_{-1.5})\times10^6$     \vspace{0.8mm}\\
		$R_{\rm Alf}\,(R_\ast)$                   & $41^{+17}_{-6}$                     & $42^{+14}_{-7}$                     \vspace{0.8mm}\\
		$R_K\,(R_\ast)$                           & $2.1^{+0.4}_{-0.7}$                 & $2.1^{+0.4}_{-0.7}$                 \vspace{0.8mm}\\
		$R_{\rm Alf}/R_K$                         & $19^{+20}_{-5}$                     & $20^{+18}_{-6}$                     \vspace{0.8mm}\\
		\hline
	\end{tabular*}
	\end{center}
\end{table}

\section{Conclusion}
\label{conclusions}

The analysis presented here demonstrates a number of new discoveries regarding the 
nature of HD~35502. The high resolution spectroscopic and spectropolarimetric 
observations obtained using ESPaDOnS, Narval, and FEROS indicate that it is an SB3 
system containing a central magnetic B-type star and two cooler A-type stars, all of 
which lie on the main sequence. We confirm that HD~35502's speckle companion reported 
by \citet{Balega2012} is indeed the A star binary system. Our analysis indicates 
that both the A stars are physically nearly identical with a mass ratio of 
$1.05\pm0.02$, masses of $2.1\,M_\odot$, and effective temperatures of 
$8.9\,{\rm kK}$.

Based on radial velocity measurements, we find that the two A stars form a binary 
system with an orbital period of $5.66866(6)\,{\rm d}$. No radial velocity variations of 
the B star were detected over the 22 year observing period, which can be explained by 
the inferred orbital period of $P_{\rm orb}\gtrsim40\,{\rm yrs}$. However, two other 
explanations can account for the lack of detected radial velocity variations: (1) the 
inclination angle associated with the A star binary's orbit about the B star may be 
$\sim0\,\degree$ or (2) the A star binary may lie along the line of sight but not be 
gravitationally bound to the B star. A number of factors favour the triple system 
description such as the consistent flux ratios between the three components derived 
here. Specifically, if the two A stars are significantly closer or further than HD 
35502's $400\pm80\,\rm{pc}$ distance, they would no longer lie on the main sequence. 
Furthermore, the radial velocity of the Orion OB1a subassociation, of which HD~35502 
is most likely a member, has a reported velocity of $\approx24\,{\rm km\,s}^{-1}$ 
\citep{Morrell1991} and is therefore consistent with both the B star's average radial 
velocity of $21\pm2\,{\rm km\,s}^{-1}$ and the radial velocity of the A star 
binary's center of mass ($27\pm3\,{\rm km\,s}^{-1}$).

Our analysis of HD~35502's central B star revealed the following:
\begin{enumerate}[leftmargin=0.5cm]
	\item it has an effective temperature of $18.4\,{\rm kK}$, a mass of 
	$5.7\,M_\odot$, and a polar radius of $3.0\,R_\odot$;
	\item it rotates relatively rapidly with a rotational period of 
	$0.853807(3)\,{\rm d}$;
	\item we detect a strong magnetic field and derive the magnitude of its dipolar 
	component ($B_p\approx14\,{\rm kG}$) and its obliquity 
	($\beta=63\,\degree$);
	\item it exhibits significant line variability in the form of emission and 
	chemical spots. The emission is predominantly observed in H$\alpha$ at a distance 
	of approximately four times the stellar radius. Strong He abundance variations 
	indicate a higher concentration near the negative pole of the magnetic field's 
	dipole component;
	\item we derive an Alfv\'{e}n and Kepler radii of $R_{\rm Alf}\approx41\,R_\ast$ 
	and $R_K\approx2.1\,R_\ast$ unambiguously indicating that HD~35502's B star hosts 
	a large centrifugally supported magnetosphere.
\end{enumerate}

Our analysis indicates that the `sn' classification appearing in HD~35502's historical 
B5IVsnp spectral type \citep{Abt1977} is most likely related to the presence of the 
sharp-lined binary companion along with the strong H emission. We therefore propose 
that this system be reclassified as a B5IVpe+A+A system.

Stars hosting strongly emitting centrifugal magnetospheres can provide useful insights 
towards our understanding of both stellar winds and stellar magnetism. Therefore, it 
is important that these rare systems be studied in detail. While this particular 
class of magnetic stars is certainly growing, the number of confirmed examples are 
still insufficient to solve various outstanding issues. For instance, the inferred 
magnetospheric material densities of CM-hosting stars are largely inconsistent with 
currently predicted values \citep[e.g.][]{Rivinius2013,Townsend2013,Shultz2014}. 
Moreover, testing the validity of theoretical models describing the physical nature of 
these magnetospheres \citep[e.g. the Rigidly Rotating Magnetosphere model derived 
by][]{Townsend2005} requires detailed comparisons with a diversity of observations such 
as those recently carried out by \citet{Oksala2012,Oksala2015}.

We conclude that the strong variability -- both in terms of $\langle B_z\rangle$ and 
the observed line variability -- makes HD~35502 a favourable subject of a 
magnetic Doppler imaging analysis \citep{Piskunov2002}. However, any attempt would 
require the three spectral components to be disentangled. Our orbital solution 
provides the necessary first step towards accomplishing this task.

\section*{Acknowledgments}

GAW acknowledges support in the form of a Discovery Grant from the Natural Science and 
Engineering Research Council (NSERC) of Canada. SIA thanks the United States National 
Science Foundation for support on his differential photometric studies. We thank Dr. 
Jason Grunhut for the helpful discussion regarding the interpretation of the 
magnetospheric emission.

\bibliography{HD35502.bib}
\bibliographystyle{mn2e}

\begin{table*}
	\caption{Longitudinal field measurements derived using the LSD profiles that were 
	generated from the ESPaDOnS and Narval spectra. Columns 3 to 6 list 
	$\langle B_z\rangle$ derived from H, He+metal, He, C, Si, and Fe LSD profiles. 
	This table will appear only in the electronic version of the paper.}
	\label{bz_full_tbl}
	\begin{center}
	\begin{tabular*}{2\columnwidth}{@{\extracolsep{\fill}}l c r r r r r r}
		\hline
		\hline
		\noalign{\vskip0.5mm}
		HJD & Instrument & \multicolumn{1}{c}{$\langle B_z\rangle_{\rm H}$} & \multicolumn{1}{c}{$\langle B_z\rangle_{\rm He+metal}$} & \multicolumn{1}{c}{$\langle B_z\rangle_{\rm He}$} & \multicolumn{1}{c}{$\langle B_z\rangle_{\rm C}$} & \multicolumn{1}{c}{$\langle B_z\rangle_{\rm Si}$} & $\langle B_z\rangle_{\rm Fe}$ \\
			&            & \multicolumn{1}{c}{$({\rm kG})$}                 & \multicolumn{1}{c}{$({\rm kG})$}                        & \multicolumn{1}{c}{$({\rm kG})$}                  & \multicolumn{1}{c}{$({\rm kG})$}                 & \multicolumn{1}{c}{$({\rm kG})$}                  & $({\rm kG})$                  \\
		\noalign{\vskip0.5mm}
		\hline	
		\noalign{\vskip0.5mm}
			$2454702.138$ & ESPaDOnS & $ 0.39\pm0.30$ & $-0.27\pm0.19$ & $-0.53\pm0.35$ & $-0.52\pm0.60$ & $ 0.32\pm0.38$ & $ 0.09\pm0.57$ \\
			$2455849.677$ & Narval   & $-0.14\pm0.26$ & $ 0.12\pm0.22$ & $-0.96\pm0.41$ & $ 0.88\pm0.60$ & $ 1.91\pm0.93$ & $ 1.44\pm0.65$ \\
			$2455893.623$ & Narval   & $-2.75\pm0.32$ & $-2.96\pm0.27$ & $-3.47\pm0.48$ & $-2.29\pm0.80$ & $-2.53\pm0.61$ & $-0.56\pm0.88$ \\
			$2455910.518$ & Narval   & $-1.26\pm0.74$ & $-1.50\pm0.55$ & $-1.63\pm1.04$ & $-2.27\pm1.60$ & $ 0.17\pm0.88$ & $-1.15\pm2.26$ \\
			$2455934.528$ & Narval   & $-2.27\pm0.28$ & $-2.86\pm0.29$ & $-3.51\pm0.56$ & $-2.64\pm0.73$ & $-6.63\pm1.57$ & $-0.57\pm0.94$ \\
			$2455936.534$ & Narval   & $-2.04\pm0.35$ & $-2.21\pm0.29$ & $-2.93\pm0.51$ & $-1.27\pm0.90$ & $-3.25\pm1.13$ & $-2.82\pm0.77$ \\
			$2455938.525$ & Narval   & $ 0.48\pm0.32$ & $-0.08\pm0.25$ & $-0.42\pm0.45$ & $ 0.11\pm0.73$ & $ 0.90\pm0.54$ & $ 1.83\pm0.85$ \\
			$2455944.500$ & Narval   & $ 0.25\pm0.37$ & $ 0.30\pm0.28$ & $ 0.25\pm0.53$ & $ 1.05\pm0.77$ & $ 1.30\pm0.54$ & $ 1.94\pm1.01$ \\
			$2455949.429$ & Narval   & $-0.89\pm0.28$ & $-1.09\pm0.24$ & $-1.69\pm0.44$ & $-1.36\pm0.81$ & $-1.08\pm0.51$ & $-1.19\pm0.81$ \\
			$2455950.472$ & Narval   & $ 0.41\pm0.40$ & $-0.10\pm0.31$ & $-0.86\pm0.60$ & $ 0.03\pm0.96$ & $ 1.57\pm0.69$ & $ 0.51\pm0.80$ \\
			$2455951.471$ & Narval   & $-0.62\pm0.42$ & $-1.37\pm0.38$ & $-2.03\pm0.74$ & $-1.34\pm0.93$ & $-1.18\pm0.91$ & $ 2.07\pm0.95$ \\
			$2455966.376$ & Narval   & $-2.45\pm0.36$ & $-2.39\pm0.24$ & $-2.52\pm0.41$ & $-2.91\pm0.78$ & $-2.70\pm0.57$ & $-1.54\pm0.64$ \\
			$2455998.332$ & Narval   & $ 0.01\pm0.44$ & $ 0.26\pm0.33$ & $ 0.01\pm0.58$ & $-0.10\pm1.07$ & $ 0.35\pm0.60$ & $ 1.85\pm1.12$ \\
			$2455999.362$ & Narval   & $-1.56\pm0.40$ & $-2.18\pm0.40$ & $-3.01\pm0.78$ & $-1.84\pm0.99$ & $-2.86\pm1.57$ & $-0.45\pm1.01$ \\
			$2456001.309$ & Narval   & $-2.95\pm0.34$ & $-2.71\pm0.24$ & $-3.18\pm0.43$ & $-3.03\pm0.79$ & $-2.86\pm0.75$ & $-1.58\pm0.80$ \\
			$2456003.329$ & Narval   & $ 0.18\pm0.29$ & $-0.30\pm0.25$ & $-0.67\pm0.44$ & $ 0.28\pm0.76$ & $ 0.23\pm0.55$ & $ 0.05\pm0.82$ \\
			$2456202.665$ & Narval   & $-2.78\pm0.32$ & $-2.97\pm0.25$ & $-3.75\pm0.49$ & $-3.26\pm0.69$ & $-3.15\pm0.61$ & $-1.24\pm0.73$ \\
			$2456205.618$ & Narval   & $-0.51\pm0.35$ & $-0.51\pm0.28$ & $-1.21\pm0.53$ & $ 0.36\pm0.76$ & $ 0.68\pm0.67$ & $-1.44\pm0.78$ \\
			$2456224.646$ & Narval   & $-0.43\pm0.37$ & $-0.09\pm0.32$ & $-0.26\pm0.57$ & $ 0.48\pm1.08$ & $-0.29\pm0.58$ & $ 0.92\pm0.95$ \\
			$2456246.505$ & Narval   & $-1.88\pm0.36$ & $-1.55\pm0.28$ & $-2.19\pm0.52$ & $-1.83\pm0.77$ & $-1.15\pm0.79$ & $-1.30\pm0.73$ \\
			$2456293.881$ & ESPaDOnS & $-1.54\pm0.30$ & $-1.71\pm0.26$ & $-2.67\pm0.56$ & $-0.77\pm0.62$ & $-1.29\pm0.70$ & $ 0.11\pm0.66$ \\
			$2456295.787$ & ESPaDOnS & $-3.78\pm1.05$ & $-1.89\pm0.73$ & $-3.29\pm1.19$ & $ 0.76\pm2.22$ & $-4.78\pm1.73$ & $ 0.55\pm2.74$ \\
			$2456295.808$ & ESPaDOnS & $-3.07\pm0.96$ & $-2.58\pm0.62$ & $-2.71\pm1.02$ & $-3.48\pm1.96$ & $-2.35\pm1.43$ & $-1.95\pm2.44$ \\
			$2456556.002$ & ESPaDOnS & $-1.11\pm0.47$ & $-1.43\pm0.30$ & $-2.00\pm0.55$ & $-0.04\pm0.87$ & $-0.46\pm0.50$ & $ 0.49\pm0.93$ \\
			$2456557.140$ & ESPaDOnS & $-2.64\pm0.39$ & $-3.00\pm0.25$ & $-3.23\pm0.44$ & $-3.73\pm0.76$ & $-6.19\pm1.16$ & $-2.18\pm0.69$ \\
			$2456560.077$ & ESPaDOnS & $ 0.31\pm0.40$ & $ 0.27\pm0.28$ & $-0.23\pm0.52$ & $ 0.69\pm0.83$ & $ 1.48\pm0.89$ & $ 1.00\pm0.65$ \\
		\noalign{\vskip0.5mm}
		\hline
	\end{tabular*}
	\end{center}
\end{table*}

\clearpage

\onecolumn
\tablecaption{Equivalent widths of H$\alpha$ calculated using all of the DAO, CFHT 
	f/8.2 spectrograph, Narval, ESPaDOnS, and FEROS spectra. The uncertainties 
	correspond to $3\sigma$. This table will appear only in the electronic version of 
	the paper.}\label{halpha_ew_tbl}
\tablefirsthead{
	\hline
	\hline
	\noalign{\vskip0.5mm}
	HJD & Instrument & ${\rm EW_H}_\alpha$ & HJD & Instrument & ${\rm EW_H}_\alpha$ \\
		&            & $({\rm \AA})$       &     &            & $({\rm \AA})$       \\
	\noalign{\vskip0.5mm}
	\hline
	\noalign{\vskip0.5mm}
}
\tablehead{
	\multicolumn{6}{l}{continued from previous page}\\
	\hline
	\hline
	\noalign{\vskip0.5mm}
	HJD & Instrument & ${\rm EW_H}_\alpha$ & HJD & Instrument & ${\rm EW_H}_\alpha$ \\
		&            & $({\rm \AA})$       &     &            & $({\rm \AA})$       \\
	\noalign{\vskip0.5mm}
	\hline	
	\noalign{\vskip0.5mm}
}
\tabletail{
	\noalign{\vskip0.5mm}\hline
	\multicolumn{6}{l}{continued on next page}\\
}
\tablelasttail{
	\noalign{\vskip0.5mm}\hline
}
\centering
\begin{mpsupertabular*}{0.8\columnwidth}{@{\extracolsep{\fill}}l c c@{\hskip 1cm}|@{\hskip1cm}c c r}
			$2448529.029$ & DAO        & $2.77\pm0.10$ & $2455910.518$ & Narval   & $2.66\pm0.08$ \\
			$2448582.005$ & CFHT f/8.2 & $3.11\pm0.09$ & $2455934.528$ & Narval   & $2.28\pm0.10$ \\
			$2448582.103$ & CFHT f/8.2 & $3.09\pm0.09$ & $2455936.534$ & Narval   & $2.49\pm0.08$ \\
			$2448582.901$ & CFHT f/8.2 & $3.09\pm0.09$ & $2455938.525$ & Narval   & $3.14\pm0.09$ \\
			$2448582.999$ & CFHT f/8.2 & $3.05\pm0.09$ & $2455944.500$ & Narval   & $3.16\pm0.09$ \\
			$2448583.112$ & CFHT f/8.2 & $2.64\pm0.08$ & $2455949.429$ & Narval   & $2.84\pm0.10$ \\
			$2448597.819$ & DAO        & $1.81\pm0.10$ & $2455950.472$ & Narval   & $3.11\pm0.10$ \\
			$2449592.079$ & CFHT f/8.2 & $3.10\pm0.10$ & $2455951.471$ & Narval   & $2.77\pm0.08$ \\
			$2449592.086$ & CFHT f/8.2 & $3.04\pm0.10$ & $2455953.653$ & DAO      & $2.66\pm0.09$ \\
			$2449593.086$ & CFHT f/8.2 & $3.14\pm0.09$ & $2455953.663$ & DAO      & $2.74\pm0.09$ \\
			$2449593.097$ & CFHT f/8.2 & $3.14\pm0.09$ & $2455953.674$ & DAO      & $2.76\pm0.09$ \\
			$2449594.055$ & CFHT f/8.2 & $2.67\pm0.08$ & $2455953.684$ & DAO      & $2.82\pm0.09$ \\
			$2449594.067$ & CFHT f/8.2 & $2.59\pm0.09$ & $2455953.695$ & DAO      & $2.85\pm0.10$ \\
			$2449682.777$ & DAO        & $2.86\pm0.09$ & $2455953.705$ & DAO      & $2.90\pm0.09$ \\
			$2449682.832$ & DAO        & $2.65\pm0.09$ & $2455953.716$ & DAO      & $2.97\pm0.10$ \\
			$2449682.879$ & DAO        & $2.48\pm0.09$ & $2455953.726$ & DAO      & $3.00\pm0.10$ \\
			$2449682.912$ & DAO        & $2.35\pm0.10$ & $2455953.737$ & DAO      & $3.00\pm0.09$ \\
			$2449682.951$ & DAO        & $2.28\pm0.10$ & $2455953.748$ & DAO      & $3.05\pm0.10$ \\
			$2449682.976$ & DAO        & $2.14\pm0.10$ & $2455953.758$ & DAO      & $3.06\pm0.10$ \\
			$2449682.999$ & DAO        & $2.09\pm0.11$ & $2455953.769$ & DAO      & $3.04\pm0.10$ \\
			$2449684.754$ & DAO        & $2.05\pm0.11$ & $2455953.779$ & DAO      & $3.04\pm0.10$ \\
			$2449684.800$ & DAO        & $2.05\pm0.10$ & $2455953.790$ & DAO      & $3.13\pm0.10$ \\
			$2449689.861$ & DAO        & $2.11\pm0.11$ & $2455959.683$ & DAO      & $2.88\pm0.10$ \\
			$2449689.908$ & DAO        & $1.99\pm0.10$ & $2455959.694$ & DAO      & $2.90\pm0.10$ \\
			$2449690.749$ & DAO        & $2.02\pm0.10$ & $2455959.704$ & DAO      & $2.88\pm0.10$ \\
			$2449690.796$ & DAO        & $2.08\pm0.10$ & $2455959.715$ & DAO      & $2.92\pm0.10$ \\
			$2449690.842$ & DAO        & $1.93\pm0.11$ & $2455959.725$ & DAO      & $2.96\pm0.10$ \\
			$2449690.864$ & DAO        & $2.29\pm0.09$ & $2455959.736$ & DAO      & $2.79\pm0.11$ \\
			$2449690.917$ & DAO        & $2.34\pm0.08$ & $2455960.660$ & DAO      & $3.00\pm0.11$ \\
			$2449690.983$ & DAO        & $2.64\pm0.09$ & $2455960.670$ & DAO      & $3.04\pm0.10$ \\
			$2449691.006$ & DAO        & $2.83\pm0.10$ & $2455960.681$ & DAO      & $3.04\pm0.11$ \\
			$2450767.881$ & DAO        & $3.07\pm0.10$ & $2455960.691$ & DAO      & $3.06\pm0.10$ \\
			$2450768.849$ & DAO        & $2.80\pm0.09$ & $2455960.702$ & DAO      & $3.06\pm0.11$ \\
			$2454702.138$ & ESPaDOnS   & $3.15\pm0.10$ & $2455960.712$ & DAO      & $3.05\pm0.10$ \\
			$2455565.739$ & DAO        & $1.92\pm0.11$ & $2455960.722$ & DAO      & $3.03\pm0.11$ \\
			$2455565.785$ & DAO        & $2.02\pm0.11$ & $2455960.733$ & DAO      & $3.05\pm0.10$ \\
			$2455569.801$ & DAO        & $2.80\pm0.09$ & $2455960.743$ & DAO      & $2.98\pm0.11$ \\
			$2455570.807$ & DAO        & $2.24\pm0.11$ & $2455960.754$ & DAO      & $3.00\pm0.10$ \\
			$2455571.815$ & DAO        & $2.02\pm0.10$ & $2455960.764$ & DAO      & $3.03\pm0.11$ \\
			$2455602.737$ & DAO        & $2.68\pm0.09$ & $2455960.775$ & DAO      & $2.97\pm0.10$ \\
			$2455849.677$ & ESPaDOnS   & $3.18\pm0.10$ & $2455960.789$ & DAO      & $3.00\pm0.11$ \\
			$2455849.934$ & DAO        & $2.50\pm0.08$ & $2455960.799$ & DAO      & $2.88\pm0.11$ \\
			$2455849.945$ & DAO        & $2.48\pm0.08$ & $2455961.644$ & DAO      & $2.97\pm0.09$ \\
			$2455849.955$ & DAO        & $2.43\pm0.09$ & $2455961.654$ & DAO      & $2.84\pm0.09$ \\
			$2455849.966$ & DAO        & $2.37\pm0.08$ & $2455966.376$ & Narval   & $2.31\pm0.08$ \\
			$2455849.976$ & DAO        & $2.36\pm0.09$ & $2455998.332$ & Narval   & $3.10\pm0.09$ \\
			$2455849.987$ & DAO        & $2.31\pm0.09$ & $2455999.362$ & Narval   & $2.46\pm0.09$ \\
			$2455849.997$ & DAO        & $2.30\pm0.09$ & $2456001.309$ & Narval   & $2.07\pm0.09$ \\
			$2455850.008$ & DAO        & $2.35\pm0.09$ & $2456003.329$ & Narval   & $3.16\pm0.10$ \\
			$2455850.018$ & DAO        & $2.23\pm0.10$ & $2456191.654$ & Narval   & $2.08\pm0.12$ \\
			$2455850.029$ & DAO        & $2.22\pm0.10$ & $2456202.665$ & Narval   & $2.15\pm0.11$ \\
			$2455850.039$ & DAO        & $2.18\pm0.10$ & $2456205.618$ & Narval   & $3.05\pm0.10$ \\
			$2455850.050$ & DAO        & $2.20\pm0.10$ & $2456224.646$ & Narval   & $2.99\pm0.09$ \\
			$2455850.062$ & DAO        & $2.15\pm0.10$ & $2456246.505$ & Narval   & $2.68\pm0.09$ \\
			$2455850.933$ & DAO        & $2.12\pm0.10$ & $2456293.881$ & ESPaDOnS & $2.65\pm0.08$ \\
			$2455850.943$ & DAO        & $2.17\pm0.10$ & $2456295.767$ & ESPaDOnS & $2.02\pm0.11$ \\
			$2455850.954$ & DAO        & $2.15\pm0.10$ & $2456295.787$ & ESPaDOnS & $2.04\pm0.11$ \\
			$2455850.964$ & DAO        & $2.09\pm0.10$ & $2456295.808$ & ESPaDOnS & $2.03\pm0.11$ \\
			$2455850.975$ & DAO        & $2.11\pm0.09$ & $2456556.002$ & ESPaDOnS & $2.71\pm0.08$ \\
			$2455850.985$ & DAO        & $2.14\pm0.10$ & $2456557.140$ & ESPaDOnS & $2.09\pm0.09$ \\
			$2455850.996$ & DAO        & $2.05\pm0.09$ & $2456560.077$ & ESPaDOnS & $3.14\pm0.09$ \\
			$2455851.006$ & DAO        & $2.06\pm0.09$ & $2456656.568$ & FEROS    & $3.08\pm0.09$ \\
			$2455851.017$ & DAO        & $2.16\pm0.09$ & $2456656.609$ & FEROS    & $3.05\pm0.09$ \\
			$2455851.027$ & DAO        & $2.16\pm0.09$ & $2456658.535$ & FEROS    & $2.29\pm0.10$ \\
			$2455851.038$ & DAO        & $2.15\pm0.08$ & $2456658.542$ & FEROS    & $2.29\pm0.10$ \\
			$2455851.048$ & DAO        & $2.23\pm0.09$ & $2456658.679$ & FEROS    & $2.00\pm0.11$ \\
			$2455880.839$ & DAO        & $2.14\pm0.11$ & $2456658.683$ & FEROS    & $1.99\pm0.10$ \\
			$2455880.850$ & DAO        & $2.18\pm0.10$ & $2456658.686$ & FEROS    & $1.94\pm0.11$ \\
			$2455880.860$ & DAO        & $2.16\pm0.10$ & $2456658.690$ & FEROS    & $1.96\pm0.10$ \\
			$2455880.871$ & DAO        & $2.20\pm0.10$ & $2456658.694$ & FEROS    & $1.99\pm0.10$ \\
			$2455880.881$ & DAO        & $2.10\pm0.10$ & $2456658.701$ & FEROS    & $1.98\pm0.10$ \\
			$2455880.892$ & DAO        & $2.22\pm0.09$ & $2456658.701$ & FEROS    & $1.98\pm0.10$ \\
			$2455880.902$ & DAO        & $2.20\pm0.09$ & $2456658.709$ & FEROS    & $1.98\pm0.10$ \\
			$2455880.913$ & DAO        & $2.26\pm0.09$ & $2456658.713$ & FEROS    & $2.00\pm0.10$ \\
			$2455880.923$ & DAO        & $2.19\pm0.09$ & $2456658.763$ & FEROS    & $2.10\pm0.09$ \\
			$2455880.933$ & DAO        & $2.25\pm0.09$ & $2456659.627$ & FEROS    & $2.16\pm0.09$ \\
			$2455880.944$ & DAO        & $2.27\pm0.09$ & $2456659.671$ & FEROS    & $2.30\pm0.09$ \\
			$2455880.954$ & DAO        & $2.32\pm0.09$ & $2456659.677$ & FEROS    & $2.33\pm0.09$ \\
			$2455880.965$ & DAO        & $2.34\pm0.08$ & $2456659.681$ & FEROS    & $2.09\pm0.11$ \\
			$2455880.975$ & DAO        & $2.42\pm0.09$ & $2456659.685$ & FEROS    & $2.40\pm0.09$ \\
			$2455880.986$ & DAO        & $2.51\pm0.09$ & $2456659.689$ & FEROS    & $2.41\pm0.08$ \\
			$2455880.996$ & DAO        & $2.44\pm0.09$ & $2456659.693$ & FEROS    & $2.39\pm0.09$ \\
			$2455881.019$ & DAO        & $2.58\pm0.08$ & $2456659.697$ & FEROS    & $2.45\pm0.09$ \\
			$2455881.029$ & DAO        & $2.62\pm0.09$ & $2456659.700$ & FEROS    & $2.42\pm0.08$ \\
			$2455881.040$ & DAO        & $2.64\pm0.09$ & $2456659.708$ & FEROS    & $2.50\pm0.09$ \\
			$2455893.623$ & Narval     & $2.06\pm0.11$ & $2456659.746$ & FEROS    & $2.66\pm0.10$ \\
			$2455903.772$ & DAO        & $2.45\pm0.10$ & $2456660.614$ & FEROS    & $2.71\pm0.09$ \\
			$2455903.782$ & DAO        & $2.48\pm0.10$ & $2456660.653$ & FEROS    & $2.88\pm0.10$ \\
			$2455903.793$ & DAO        & $2.40\pm0.10$ & $2456660.724$ & FEROS    & $3.07\pm0.10$ \\
			$2455903.803$ & DAO        & $2.33\pm0.10$ & $2456660.763$ & FEROS    & $3.14\pm0.11$ \\
			$2455903.814$ & DAO        & $2.30\pm0.10$ & $2456660.801$ & FEROS    & $3.17\pm0.10$ \\
\end{mpsupertabular*}

\begin{table*}
	\caption{Equivalent widths of various spectral lines calculated using the Narval, 
	ESPaDOnS, and FEROS spectra. The uncertainties correspond to $3\sigma$. This table 
	will appear only in the electronic version of the paper.}
	\label{ew_tbl}
	\begin{center}
	\begin{tabular*}{1\columnwidth}{@{\extracolsep{\fill}}l c r r r r r r}
		\hline
		\hline
		\noalign{\vskip0.5mm}
		HJD & Instrument & EW$_{\text{C\,{\sc ii}}\,\lambda4267}$ & EW$_{\text{He\,{\sc i}}\,\lambda4713}$ & EW$_{\text{Si\,{\sc iii}}\,\lambda4553}$ & EW$_{\text{N\,{\sc ii}}\,\lambda4643}$ & ${\rm EW_H}_\beta$      & ${\rm EW_H}_\gamma$     \\
			&            & $(\times10\,{\rm \AA})$                & $(\times10\,{\rm \AA})$                & $(\times10\,{\rm \AA})$                  & $(\times10\,{\rm \AA})$                & $(\times10\,{\rm \AA})$ & $(\times10\,{\rm \AA})$ \\
		\noalign{\vskip0.5mm}
		\hline	
		\noalign{\vskip0.5mm}
			$2454702.138$ & ESPaDOnS & $1.53\pm0.07$ & $1.23\pm0.05$ & $0.89\pm0.02$ & $0.25\pm0.01$ & $59.38\pm0.70$ & $56.40\pm0.67$ \\
			$2455849.677$ & Narval   & $1.49\pm0.07$ & $1.30\pm0.05$ & $0.84\pm0.03$ & $0.26\pm0.01$ & $58.36\pm0.71$ & $56.13\pm0.69$ \\
			$2455893.623$ & Narval   & $1.47\pm0.07$ & $1.98\pm0.06$ & $0.78\pm0.04$ & $0.28\pm0.02$ & $56.84\pm0.72$ & $56.01\pm0.69$ \\
			$2455910.518$ & Narval   & $1.46\pm0.07$ & $1.49\pm0.05$ & $1.00\pm0.04$ & $0.31\pm0.02$ & $57.59\pm0.69$ & $55.84\pm0.67$ \\
			$2455934.528$ & Narval   & $1.49\pm0.07$ & $1.78\pm0.06$ & $0.87\pm0.04$ & $0.24\pm0.02$ & $57.09\pm0.69$ & $55.69\pm0.65$ \\
			$2455936.534$ & Narval   & $1.50\pm0.07$ & $1.63\pm0.06$ & $0.78\pm0.03$ & $0.34\pm0.02$ & $57.33\pm0.66$ & $55.97\pm0.65$ \\
			$2455938.525$ & Narval   & $1.50\pm0.07$ & $1.24\pm0.04$ & $0.81\pm0.03$ & $0.30\pm0.01$ & $58.16\pm0.70$ & $55.76\pm0.67$ \\
			$2455944.500$ & Narval   & $1.50\pm0.07$ & $1.18\pm0.05$ & $0.84\pm0.03$ & $0.31\pm0.01$ & $58.23\pm0.70$ & $55.93\pm0.67$ \\
			$2455949.429$ & Narval   & $1.55\pm0.07$ & $1.34\pm0.05$ & $0.85\pm0.03$ & $0.32\pm0.02$ & $58.42\pm0.70$ & $55.74\pm0.67$ \\
			$2455950.472$ & Narval   & $1.51\pm0.07$ & $1.16\pm0.05$ & $0.93\pm0.03$ & $0.31\pm0.02$ & $58.64\pm0.70$ & $55.89\pm0.67$ \\
			$2455951.471$ & Narval   & $1.51\pm0.07$ & $1.35\pm0.05$ & $0.98\pm0.04$ & $0.32\pm0.01$ & $58.15\pm0.68$ & $55.39\pm0.67$ \\
			$2455966.376$ & Narval   & $1.54\pm0.07$ & $1.76\pm0.06$ & $0.68\pm0.03$ & $0.32\pm0.02$ & $57.45\pm0.67$ & $56.00\pm0.65$ \\
			$2455998.332$ & Narval   & $1.47\pm0.08$ & $1.23\pm0.04$ & $0.87\pm0.03$ & $0.34\pm0.02$ & $58.02\pm0.69$ & $56.05\pm0.66$ \\
			$2455999.362$ & Narval   & $1.50\pm0.07$ & $1.55\pm0.05$ & $0.87\pm0.04$ & $0.34\pm0.02$ & $56.27\pm0.69$ & $55.95\pm0.67$ \\
			$2456001.309$ & Narval   & $1.51\pm0.07$ & $1.91\pm0.06$ & $0.75\pm0.02$ & $0.32\pm0.02$ & $56.69\pm0.68$ & $56.14\pm0.65$ \\
			$2456003.329$ & Narval   & $1.54\pm0.07$ & $1.24\pm0.05$ & $0.79\pm0.03$ & $0.30\pm0.02$ & $58.24\pm0.70$ & $56.57\pm0.66$ \\
			$2456191.654$ & Narval   & $1.51\pm0.07$ & $2.01\pm0.06$ & $0.78\pm0.03$ & $0.31\pm0.02$ & $56.86\pm0.70$ & $56.08\pm0.66$ \\
			$2456202.665$ & Narval   & $1.50\pm0.07$ & $1.90\pm0.06$ & $0.84\pm0.04$ & $0.30\pm0.02$ & $56.72\pm0.69$ & $55.84\pm0.67$ \\
			$2456205.618$ & Narval   & $1.54\pm0.07$ & $1.33\pm0.05$ & $0.89\pm0.03$ & $0.31\pm0.02$ & $57.91\pm0.70$ & $56.05\pm0.67$ \\
			$2456224.646$ & Narval   & $1.49\pm0.07$ & $1.33\pm0.04$ & $0.92\pm0.03$ & $0.32\pm0.02$ & $57.88\pm0.68$ & $55.72\pm0.67$ \\
			$2456246.505$ & Narval   & $1.55\pm0.07$ & $1.51\pm0.06$ & $0.81\pm0.04$ & $0.27\pm0.02$ & $57.70\pm0.67$ & $55.84\pm0.65$ \\
			$2456293.881$ & ESPaDOnS & $1.47\pm0.07$ & $1.49\pm0.05$ & $1.03\pm0.03$ & $0.30\pm0.01$ & $58.46\pm0.68$ & $56.06\pm0.66$ \\
			$2456295.767$ & ESPaDOnS & $1.53\pm0.09$ & $2.01\pm0.08$ & $0.80\pm0.04$ & $0.19\pm0.07$ & $57.20\pm0.73$ & $56.02\pm0.68$ \\
			$2456295.787$ & ESPaDOnS & $1.45\pm0.08$ & $1.97\pm0.07$ & $0.86\pm0.03$ & $0.43\pm0.04$ & $57.01\pm0.71$ & $56.34\pm0.66$ \\
			$2456295.808$ & ESPaDOnS & $1.49\pm0.08$ & $2.02\pm0.07$ & $0.85\pm0.03$ & $0.30\pm0.03$ & $56.44\pm0.70$ & $56.22\pm0.67$ \\
			$2456556.002$ & ESPaDOnS & $1.48\pm0.07$ & $1.46\pm0.05$ & $1.02\pm0.03$ & $0.25\pm0.02$ & $58.17\pm0.68$ & $56.18\pm0.68$ \\
			$2456557.140$ & ESPaDOnS & $1.52\pm0.07$ & $1.94\pm0.06$ & $0.92\pm0.03$ & $0.21\pm0.01$ & $57.91\pm0.67$ & $56.44\pm0.67$ \\
			$2456560.077$ & ESPaDOnS & $1.53\pm0.07$ & $1.19\pm0.05$ & $1.01\pm0.03$ & $0.21\pm0.02$ & $58.63\pm0.70$ & $56.34\pm0.68$ \\
			$2456656.568$ & FEROS    & $1.55\pm0.07$ & $1.23\pm0.05$ & $0.81\pm0.03$ & $0.38\pm0.02$ & $58.40\pm0.71$ & $55.16\pm0.66$ \\
			$2456656.609$ & FEROS    & $1.55\pm0.07$ & $1.15\pm0.05$ & $0.83\pm0.03$ & $0.33\pm0.02$ & $58.48\pm0.72$ & $55.15\pm0.68$ \\
			$2456658.535$ & FEROS    & $1.50\pm0.07$ & $1.74\pm0.06$ & $0.70\pm0.03$ & $0.36\pm0.02$ & $57.26\pm0.78$ & $55.12\pm0.68$ \\
			$2456658.542$ & FEROS    & $1.50\pm0.08$ & $1.81\pm0.07$ & $0.79\pm0.04$ & $0.30\pm0.03$ & $56.68\pm0.73$ & $55.32\pm0.68$ \\
			$2456658.679$ & FEROS    & $1.50\pm0.07$ & $2.01\pm0.07$ & $0.65\pm0.03$ & $0.31\pm0.03$ & $57.00\pm0.77$ & $55.54\pm0.68$ \\
			$2456658.683$ & FEROS    & $1.53\pm0.07$ & $2.04\pm0.07$ & $0.74\pm0.03$ & $0.35\pm0.02$ & $56.77\pm0.75$ & $55.49\pm0.68$ \\
			$2456658.686$ & FEROS    & $1.51\pm0.08$ & $1.99\pm0.07$ & $0.69\pm0.03$ & $0.29\pm0.02$ & $56.81\pm0.72$ & $55.35\pm0.67$ \\
			$2456658.690$ & FEROS    & $1.45\pm0.07$ & $1.91\pm0.07$ & $0.66\pm0.02$ & $0.30\pm0.02$ & $57.14\pm0.82$ & $55.47\pm0.68$ \\
			$2456658.694$ & FEROS    & $1.46\pm0.07$ & $1.97\pm0.07$ & $0.66\pm0.02$ & $0.33\pm0.02$ & $56.74\pm0.70$ & $55.34\pm0.66$ \\
			$2456658.698$ & FEROS    & $1.52\pm0.07$ & $1.91\pm0.07$ & $0.70\pm0.02$ & $0.33\pm0.02$ & $56.87\pm0.71$ & $55.40\pm0.67$ \\
			$2456658.701$ & FEROS    & $1.51\pm0.07$ & $1.94\pm0.07$ & $0.68\pm0.03$ & $0.28\pm0.02$ & $56.93\pm0.70$ & $55.34\pm0.66$ \\
			$2456658.702$ & FEROS    & $1.51\pm0.07$ & $1.99\pm0.07$ & $0.66\pm0.03$ & $0.35\pm0.02$ & $56.77\pm0.72$ & $55.36\pm0.67$ \\
			$2456658.709$ & FEROS    & $1.46\pm0.07$ & $1.95\pm0.06$ & $0.68\pm0.02$ & $0.27\pm0.02$ & $56.76\pm0.72$ & $55.32\pm0.68$ \\
			$2456658.713$ & FEROS    & $1.44\pm0.08$ & $1.94\pm0.06$ & $0.70\pm0.02$ & $0.31\pm0.02$ & $56.97\pm0.71$ & $55.43\pm0.65$ \\
			$2456658.763$ & FEROS    & $1.55\pm0.07$ & $1.85\pm0.06$ & $0.63\pm0.03$ & $0.28\pm0.02$ & $57.23\pm0.72$ & $55.46\pm0.67$ \\
			$2456659.628$ & FEROS    & $1.58\pm0.07$ & $1.79\pm0.07$ & $0.68\pm0.04$ & $0.30\pm0.02$ & $57.52\pm0.74$ & $55.33\pm0.66$ \\
			$2456659.671$ & FEROS    & $1.56\pm0.07$ & $1.64\pm0.07$ & $0.70\pm0.04$ & $0.28\pm0.02$ & $57.43\pm0.68$ & $55.13\pm0.66$ \\
			$2456659.677$ & FEROS    & $1.59\pm0.07$ & $1.58\pm0.06$ & $0.73\pm0.03$ & $0.33\pm0.03$ & $57.69\pm0.70$ & $55.26\pm0.65$ \\
			$2456659.681$ & FEROS    & $1.56\pm0.06$ & $1.62\pm0.07$ & $0.74\pm0.04$ & $0.29\pm0.03$ & $57.62\pm0.71$ & $55.21\pm0.65$ \\
			$2456659.685$ & FEROS    & $1.60\pm0.07$ & $1.56\pm0.07$ & $0.73\pm0.04$ & $0.28\pm0.02$ & $57.49\pm0.69$ & $55.55\pm0.68$ \\
			$2456659.689$ & FEROS    & $1.58\pm0.07$ & $1.62\pm0.07$ & $0.77\pm0.04$ & $0.30\pm0.02$ & $57.77\pm0.70$ & $55.27\pm0.64$ \\
			$2456659.693$ & FEROS    & $1.53\pm0.08$ & $1.51\pm0.07$ & $0.76\pm0.04$ & $0.29\pm0.03$ & $57.65\pm0.70$ & $55.16\pm0.67$ \\
			$2456659.697$ & FEROS    & $1.60\pm0.07$ & $1.55\pm0.07$ & $0.75\pm0.04$ & $0.30\pm0.03$ & $57.58\pm0.68$ & $55.15\pm0.66$ \\
			$2456659.700$ & FEROS    & $1.60\pm0.07$ & $1.57\pm0.06$ & $0.74\pm0.04$ & $0.29\pm0.03$ & $57.59\pm0.69$ & $55.25\pm0.66$ \\
			$2456659.704$ & FEROS    & $1.53\pm0.06$ & $1.48\pm0.07$ & $0.76\pm0.04$ & $0.32\pm0.02$ & $57.75\pm0.68$ & $55.20\pm0.65$ \\
			$2456659.708$ & FEROS    & $1.62\pm0.07$ & $1.54\pm0.06$ & $0.80\pm0.04$ & $0.29\pm0.02$ & $57.65\pm0.69$ & $55.05\pm0.66$ \\
			$2456659.746$ & FEROS    & $1.59\pm0.07$ & $1.39\pm0.06$ & $0.78\pm0.04$ & $0.31\pm0.02$ & $58.13\pm0.73$ & $55.13\pm0.66$ \\
			$2456660.614$ & FEROS    & $1.52\pm0.07$ & $1.50\pm0.06$ & $0.76\pm0.03$ & $0.32\pm0.02$ & $58.03\pm0.76$ & $55.16\pm0.67$ \\
			$2456660.653$ & FEROS    & $1.55\pm0.07$ & $1.43\pm0.05$ & $0.78\pm0.03$ & $0.33\pm0.02$ & $58.04\pm0.71$ & $55.21\pm0.68$ \\
			$2456660.724$ & FEROS    & $1.55\pm0.07$ & $1.29\pm0.05$ & $0.79\pm0.03$ & $0.27\pm0.01$ & $58.33\pm0.74$ & $55.25\pm0.67$ \\
			$2456660.763$ & FEROS    & $1.54\pm0.07$ & $1.21\pm0.05$ & $0.75\pm0.03$ & $0.36\pm0.02$ & $58.34\pm0.71$ & $55.31\pm0.69$ \\
			$2456660.801$ & FEROS    & $1.52\pm0.07$ & $1.22\pm0.05$ & $0.79\pm0.03$ & $0.33\pm0.02$ & $58.30\pm0.74$ & $55.37\pm0.67$ \\
		\noalign{\vskip0.5mm}
		\hline
	\end{tabular*}
	\end{center}
\end{table*}

\clearpage

\onecolumn
\tablecaption{$uvby$ photometry obtained from Jan. 1992 to Mar. 1994. Each 
measurement is given as a magnitude difference between the variable source (``v", 
HD~35502) and a non-variable comparison star (``c", HD~35575). This table will appear 
only in the electronic version of the paper.}\label{uvby_tbl}
\tablefirsthead{
	\hline
	\hline
	\noalign{\vskip0.5mm}
	HJD & $u$(v-c) & $v$(v-c) & $b$(v-c) & $y$(v-c) & HJD & $u$(v-c) & $v$(v-c) & $b$(v-c) & $y$(v-c) \\
	    &    (mag) &    (mag) &    (mag) &    (mag) &     &    (mag) &    (mag) &    (mag) &    (mag) \\
	\noalign{\vskip0.5mm}
	\hline	
\noalign{\vskip0.5mm}
}
\tablehead{
	\multicolumn{10}{l}{continued from previous page}\\
	\hline
	\hline
	\noalign{\vskip0.5mm}
	HJD & $u$(v-c) & $v$(v-c) & $b$(v-c) & $y$(v-c) & HJD & $u$(v-c) & $v$(v-c) & $b$(v-c) & $y$(v-c) \\
	    &    (mag) &    (mag) &    (mag) &    (mag) &     &    (mag) &    (mag) &    (mag) &    (mag) \\
	\noalign{\vskip0.5mm}
	\hline	
\noalign{\vskip0.5mm}
}
\tabletail{
	\noalign{\vskip0.5mm}\hline
	\multicolumn{10}{l}{continued on next page}\\
}
\tablelasttail{
	\noalign{\vskip0.5mm}\hline
}
\centering
\begin{mpsupertabular*}{1.0\columnwidth}{@{\extracolsep{\fill}}l c c c c@{\hskip 1cm}|@{\hskip 1cm}c c c c r}
	$2448648.764$ & $1.25$ & $1.04$ & $0.96$ & $0.89$ & $2449291.008$ & $1.27$ & $1.06$ & $0.98$ & $0.91$ \\
	$2448657.697$ & $1.27$ & $1.06$ & $0.97$ & $0.89$ & $2449292.006$ & $1.28$ & $1.08$ & $0.99$ & $0.91$ \\
	$2448676.640$ & $1.27$ & $1.07$ & $0.98$ & $0.91$ & $2449294.930$ & $1.27$ & $1.07$ & $0.99$ & $0.92$ \\
	$2448678.635$ & $1.26$ & $1.05$ & $0.97$ & $0.90$ & $2449297.965$ & $1.28$ & $1.08$ & $0.99$ & $0.91$ \\
	$2448681.635$ & $1.27$ & $1.07$ & $0.97$ & $0.90$ & $2449311.949$ & $1.27$ & $1.06$ & $0.98$ & $0.91$ \\
	$2448687.621$ & $1.28$ & $1.07$ & $0.98$ & $0.90$ & $2449312.945$ & $1.26$ & $1.06$ & $0.97$ & $0.90$ \\
	$2448692.632$ & $1.27$ & $1.08$ & $0.99$ & $0.91$ & $2449318.933$ & $1.26$ & $1.05$ & $0.98$ & $0.90$ \\
	$2448695.633$ & $1.27$ & $1.06$ & $0.98$ & $0.91$ & $2449324.920$ & $1.26$ & $1.06$ & $0.97$ & $0.90$ \\
	$2448696.605$ & $1.26$ & $1.06$ & $0.97$ & $0.90$ & $2449325.917$ & $1.26$ & $1.05$ & $0.97$ & $0.90$ \\
	$2448697.604$ & $1.26$ & $1.06$ & $0.97$ & $0.90$ & $2449328.909$ & $1.27$ & $1.06$ & $0.97$ & $0.91$ \\
	$2448701.606$ & $1.27$ & $1.07$ & $0.98$ & $0.91$ & $2449336.881$ & $1.26$ & $1.06$ & $0.99$ & $0.92$ \\
	$2448701.621$ & $1.27$ & $1.06$ & $0.98$ & $0.90$ & $2449337.881$ & $1.27$ & $1.06$ & $0.97$ & $0.91$ \\
	$2448702.606$ & $1.27$ & $1.07$ & $0.98$ & $0.91$ & $2449338.877$ & $1.27$ & $1.07$ & $0.99$ & $0.91$ \\
	$2448705.607$ & $1.29$ & $1.08$ & $1.00$ & $0.92$ & $2449346.853$ & $1.28$ & $1.08$ & $0.99$ & $0.92$ \\
	$2448873.946$ & $1.26$ & $1.06$ & $0.98$ & $0.90$ & $2449351.841$ & $1.29$ & $1.08$ & $0.99$ & $0.92$ \\
	$2448874.944$ & $1.26$ & $1.05$ & $0.97$ & $0.90$ & $2449359.819$ & $1.28$ & $1.07$ & $0.99$ & $0.92$ \\
	$2448875.941$ & $1.25$ & $1.05$ & $0.97$ & $0.90$ & $2449362.654$ & $1.26$ & $1.07$ & $0.98$ & $0.91$ \\
	$2448876.939$ & $1.26$ & $1.06$ & $0.97$ & $0.89$ & $2449362.686$ & $1.27$ & $1.06$ & $0.98$ & $0.91$ \\
	$2448889.904$ & $1.26$ & $1.06$ & $0.96$ & $0.89$ & $2449362.710$ & $1.26$ & $1.07$ & $0.98$ & $0.91$ \\
	$2448888.907$ & $1.26$ & $1.06$ & $0.97$ & $0.89$ & $2449362.726$ & $1.27$ & $1.07$ & $0.98$ & $0.91$ \\
	$2448889.904$ & $1.26$ & $1.06$ & $0.96$ & $0.89$ & $2449362.740$ & $1.26$ & $1.07$ & $0.98$ & $0.91$ \\
	$2448890.902$ & $1.26$ & $1.06$ & $0.97$ & $0.90$ & $2449362.756$ & $1.27$ & $1.07$ & $0.99$ & $0.91$ \\
	$2448891.899$ & $1.26$ & $1.06$ & $0.98$ & $0.90$ & $2449362.770$ & $1.27$ & $1.07$ & $0.98$ & $0.91$ \\
	$2448897.970$ & $1.26$ & $1.06$ & $0.97$ & $0.90$ & $2449362.786$ & $1.27$ & $1.08$ & $0.98$ & $0.91$ \\
	$2448898.968$ & $1.25$ & $1.05$ & $0.96$ & $0.89$ & $2449362.800$ & $1.27$ & $1.07$ & $0.98$ & $0.91$ \\
	$2448899.965$ & $1.26$ & $1.06$ & $0.97$ & $0.90$ & $2449362.816$ & $1.27$ & $1.07$ & $0.99$ & $0.91$ \\
	$2448900.942$ & $1.29$ & $1.08$ & $0.98$ & $0.90$ & $2449363.650$ & $1.27$ & $1.07$ & $0.99$ & $0.91$ \\
	$2448901.960$ & $1.27$ & $1.07$ & $0.97$ & $0.90$ & $2449363.683$ & $1.27$ & $1.08$ & $0.99$ & $0.91$ \\
	$2448903.954$ & $1.25$ & $1.05$ & $0.96$ & $0.90$ & $2449363.708$ & $1.27$ & $1.08$ & $0.99$ & $0.91$ \\
	$2448904.935$ & $1.24$ & $1.05$ & $0.96$ & $0.89$ & $2449363.724$ & $1.28$ & $1.08$ & $0.99$ & $0.91$ \\
	$2448905.949$ & $1.26$ & $1.06$ & $0.97$ & $0.89$ & $2449363.738$ & $1.28$ & $1.08$ & $0.99$ & $0.91$ \\
	$2448906.946$ & $1.28$ & $1.07$ & $0.98$ & $0.90$ & $2449363.754$ & $1.28$ & $1.08$ & $0.99$ & $0.91$ \\
	$2448908.025$ & $1.28$ & $1.07$ & $0.98$ & $0.91$ & $2449363.769$ & $1.28$ & $1.08$ & $0.99$ & $0.91$ \\
	$2448910.921$ & $1.26$ & $1.06$ & $0.98$ & $0.91$ & $2449363.784$ & $1.28$ & $1.08$ & $0.99$ & $0.92$ \\
	$2448911.989$ & $1.26$ & $1.07$ & $0.98$ & $0.91$ & $2449363.799$ & $1.28$ & $1.08$ & $0.99$ & $0.92$ \\
	$2448913.005$ & $1.28$ & $1.08$ & $0.99$ & $0.91$ & $2449364.649$ & $1.27$ & $1.08$ & $0.99$ & $0.92$ \\
	$2448918.998$ & $1.28$ & $1.08$ & $0.98$ & $0.91$ & $2449364.681$ & $1.28$ & $1.08$ & $0.99$ & $0.92$ \\
	$2448920.993$ & $1.26$ & $1.06$ & $0.98$ & $0.91$ & $2449364.707$ & $1.29$ & $1.08$ & $0.99$ & $0.92$ \\
	$2448921.990$ & $1.25$ & $1.05$ & $0.98$ & $0.90$ & $2449364.720$ & $1.29$ & $1.08$ & $0.99$ & $0.91$ \\
	$2448928.970$ & $1.26$ & $1.06$ & $0.97$ & $0.90$ & $2449364.736$ & $1.28$ & $1.08$ & $0.99$ & $0.92$ \\
	$2448937.973$ & $1.26$ & $1.06$ & $0.98$ & $0.91$ & $2449364.750$ & $1.28$ & $1.08$ & $1.00$ & $0.92$ \\
	$2448938.839$ & $1.27$ & $1.06$ & $0.98$ & $0.92$ & $2449364.767$ & $1.28$ & $1.08$ & $0.99$ & $0.92$ \\
	$2448939.861$ & $1.25$ & $1.04$ & $0.96$ & $0.89$ & $2449364.781$ & $1.28$ & $1.08$ & $0.99$ & $0.92$ \\
	$2448940.908$ & $1.25$ & $1.05$ & $0.98$ & $0.90$ & $2449364.797$ & $1.28$ & $1.07$ & $0.99$ & $0.91$ \\
	$2448941.864$ & $1.26$ & $1.07$ & $0.98$ & $0.90$ & $2449364.810$ & $1.28$ & $1.07$ & $0.99$ & $0.92$ \\
	$2448943.867$ & $1.27$ & $1.06$ & $0.97$ & $0.90$ & $2449365.650$ & $1.28$ & $1.07$ & $0.99$ & $0.92$ \\
	$2448944.862$ & $1.26$ & $1.06$ & $0.97$ & $0.90$ & $2449365.683$ & $1.28$ & $1.08$ & $0.99$ & $0.93$ \\
	$2448954.890$ & $1.27$ & $1.07$ & $0.99$ & $0.91$ & $2449365.723$ & $1.28$ & $1.08$ & $0.99$ & $0.92$ \\
	$2448955.883$ & $1.27$ & $1.07$ & $0.98$ & $0.91$ & $2449365.738$ & $1.27$ & $1.07$ & $0.99$ & $0.92$ \\
	$2448957.878$ & $1.25$ & $1.04$ & $0.96$ & $0.89$ & $2449365.753$ & $1.27$ & $1.07$ & $0.99$ & $0.91$ \\
	$2448963.856$ & $1.25$ & $1.05$ & $0.96$ & $0.90$ & $2449365.798$ & $1.27$ & $1.06$ & $0.99$ & $0.91$ \\
	$2448966.712$ & $1.27$ & $1.07$ & $0.98$ & $0.92$ & $2449366.651$ & $1.27$ & $1.06$ & $0.98$ & $0.91$ \\
	$2448968.857$ & $1.26$ & $1.06$ & $0.97$ & $0.90$ & $2449369.644$ & $1.28$ & $1.07$ & $0.99$ & $0.92$ \\
	$2449008.659$ & $1.28$ & $1.07$ & $0.98$ & $0.91$ & $2449370.653$ & $1.29$ & $1.08$ & $0.99$ & $0.91$ \\
	$2449010.625$ & $1.27$ & $1.06$ & $0.98$ & $0.90$ & $2449371.670$ & $1.28$ & $1.07$ & $0.99$ & $0.92$ \\
	$2449011.617$ & $1.25$ & $1.05$ & $0.97$ & $0.90$ & $2449373.677$ & $1.26$ & $1.06$ & $0.98$ & $0.91$ \\
	$2449012.625$ & $1.25$ & $1.05$ & $0.97$ & $0.89$ & $2449374.661$ & $1.26$ & $1.06$ & $0.98$ & $0.91$ \\
	$2449013.619$ & $1.27$ & $1.07$ & $0.99$ & $0.91$ & $2449382.700$ & $1.27$ & $1.07$ & $0.99$ & $0.91$ \\
	$2449017.669$ & $1.25$ & $1.04$ & $0.97$ & $0.90$ & $2449383.698$ & $1.27$ & $1.06$ & $0.98$ & $0.91$ \\
	$2449019.638$ & $1.27$ & $1.07$ & $0.97$ & $0.90$ & $2449384.695$ & $1.26$ & $1.06$ & $0.97$ & $0.91$ \\
	$2449021.679$ & $1.26$ & $1.05$ & $0.97$ & $0.90$ & $2449389.687$ & $1.26$ & $1.06$ & $0.98$ & $0.91$ \\
	$2449022.690$ & $1.25$ & $1.05$ & $0.97$ & $0.90$ & $2449393.690$ & $1.27$ & $1.07$ & $0.99$ & $0.91$ \\
	$2449023.688$ & $1.25$ & $1.05$ & $0.96$ & $0.90$ & $2449396.667$ & $1.25$ & $1.05$ & $0.96$ & $0.90$ \\
	$2449025.683$ & $1.27$ & $1.07$ & $0.97$ & $0.90$ & $2449398.660$ & $1.27$ & $1.07$ & $0.97$ & $0.90$ \\
	$2449052.654$ & $1.26$ & $1.06$ & $0.98$ & $0.91$ & $2449399.659$ & $1.27$ & $1.07$ & $0.99$ & $0.91$ \\
	$2449053.655$ & $1.26$ & $1.06$ & $0.98$ & $0.90$ & $2449406.639$ & $1.27$ & $1.07$ & $0.98$ & $0.91$ \\
	$2449054.655$ & $1.28$ & $1.07$ & $0.98$ & $0.90$ & $2449408.634$ & $1.26$ & $1.06$ & $0.98$ & $0.91$ \\
	$2449057.639$ & $1.27$ & $1.06$ & $0.98$ & $0.90$ & $2449410.628$ & $1.27$ & $1.07$ & $0.98$ & $0.91$ \\
	$2449058.638$ & $1.26$ & $1.06$ & $0.98$ & $0.90$ & $2449411.625$ & $1.28$ & $1.08$ & $0.99$ & $0.91$ \\
	$2449063.622$ & $1.26$ & $1.06$ & $0.98$ & $0.91$ & $2449416.610$ & $1.28$ & $1.08$ & $0.99$ & $0.91$ \\
	$2449065.620$ & $1.26$ & $1.06$ & $0.97$ & $0.90$ & $2449418.623$ & $1.28$ & $1.08$ & $0.99$ & $0.92$ \\
	$2449253.906$ & $1.26$ & $1.06$ & $0.99$ & $0.90$ & $2449421.607$ & $1.26$ & $1.06$ & $0.98$ & $0.91$ \\
	$2449257.895$ & $1.26$ & $1.03$ & $0.98$ & $0.91$ & $2449423.613$ & $1.30$ & $1.09$ & $1.00$ & $0.92$ \\
	$2449258.893$ & $1.26$ & $1.06$ & $0.98$ & $0.90$ & $2449424.613$ & $1.28$ & $1.07$ & $0.99$ & $0.93$ \\
	$2449259.890$ & $1.28$ & $1.05$ & $0.98$ & $0.91$ &               &        &        &        &        \\
\end{mpsupertabular*}

\end{document}